\documentclass[10pt,aps,prl,amsmath,amssymb,superscriptaddress,
onecolumn,notitlepage,nofootinbib,showpacs,preprintnumbers,
amsfonts]{revtex4-1}

\usepackage[usenames,dvipsnames,svgnames,table]{xcolor} 
\usepackage[utf8]{inputenc}
\usepackage{subfigure,lineno,wrapfig}

\usepackage{graphicx}% Include figure files
\usepackage{dcolumn}% Align table columns on decimal point
\usepackage{bm}% bold math
\usepackage{hyperref}% add hypertext capabilities
\hypersetup{ 
    pdfnewwindow=true,      % links in new window
    colorlinks=true,       % false: boxed links; true: colored links
    allcolors=[RGB]{31 119 180}
} 
\usepackage{dsfont}
\usepackage{placeins}
\usepackage{todonotes}
\usepackage[utf8]{inputenc}
\usepackage{gensymb}
\usepackage{multirow}
\usepackage{amsmath}
\usepackage{amsfonts}
\usepackage{amssymb}
\usepackage{cleveref}
\usepackage{enumitem}
\usepackage{physics}
\usepackage{array}   % for \newcolumntype macro
\newcolumntype{L}{>{$}l<{$}} % math-mode version of "l" column type
\newcolumntype{R}{>{$}r<{$}}
\newcolumntype{C}{>{$}c<{$}}
\usepackage{xspace}
\usepackage{braket}
\usepackage{units}
\usepackage{slashed}
\usepackage[normalem]{ulem}

\usepackage[format=plain,justification=RaggedRight,singlelinecheck=false]{caption}

\graphicspath{{figures/}}

\newcommand{\mevnospace}{\ensuremath{{\mathrm{\,Me\kern -0.1em V}}}}
\newcommand{\gevnospace}{\ensuremath{{\mathrm{\,Ge\kern -0.1em V}}}}
\newcommand{\tevnospace}{\ensuremath{{\mathrm{\,Te\kern -0.1em V}}}}
\newcommand{\kev}{\ensuremath{{\mathrm{\,ke\kern -0.1em V}}}\xspace}
\newcommand{\mev}{\mevnospace\xspace}
\newcommand{\gev}{\gevnospace\xspace}

\newcommand{\mevp}{\ensuremath{{\mathrm{Me\kern -0.1em V}}}}
\newcommand{\gevp}{\ensuremath{{\mathrm{Ge\kern -0.1em V}}}}
\newcommand{\tevp}{\ensuremath{{\mathrm{Te\kern -0.1em V}}}}

\newcommand{\eV}{\ensuremath{{\mathrm{\,e\kern -0.1em V}}}\xspace}
\newcommand{\gevsq}{\ensuremath{{\mathrm{\,Ge\kern -0.1em V}^2}}\xspace}
\newcommand{\nsgev}{\ensuremath{{\mathrm{Ge\kern -0.1em V}}}\xspace}
\newcommand{\nsmev}{\ensuremath{{\mathrm{Me\kern -0.1em V}}}\xspace}
\newcommand{\nskev}{\ensuremath{{\mathrm{ke\kern -0.1em V}}}\xspace}

\newcommand{\babar}{BaBar\xspace}
\newcommand{\belle}{Belle\xspace}

\newcommand{\lhcb}{LHCb\xspace}
\newcommand{\bes}{BESIII\xspace}

\newcommand{\etapi}{\ensuremath{\eta^{(\prime)}\pi}\xspace}
\newcommand{\pione}{\ensuremath{\pi_1}\xspace}

\newcommand{\etap}{\ensuremath{\eta^{\prime}}}

\newlist{todolist}{itemize}{2}
\setlist[todolist]{label=$\square$}
\usepackage{pifont}
\usepackage{mfirstuc} % to uppercase the name
\newcommand{\addReviewer}[2]{
  \expandafter\newcommand\csname #1\endcsname[1]{{\bf \color{#2} \capitalisewords{#1}:\,##1}}
  \expandafter\newcommand\csname #1cor\endcsname[2]{{\color{#2} \capitalisewords{#1}:\,\st{##1}{\bf ##2}}}
  \expandafter\newcommand\csname #1color\endcsname{#2}
}

\usepackage{tikz,marginnote} % margin notes with circles
\newcommand{\checkedby}[1]{
\ifdefined\CROSSCHECKS
  \marginnote{
    \begin{tikzpicture}
      \foreach \x [count=\xi] in {#1} {
         \node[shape=circle,inner sep=0mm,
         minimum size=2mm,
         fill=\csname \x color\endcsname] at (\xi*3mm,0) {};
      }
    \end{tikzpicture}
  }
\else
\fi
}

\usepackage{soul}

\definecolor{chromeyellow}{rgb}{1.0, 0.65, 0.0}
\definecolor{DodgeBlue}{rgb}{0.118, 0.565,1.000}
\definecolor{asparagus}{rgb}{0.53, 0.66, 0.42}
\definecolor{cadmiumgreen}{rgb}{0.0, 0.42, 0.24}
\definecolor{cardinal}{rgb}{0.77, 0.12, 0.23}

\newcommand{\jpsi}{\ensuremath{J/\psi}\xspace}

\newcommand{\KSKS}{\ensuremath{K_S^0 K_S^0}\xspace}

\def\Dbar    {\kern 0.2em\bar{\kern -0.2em D}{}\xspace}

\def\Dz      {\ensuremath{D^0}\xspace}
\def\Dzb     {\ensuremath{\Dbar^0}\xspace}
\def\DzDzb   {\ensuremath{\Dz {\kern -0.16em \Dzb}}\xspace}
\def\Dp      {\ensuremath{D^+}\xspace}
\def\Dm      {\ensuremath{D^-}\xspace}

\def\DpDm    {\ensuremath{\Dp {\kern -0.16em \Dm}}\xspace}

\def\Bbar    {\kern 0.2em\bar{\kern -0.2em B}{}\xspace}

\def\Lbar     {\kern 0.2em\overline{\kern -0.2em\Lambda\kern 0.05em}\kern-0.05em{}\xspace}

\newcommand{\SigmaD}{\ensuremath{\Sigma_c^+\bar{D}^0}\xspace}

\newcommand{\XYZ}{\ensuremath{XY\!Z}\xspace}

\newcommand\snowmass{\begin{center}\rule[-0.2in]{\hsize}{0.01in}\\\rule{\hsize}{0.01in}\\
\vskip 0.1in Submitted to the  Proceedings of the US Community Study\\ 
on the Future of Particle Physics (Snowmass 2021)\\ 
\rule{\hsize}{0.01in}\\\rule[+0.2in]{\hsize}{0.01in} \end{center}}

\begin{document}

\snowmass
%% Preamble
\title{Need for amplitude analysis in the discovery of new hadrons
}

%%%%%%%%%%%%%%%%%%%%%%%%%%%%%%%%%%

\author{Miguel~Albaladejo}
\affiliation{\ific}

\author{Marco~Battaglieri}
%\affiliation{\jlabauth}
\affiliation{\genova}

\author{Łukasz~Bibrzycki}
\affiliation{\icsup}

\author{Andrea~Celentano}
\affiliation{\genova}

\author{Igor~V.~Danilkin}
\affiliation{\mainz}

\author{Sebastian~M.~Dawid}
\affiliation{\ceem}
\affiliation{\indiana}

\author{Michael~D\"oring}
\affiliation{\gwu}

\author{Cristiano~Fanelli}
\affiliation{\mmit}

\author{C\'esar~Fern\'andez-Ram\'irez}
\email{cesar@jlab.org}
\affiliation{\uned}
\affiliation{\icn}

\author{Sergi~Gonz\`{a}lez-Sol\'is}
\affiliation{\lanl}

\author{Astrid~N.~Hiller~Blin}
\affiliation{\tubingen}

\author{Andrew~W.~Jackura}
\affiliation{\jlabth}
\affiliation{\odu}

\author{Vincent~Mathieu}
\affiliation{\ub}
\affiliation{\ucm}

\author{Mikhail~Mikhasenko}
\affiliation{\origins}
\affiliation{\lmu}

\author{Victor~I.~Mokeev}
\affiliation{\jlabauth}

\author{Emilie~Passemar}
\affiliation{\ceem}
\affiliation{\indiana}
\affiliation{\jlabth}

\author{Robert~J.~Perry}
\affiliation{\chiaotung}

\author{Alessandro~Pilloni}
\email{alessandro.pilloni@unime.it}
\affiliation{\messina}
\affiliation{\catania}

\author{Arkaitz~Rodas}
\email{arodas@wm.edu}
\affiliation{\jlabth}
\affiliation{\wandm}

\author{Matthew~R.~Shepherd}
\affiliation{\indiana}

\author{Nathaniel~Sherrill}
\affiliation{\sussex}

\author{Jorge~A.~Silva-Castro}
\affiliation{\icn}

\author{Tomasz~Skwarnicki}
\affiliation{\syracuse}

\author{Adam~P.~Szczepaniak}
\email{aszczepa@indiana.edu}
\affiliation{\ceem}
\affiliation{\indiana}
\affiliation{\jlabth}

\author{Daniel~Winney}
\affiliation{\ceem}
\affiliation{\indiana}
\affiliation{\scnuJLQM}
\affiliation{\scnuIQM}

\collaboration{Joint Physics Analysis Center}

%%%%%%%%%%%%%%%%%%%%%%%%%%%%%%%%%%

%%Affiliations
\newcommand{\catania}{INFN Sezione di Catania, I-95123 Catania, Italy}

\newcommand{\ceem}{Center for  Exploration  of  Energy  and  Matter,
Indiana  University,
Bloomington,  IN  47403,  USA}
\newcommand{\cern}{CERN, 1211 Geneva 23, Switzerland}
\newcommand{\chiaotung}{Institute of Physics, National Chiao-Tung University, 1001 Ta-Hsueh Road, Hsinchu 30010, Taiwan
}
\newcommand{\ect}{European Centre for Theoretical Studies in Nuclear Physics and related Areas (ECT$^*$) and Fondazione Bruno Kessler, Villazzano (Trento), I-38123, Italy}
\newcommand{\genova}{INFN Sezione di Genova, Genova, I-16146, Italy}
\newcommand{\gwu}{George Washington University, Washington, DC 20052, USA}
\newcommand{\hiskp}{Universit\"at Bonn,
Helmholtz-Institut f\"ur Strahlen- und Kernphysik, 53115 Bonn, Germany}
\newcommand{\icn}{Instituto de Ciencias Nucleares, 
Universidad Nacional Aut\'onoma de M\'exico, Ciudad de M\'exico 04510, Mexico}
\newcommand{\icsup}{Pedagogical University of Kraków, 30-084 Krak\'ow, Poland}
\newcommand{\indiana}{Department of Physics,
Indiana  University, Bloomington,  IN  47405,  USA}
\newcommand{\jlabth}{Theory Center, Thomas  Jefferson  National  Accelerator  Facility, Newport  News,  VA  23606,  USA}
\newcommand{\jlabauth}{Thomas  Jefferson  National  Accelerator  Facility, Newport  News,  VA  23606,  USA}

\newcommand{\mainz}{Institut f\"ur Kernphysik \& PRISMA$^+$  Cluster of Excellence, Johannes Gutenberg Universit\"at,  D-55099 Mainz, Germany}
\newcommand{\mmit}{Laboratory for Nuclear Science, Massachusetts Institute of Technology, Cambridge, MA 02139, USA}
\newcommand{\syracuse}{Syracuse University, Syracuse, NY 13244, USA}
\newcommand{\odu}{Department of Physics, Old Dominion University, Norfolk, Virginia 23529, USA}
\newcommand{\ucm}{Departamento de F\'isica Te\'orica, Universidad Complutense de Madrid and IPARCOS, 28040 Madrid, Spain}
\newcommand{\uned}{Departamento de F\'isica Interdisciplinar, Universidad Nacional de Educaci\'on a Distancia (UNED), Madrid E-28040, Spain}
\newcommand{\wandm}{Department of Physics, College of William and Mary, Williamsburg, VA 23187, USA}
\newcommand{\lanl}{Theoretical Division, Los Alamos National Laboratory, Los Alamos, NM 87545, USA}
\newcommand{\lmu}{Ludwig-Maximilian University of Munich, Germany}
\newcommand{\messina}{Dipartimento di Scienze Matematiche e Informatiche, Scienze Fisiche e Scienze della Terra, 
Universit\`a degli Studi di Messina, I-98122 Messina, Italy}
\newcommand{\origins}{ORIGINS Excellence Cluster, 80939 Munich, Germany}
\newcommand{\scnuIQM}{Guangdong Provincial Key Laboratory of Nuclear Science, Institute of Quantum Matter, South China Normal University, Guangzhou 510006, China}
\newcommand{\scnuJLQM}{Guangdong-Hong Kong Joint Laboratory of Quantum Matter, Southern Nuclear Science Computing Center, South China Normal University, Guangzhou 510006, China}
\newcommand{\ub}{Departament de F\'isica Qu\`antica i Astrof\'isica and Institut de Ci\`encies del Cosmos, Universitat de Barcelona, E-08028, Spain}
\newcommand{\ific}{Instituto de F\'isica Corpuscular (IFIC), Centro Mixto CSIC-Universidad de Valencia, E-46071 Valencia, Spain}
\newcommand{\tubingen}{Institute for Theoretical Physics, T\"ubingen University, 72076 T\"ubingen, Germany}
\newcommand{\sussex}{Department of Physics and Astronomy, University of Sussex, Brighton BN1 9QH, UK}

\begin{abstract}
We highlight the need for the development of comprehensive amplitude analysis methods to further our understanding of 
 hadron spectroscopy. 
 Reaction amplitudes constrained by first principles of $S$-matrix theory and by QCD phenomenology are needed to extract robust interpretations of the data from experiments and from lattice calculations. 
\end{abstract}

% \begin{keyword}
% {\it Preprint numbers: LA-UR-22-22335}
% \end{keyword}
\preprint{LA-UR-22-22335}
\preprint{JLAB-THY-22-3584}

\maketitle

In the last two decades, high-energy physics experiments 
have delivered a lot of unexpected exotic
hadron resonances, that challenge  the minimal quark model lore of baryons with three quarks and mesons with a quark-antiquark
pair. Candidates for tetraquarks, pentaquarks, molecules, and hadrons with gluonic degrees of freedom have been found~\cite{pdg,Esposito:2016noz,*Olsen:2017bmm,*Guo:2017jvc,*Lebed:2016hpi,*Karliner:2017qhf,*Guo:2019twa,*ali2019multiquark,*Brambilla:2019esw}.
Establishing the existence of isolated resonances that go beyond the minimal quark model is just the first step: one needs to identify the complete
multiplets and study the differences and similarities among their members. 
This would provide insights into the nature of exotic resonances and 
the inner workings of QCD in the nonperturbative regime.
However, a comprehensive and consistent picture of this sector of the spectrum is still missing. Many of these resonance candidates have been seen in just a single production and decay channel. 
The analyses at lepton colliders have been limited by statistics so far. Moreover, measurements often face complications due to the presence of multibody final states, which makes a model-independent determination of an exotic candidate difficult. 

Despite tremendous progress in understanding gauge theories, an analytic solution of QCD in the nonperturbative regime will not be available in the foreseeable future. 
At the moment, Lattice QCD represents the most rigorous tool to calculate observables from first principles, albeit numerically~\cite{Shepherd:2016dni,*Briceno:2017max}. 
However, it does not answer how the specific properties of QCD, as confinement and mass generation, emerge. Ultimately, it does not explain \emph{why} quarks and gluons organize themselves in the hadron spectrum in the way we observe it.
For this, using other approximate tools (such as functional methods~\cite{Polonyi:2001se,*Maris:2003vk}) and models of QCD (as the quark model, or the holography-inspired description~\cite{Plessas:2015mpa,*Brodsky:2014yha}) is required. 
Together with these top-down approaches, bottom-up strategies are also feasible. We know indeed that any reaction amplitude in QCD must satisfy a set of general principles, such as unitarity, analyticity, crossing, and Lorentz symmetries, as well as the specific symmetries of the strong interactions~\cite{Eden:1966dnq,*Martin:1970,*Gribov:2009zz}. One can thus write ans\"atze that follow these principles as much as possible, at least in a given kinematical domain, and fit to data. 
If the amplitude model space is large enough, the resonance properties obtained will be as unbiased as possible.

In this White Paper, we will discuss what is needed to identify and obtain the physical properties of hadron resonances within the wealth of experimental data that has been produced in recent years and is expected to be produced in the current and forthcoming experiments. We will focus on the activity of the Joint Physics Analysis Center (JPAC) as an example of collaboration between theorists and experimentalists, and highlight the most interesting lines of development for the field in the future.

%%%%%%%%%%%%%%%%%%%%%%%%%%%%%%%%%%%%
%	Reaction/Lineshape
%%%%%%%%%%%%%%%%%%%%%%%%%%%%%%%%%%%%
\section{Reaction theory and lineshape studies} \label{sec:reaction}

The excited spectrum of QCD is composed of states with lifetimes $\lesssim 10^{-21}\text{~s}$, which need to be reconstructed from the energy and angular dependence of their decay products. 
The measured rates are proportional to the modulus squared of the {\em reaction amplitude}, which encodes the information at the quantum level.
While the reaction amplitude's angular dependence is determined by the spin of the particles involved, the energy behavior is dynamical. 

Despite fifty years of efforts, we do not have a constructive solution of QCD, %\arkaitz{Models ppl will complain about what's next}
nor a simple connection between the interaction at the quark- and hadron-level. 
Nevertheless, even if no theory of strong interactions was available, the underlying $S$-matrix must satisfy certain properties. Lorentz invariance requires that the $S$-matrix elements, and therefore amplitudes, depend on particle momenta only through the Mandelstam invariants.
Analyticity (stemming from causality), unitarity (from probability conservation), and crossing symmetry (proper of relativistic quantum theories) constitute the so-called $S$-matrix principles. 
There is a renewed interest in what one can learn from amplitude properties alone, and if possible, to constrain the space of feasible solutions rather than to look for a unique one. 
The new program is thus to postulate {\it ans\"atze} for the amplitudes that depend on a finite number of parameters and fit them to data.
Ideally, one requires the amplitudes to fulfill the constraints given by the $S$-matrix principles, to obtain physical results as sound as possible. It should be stressed, however, that implementing all the constraints simultaneously is extremely difficult, and the problem has to be approached on a case-by-case basis, in order to enforce the constraints that are most relevant for the physics at hand. 

%\ale{lineshape}

\subsection{The light sector}
The light hadron sector
has been subject to fierce debate for many decades. Resonances are generally broad and overlap each other; experimental analyses were limited by statistics and often implemented simplistic methods. All these issues hindered the extraction of robust information. The natures and in some cases even the existence of some states are still under debate~\cite{pdg}.

Quark models play a crucial role in guiding the analysis, predicting the number and properties of states to search for~\cite{Godfrey:1985xj,*Capstick:1985xss}. However, since we are entering an era of high-statistics experiments, we are now facing the limits of such models. A complementary path was followed with effective field theories having hadrons as degrees of freedom, in particular Chiral Perturbation Theory ($\chi$PT)\cite{Weinberg:1978kz,*Weinberg:1990rz,*Jenkins:1990jv}.
The low energy constants at a given order can be fixed from experimental data~\cite{Gasser:1983yg, *Gasser:1984gg,*Epelbaum:2008ga}.
However, fixed order effective theories respect unitarity only perturbatively, and cannot produce resonance poles, if not explicitly incorporated. This problem was circumvented by various unitarization methods (U$\chi$PT), for a recent review see~\cite{Oller:2020guq}),
at least in the low-energy region. 
Nevertheless, these methods still suffer from several model dependencies and approximations. This becomes particularly clear when dealing with light scalars, where all the $S$-matrix principles play a significant role. This is the main reason why dispersive approaches~\cite{Roy:1971tc,Hite:1973pm} have been gaining attention in recent years~\cite{Pelaez:2015qba}. The combination of dispersion relations with experimental data is able to provide us the most robust information about the lightest mesons~\cite{Caprini:2005zr,GarciaMartin:2011nna,Moussallam:2011zg, Pelaez:2020uiw}. Unfortunately, partial wave dispersive analyses are usually applicable only up to $\sim 1\gev$. At a practical level, most of the data at higher energies come from photo-, electro- and hadroproduction, heavy meson decays, peripheral production, or $e^+e^-$ annihilations. Furthermore, the large number of open channels available makes the rigorous application of unitarity unfeasible.
For these reasons, loosening the $S$-matrix constraints, and studying a number of phenomenological amplitudes to assess the systematic uncertainties and reduce the model bias seems the appropriate path to follow. In the following, we will discuss some recent successful examples that follow this philosophy and highlight some of the open questions that we need to address in the future.

\begin{figure}[t]
\centering
\includegraphics[width=0.9\textwidth]{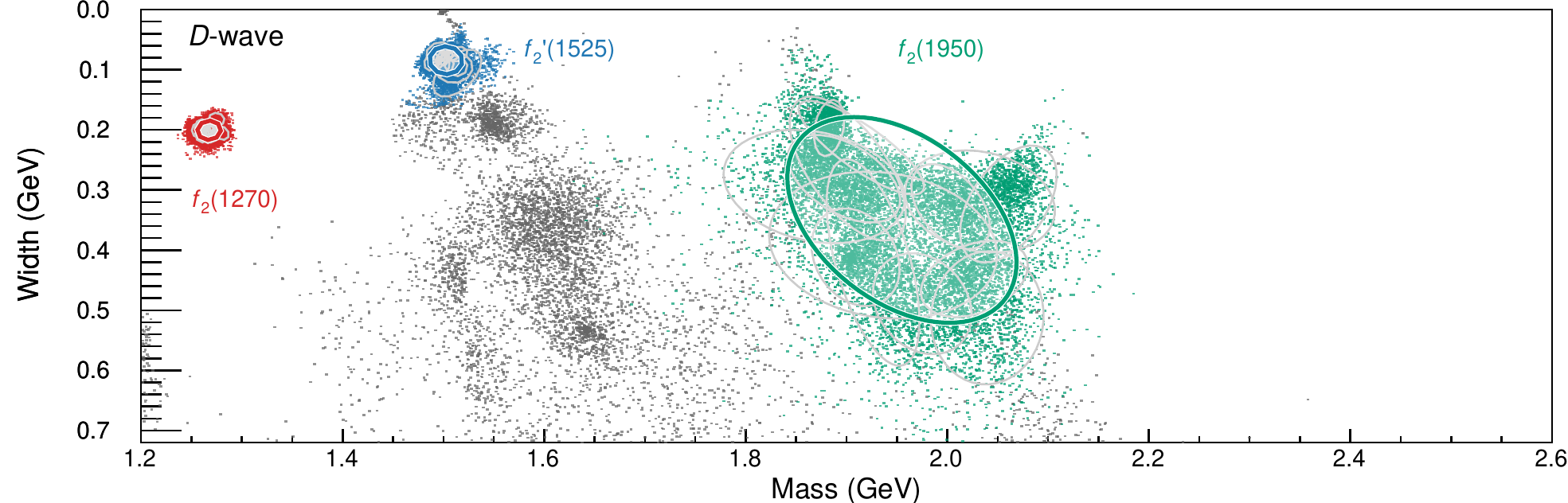} \\ \includegraphics[width=0.9\textwidth]{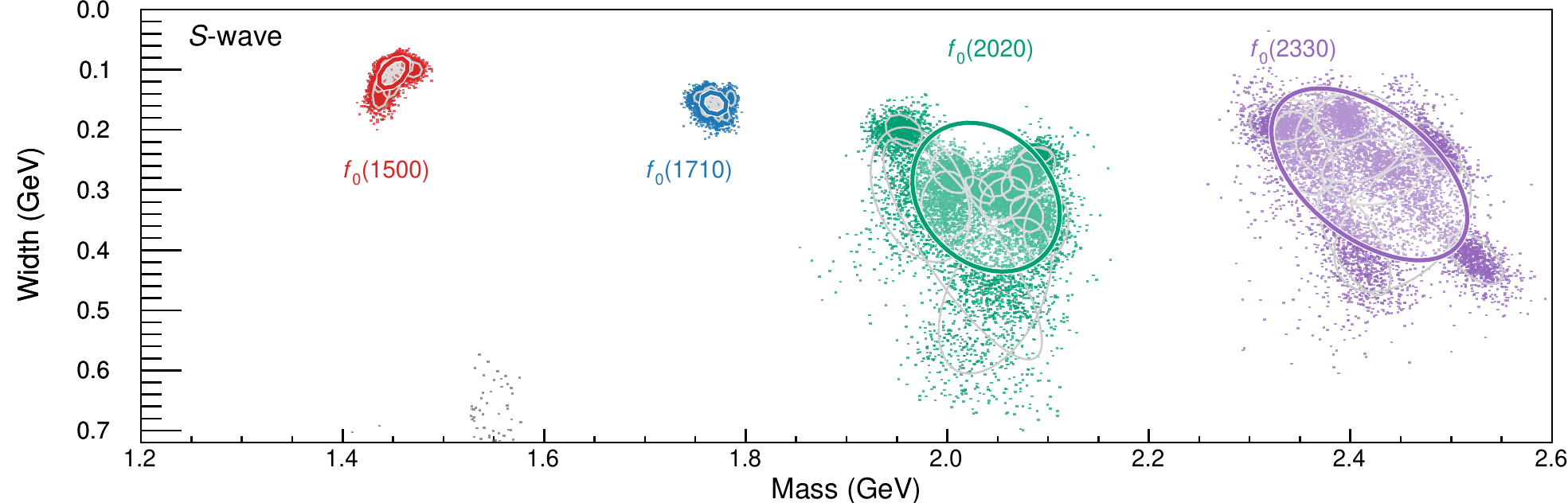} 
\caption{Scalar and tensor resonances in $J/\psi$ radiative decays. Gray points are identified as spurious model artefacts. For each physical resonance and systematic, gray ellipses show the $68\%$ confidence region. Colored ellipses show the final average of all systematic model variations. Figure from~\cite{Rodas:2021tyb}.
}
\label{fig:finalpoles}
\end{figure}

There are several interesting topics in the light sector. The most fundamental questions concern the existence of resonances where gluons play the role of constituents, as glueball or hybrid mesons~\cite{Mathieu:2008me,Llanes-Estrada:2021evz,Meyer:2015eta}. 
The isoscalar-scalar mesons, and -tensor mesons to some extent, have played a central role in this. 
They can mix with the lightest glueball with the same quantum numbers. In pure Yang-Mills, the spectrum is populated by glueballs, the lightest one expected to be around $1.5$--$2\gev$~\cite{Morningstar:1999rf,Szczepaniak:2003mr,Athenodorou:2020ani}. In nature, glueball production is expected to be enhanced in processes where quarks annihilate into gluons, like $p\bar p$ collisions or \jpsi radiative decays. 
Most of the literature traces the existence of a significant glueball component with the emergence of a supernumerary state with respect to how many are predicted by the quark model~\cite{Mathieu:2008me}.
In particular the $f_0(1370)$, $f_0(1500)$, $f_0(1710)$ in the $1.2$--$2\gev$ region are one more than expected by the quark model, which stimulated an intense work to identify one of them as the glueball.

The \jpsi radiative decays to $\pi^0\pi^0$ and $\KSKS$ were measured with high precision by BESIII~\cite{BESIII:2015rug,BESIII:2018ubj}. The partial waves have been analyzed by JPAC according to the bottom-up philosophy mentioned earlier~\cite{Rodas:2021tyb}. 
To assess the model dependence realistically, results were given for 14 different amplitude parametrizations respecting the $S$-matrix principles as much as possible.  Unitarity was first enforced strictly on the two channels available, then extended to a third unconstrained $\rho\rho$ channel, which is known to contribute substantially to the resonances in this region. Four scalar and three tensor resonances were identified, and a robust estimate of their statistic and systematic uncertainties was given. The situation is summarized in Fig.~\ref{fig:finalpoles}.
The four lightest resonances are determined with great accuracy, which makes it possible to study their couplings. The $f_2(1270)$ and $f_2'(1525)$ couple largely to $\pi\pi$ and $K\bar K$, respectively, as expected by their quark model assignments. In the scalar sector, it seems that the $f_0(1710)$ appears in $\jpsi \to \gamma f_0$ more strongly than the $f_0(1500)$. This affinity of the $f_0(1710)$ to the gluon-rich initial state, together with a coupling to $K\bar K$ larger by one order of magnitude, are hints for a sizeable glueball component.

\begin{figure*}[t]
\includegraphics[width=0.45\textwidth]{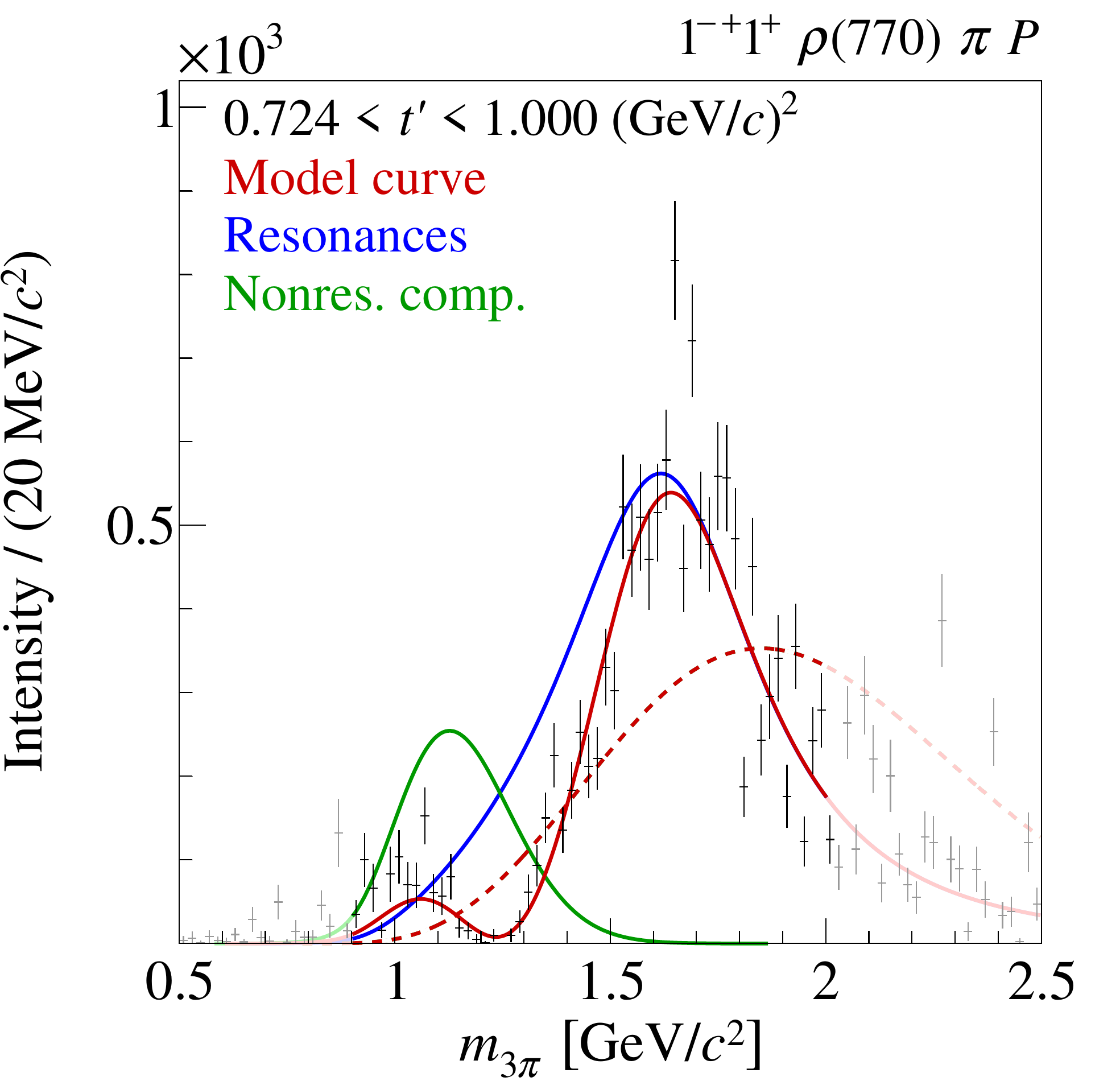} \includegraphics[width=0.45\textwidth]{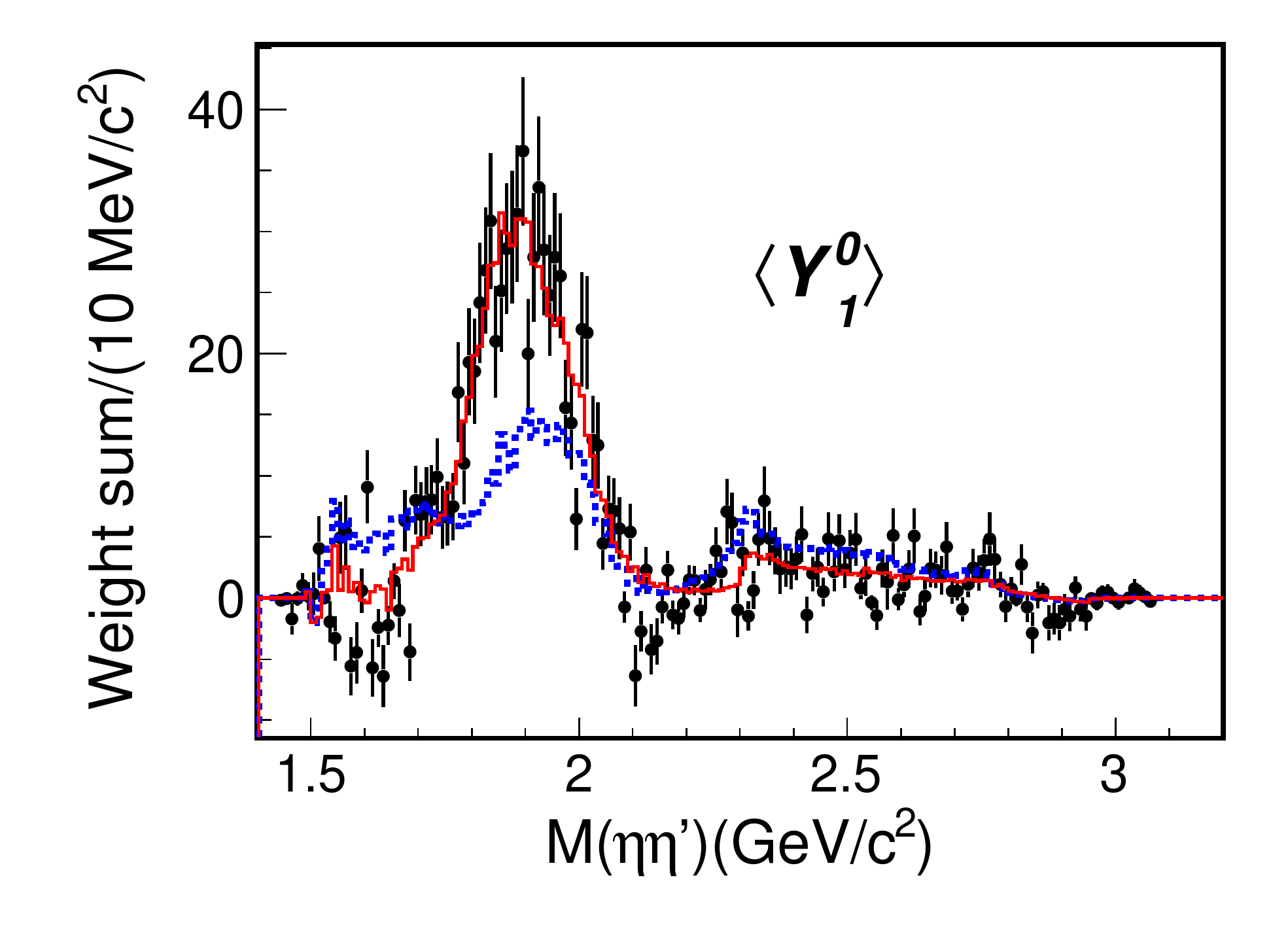}
  
    \caption{Left panel: Intensity of the exotic $1^{-+}$ $\rho \pi$ $P$-wave from COMPASS~\cite{COMPASS:2021ogp}. A resonant $\pi_1(1600)$ is required to describe data at high $t'$.
    Right panel: $\langle Y^{0}_{1}\rangle$ moment of the $\jpsi\rightarrow\gamma \eta \etap$ decay as a function of the $\eta \etap$ mass, from BESIII~\cite{BESIII:2022riz}. The strength of the signal is driven by an exotic isoscalar $\eta_1(1855)$.}

  \label{fig:hybridscompassbes}
\end{figure*}

\begin{figure}
\includegraphics[width=0.95\textwidth]{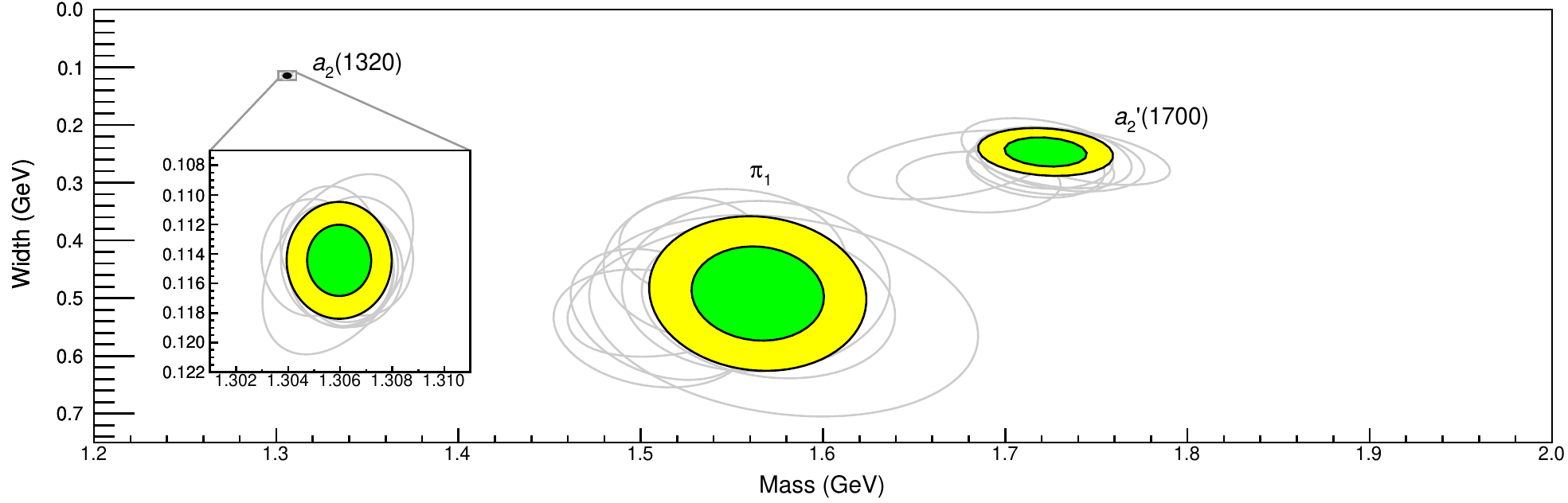}
\caption{Positions of the poles identified as the $a_2(1320)$, \pione, and $a_2'(1700)$. The inset shows the position of the $a_2(1320)$. The green and yellow ellipses show the $1\sigma$ and $2\sigma$ confidence levels, respectively. The gray ellipses in the background show, 
 within $2\sigma$, the different pole positions produced by model variations.
Figure from~\cite{JPAC:2018zyd}. 
 }
 \label{fig:pionepoles}
\end{figure}

Hybrid mesons also include gluonic degrees of freedom, which permits to reach quantum numbers forbidden for $q\bar q$ states, as $J^{PC} = 1^{-+}$. The isovector is supposed to couple to the $\eta^{(\prime)} \pi$ system. 
 The first reported hybrid candidate was the $\pi_1(1400)$ decaying into $\eta \pi$~\cite{pdg}
was claimed to appear $\sim200\mev$ heavier in the $\rho \pi$ and $\eta'\pi$ channels
While the $\pi_1(1600)$ is closer to  theoretical expectations, having two nearby $1^{-+}$ hybrids below 2\gev is problematic~\cite{Close:1987aw}.
Establishing whether there exist one or two exotic states in this mass region 
is thus a stringent test for our understanding of QCD in the non-perturbative regime.
The lowest partial waves of  $\pi p\to \etapi p$ by COMPASS~\cite{COMPASS:2014vkj} were analyzed by JPAC in~\cite{JPAC:2017dbi,JPAC:2018zyd}.
The production amplitude was parametrized following the $N/D$ formalism~\cite{Chew:1960iv}
, which allows to separate the resonance physics from the background processes in a way consistent with the $S$-matrix constraints.
The dominant $D$-wave of $\eta \pi$ was first studied in the single-channel analysis~\cite{JPAC:2017dbi}, 
that identified the $a_2(1320)$ and its radial excitation $a_2'(1700)$.
The analysis was then extended to coupled channels and to the exotic $P$-wave, to investigate the $\pi_1$ hybrid candidates.
with the $\eta^{(\prime)}\pi$ data from COMPASS. 
The best fit of the nominal model describes well both the $\pi_1(1400)$ and $\pi_1(1600)$ peaks. However, the $P$-wave amplitude exhibits a single pole, that corresponds to a single hybrid state as expected by theoretical arguments. Possible additional poles have scarce significance and unstable upon systematic checks. The result of the nominal model and of the systematic variations is shown in Fig.~\ref{fig:pionepoles}. The same conclusion is reached by a recent Lattice QCD calculation, albeit at unphysical pion masses~\cite{Woss:2020ayi}. A similar parametrization was also adopted by a subsequent analysis that combines COMPASS and Crystal Barrel data, confirming JPAC's findings~\cite{Kopf:2020yoa}. These analyses reconcile a longstanding disagreement between our theoretical understanding of the light sector and the experimental data.

In the future, these coupled channel analyses will become the standard approach to
unravel physics from data. Focusing on the hybrid sector, a major step forward will be studying the high statistics $3\pi$ system by COMPASS in a complete analytic and unitary framework~\cite{COMPASS:2018uzl,COMPASS:2021ogp}. 
This will require to understand  
nontrivial production mechanisms, that strongly depend on the momentum transferred, and can distort the resonance line shapes (see Sec. \ref{sec:production}). Completing the study of the hybrid multiplets is also a long-term goal, and the claim by BESIII of the isosinglet partner of the $\pi_1$ opens this new direction (see Fig.~\ref{fig:hybridscompassbes}).

\subsection{The heavy sector}

\begin{figure}[t]
\begin{center}
\includegraphics[width=0.8\textwidth]{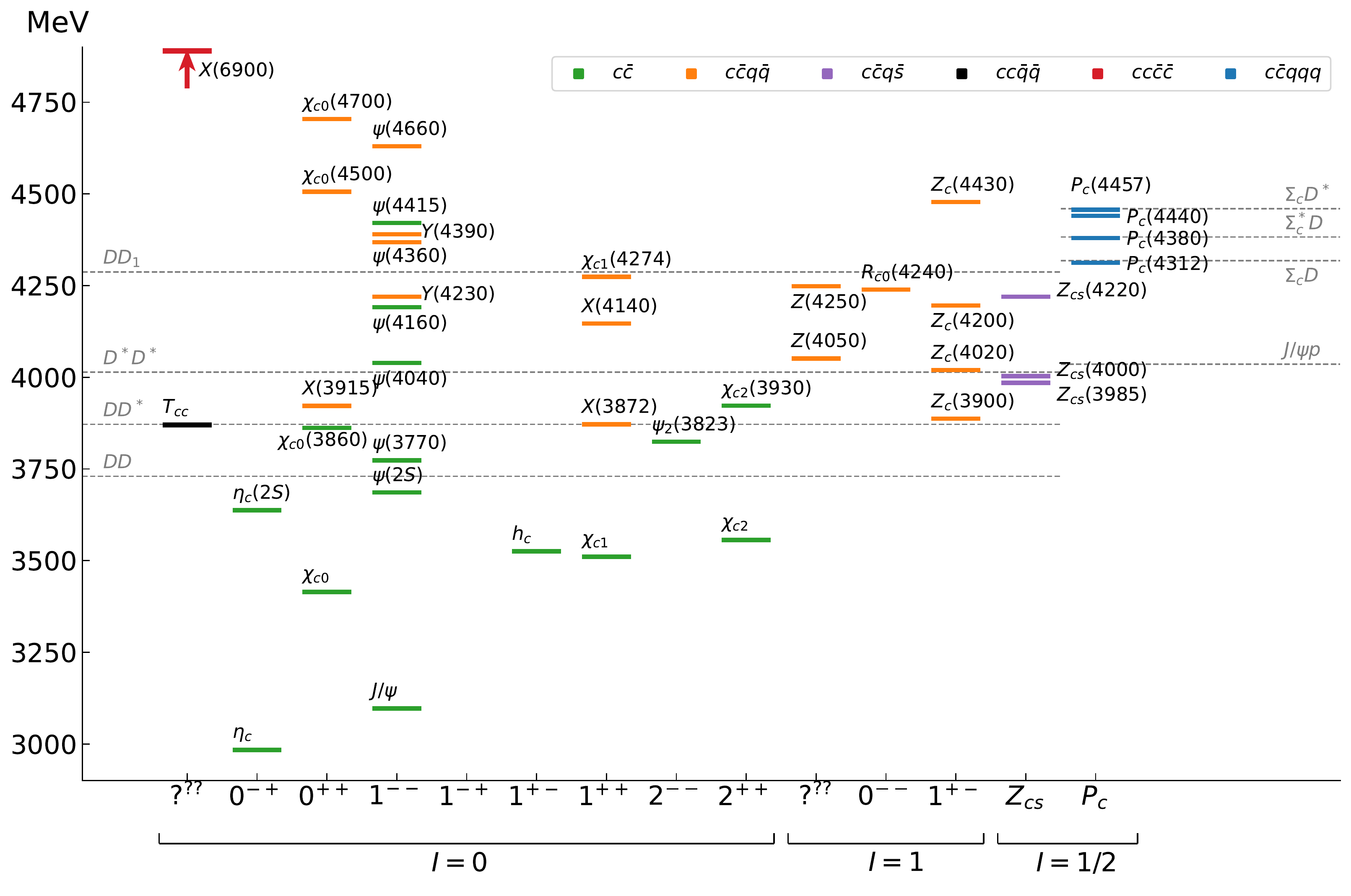}
\end{center}
\caption{Summary of ordinary charmonia, \XYZ and pentaquarks from PDG~\cite{pdg}. Figure from~\cite{JPAC:2021rxu}.}\label{fig:charmonia_et_exotica}
\end{figure}

The unexpected discovery of the $X(3872)$ in 2003 ushered in a new era in hadron spectroscopy~\cite{Belle:2003nnu}. Experiments have claimed a long list of states, collectively called \XYZ, that appear mostly in the charmonium sector, but do not respect the expectations for ordinary $Q\bar Q$ states. 
An exotic composition is thus likely required~\cite{Olsen:2017bmm,Brambilla:2019esw}. Several of these states appear as relatively narrow peaks in the proximity of open charm thresholds, suggesting that hadron-hadron dynamics can play a role in their formation~\cite{Guo:2017jvc}. Alternatively, QCD-like models also predict the existence of supernumerary states, by increasing the number of quark/gluon constituents~\cite{Esposito:2016noz}. The recent discovery of a doubly-heavy $T_{cc}^+$~\cite{LHCb:2021vvq,LHCb:2021auc} and of a fully-heavy $X(6900)$~\cite{LHCb:2020bwg} states make the whole picture extremely rich. The current status of charmonia is summarized in Fig.~\ref{fig:charmonia_et_exotica}.   
Having a comprehensive description of these states  
   will improve our
understanding of the nonperturbative features of QCD. Most of the analyses from \belle and \babar suffered from limited statistics, and strong claims were sometimes made with simplistic models on a handful of events. Currently running experiments like \lhcb and \bes have overcome this issue, providing extremely precise datasets,
that also require more sophisticated analysis methods and theory inputs. 
Depending on their width and the production mechanism, the states can roughly be classified in narrow states produced in  $b$-hadron decays and at $e^+e^-$ colliders, broad states produced in $b$-hadron decays, and states produced promptly at hadron machines.
The narrow signals do not require a thorough understanding of interferences with the background. Since they often appear close to some open flavor threshold they call for analysis methods that incorporate such information and, to some extent, it is possible to give model-independent statements.

The $X(3872)$ is very special. It has $J^{PC}=1^{++}$, violates isospin substantially decaying into $\jpsi \rho$ and $\jpsi \omega$ with similar rates, and lies exactly at the $\bar D^0 D^{*0}$ threshold. Its lineshape was recently studied by \lhcb~\cite{LHCb:2020xds}, which triggered several discussions~\cite{Esposito:2021vhu,*Baru:2021ldu}. The $Z_c(3900)$ (with $=1^{+-}$) was seen as a peak in the $\jpsi\,\pi$ invariant mass in the $e^+ e^- \to  \jpsi\,\pi \pi$ process, and as an enhancement at the $D\bar D^*$ threshold in $e^+ e^- \to  \pi D \bar D^*$. Similarly, a $Z_c'(4020)$ with same quantum numbers peaks in $h_c\,\pi$ invariant mass in the $e^+ e^- \to  h_c\,\pi \pi$ process, and enhances the cross section at the $D^*\bar D^*$ threshold in $e^+ e^- \to  \pi D \bar D^*$. The system of two $1^{+-}$ at the two thresholds seems replicated in the bottomonium sector, by the $Z_b(10610)$ and $Z_b^\prime(10650)$. The proximity to threshold motivated their identification as hadron molecules~\cite{Tornqvist:1993ng,*Braaten:2003he,*Voloshin:2003nt,*Close:2003sg,*Swanson:2006st}, but tetraquark interpretations are also viable~\cite{Maiani:2004vq,*Ali:2011ug,*Ali:2014dva}. 
The discovery of pentaquark candidates in $\Lambda_b^0 \to J/\psi p K^-$ decay in 2015 also produced effervescent theoretical efforts. The \lhcb collaboration reported a narrow and a broad state, the $P_c(4450)$ and the $P_c(4380)$, with likely opposite parities~\cite{Aaij:2015tga}. The subsequent 1D analysis in 2019, with ten-times higher statistics, reported a composite structure of the narrow peak, that splits into $P_c(4440)$ and $P_c(4457)$, and found a new isolated peak, the $P_c(4312)$~\cite{LHCb:2019kea}. The signals have been explained as compact five-quark states~\cite{Maiani:2015vwa,*Lebed:2015tna,*Anisovich:2015cia,*Ali:2019npk} or weakly bound meson-baryon molecules~\cite{Chen:2015loa,*Chen:2015moa,*Roca:2015dva,*Guo:2019fdo,*Guo:2019kdc,*Liu:2019tjn}. The effect of rescattering with the bachelor particle in a 3-body decay, enhanced by the so-called ``triangle singularities'', can also mimic the presence of a resonance~\cite{Szczepaniak:2015hya,*Meissner:2015mza,*Mikhasenko:2015vca,*Guo:2016bkl,*Bayar:2016ftu,*Guo:2019twa,*Nakamura:2021qvy}.

\begin{figure*}[t]
\includegraphics[width=0.40\textwidth]{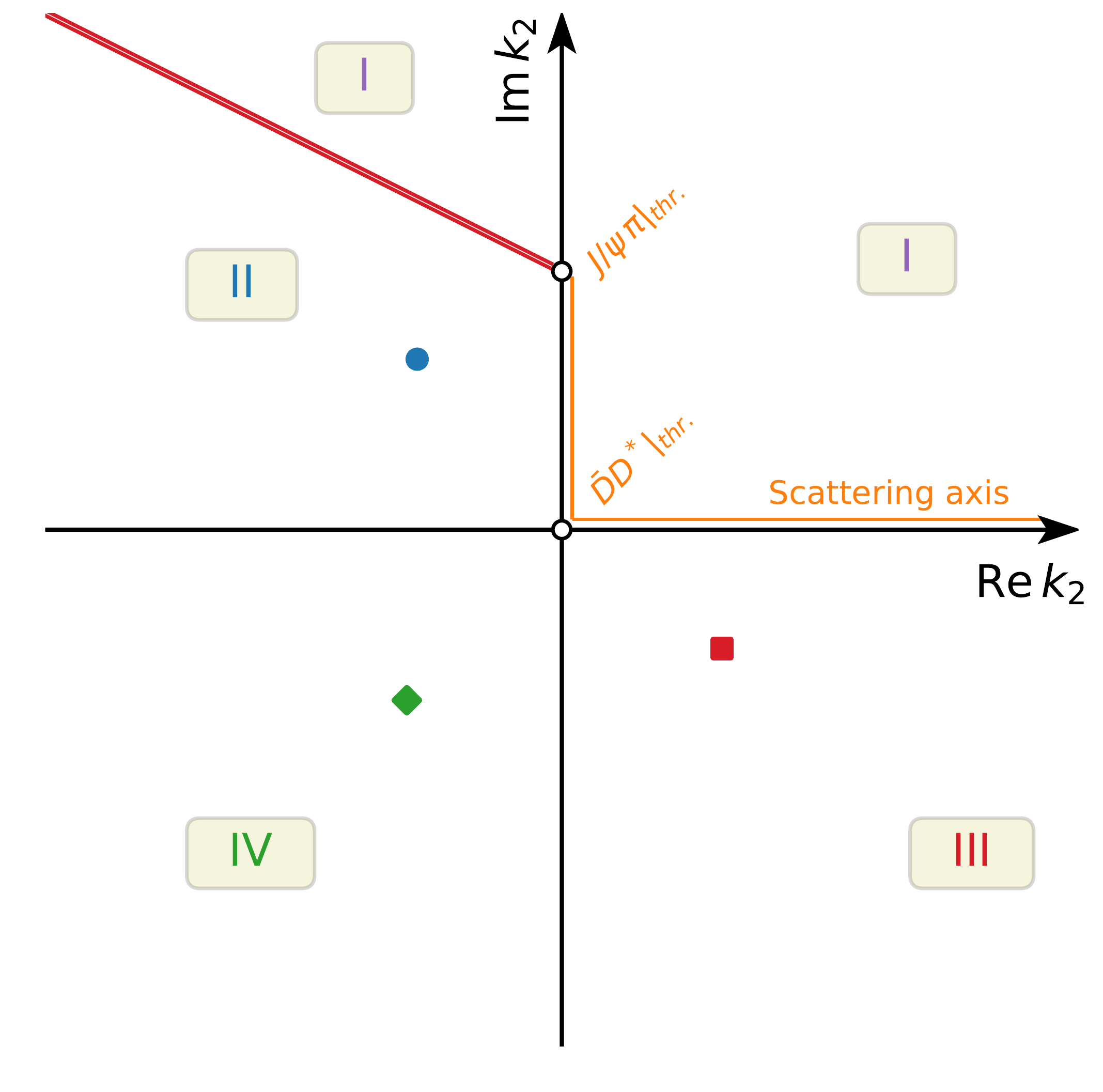}\hspace{.5cm}
\includegraphics[width=0.54\textwidth]{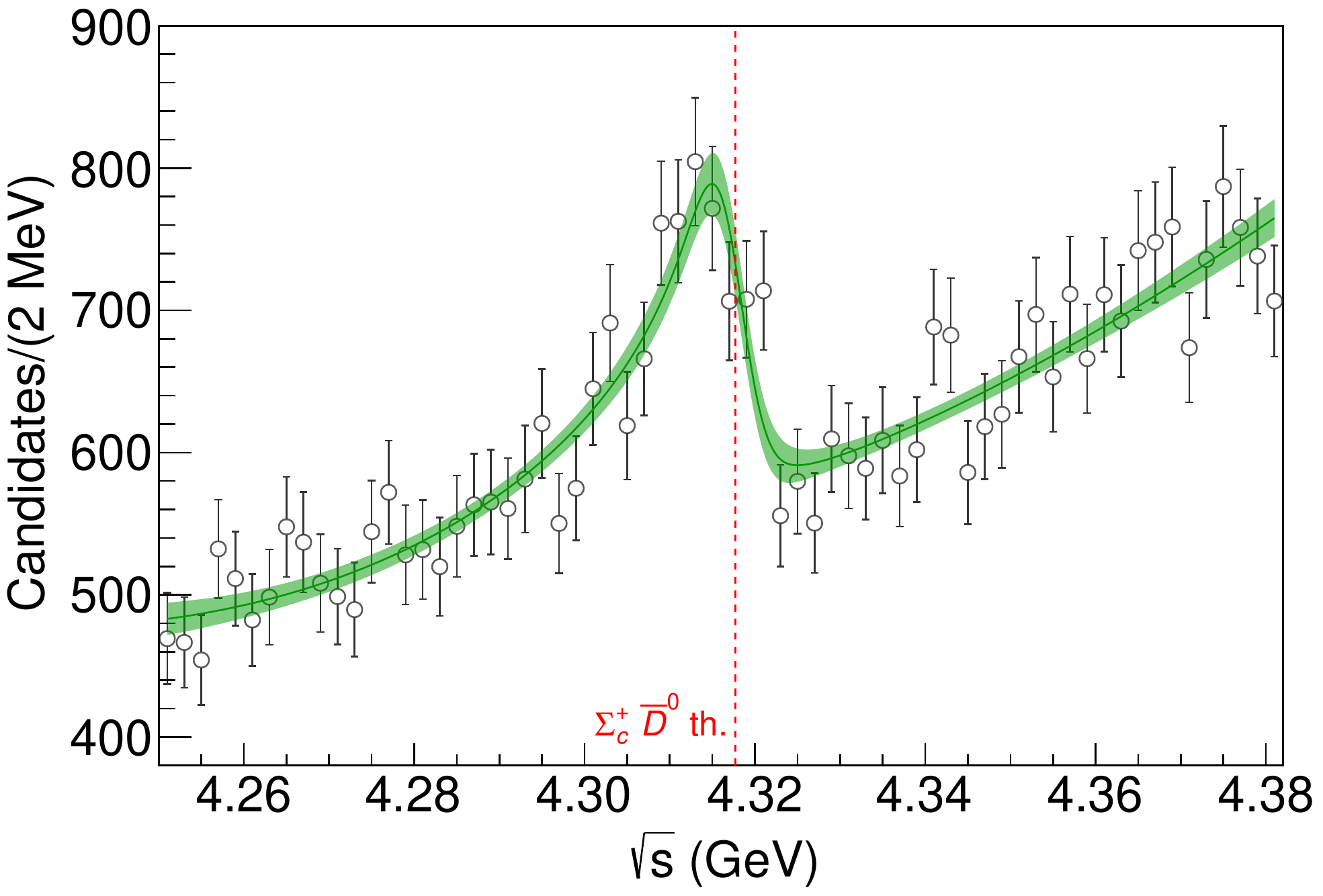}

     \caption{Left panel: 
     Schematic representation of the $Z_c(3900)$ amplitude near the 
    $\bar D D^*$ threshold for complex values of energy. The adjacent Riemann sheets are continuously connected along the axes. Four possibilities for a resonant pole structure are depicted. 
    A pole on the III sheet above the $DD^*$ threshold (red square) generates a usual Breit-Wigner-like lineshape and is likely due to a genuine QCD resonance. A pole on the II sheet below threshold (blue circle) is likely due to a bound state of $\bar DD^*$. Similarly, a pole on the IV sheet is not immediately visible on the physical region (orange), but enhances the threshold cusp. This is likely due to a virtual state. Figure from~\cite{JPAC:2021rxu}. Right panel: Fits to the $J/\psi\,p$ mass distribution from LHCb~\cite{LHCb:2019kea} in the scattering length approximation. The solid line and green band show the result of the fit and the $1\sigma$
confidence level provided by the bootstrap analysis, respectively. Figure from~\cite{Fernandez-Ramirez:2019koa}. }\label{fig:complexplaneZc}
\end{figure*}

The $Z_c(3900)$ and $P_c(4312)$ provide excellent examples of how to
approach the study of signals close to thresholds in the heavy sector.
The latter is particularly interesting as it appears as a very clean narrow isolated  structure
that peaks approximately $5\mev$ below the \SigmaD\ threshold,
making it a prime candidate
for a hadron molecule composed 
of the two particles. 
The opening of a threshold can also generate a virtual state~\cite{Eden:1964zz},
when the interaction is attractive and generates a signal
in the cross section, but is not strong enough to bind a state.
A well-known example is in neutron-neutron scattering, where the cross section is enhanced at threshold, despite no dineutron bound state existing~\cite{Hammer:2014rba}. Compact pentaquark assignments cannot be excluded either.
Each microscopic interpretation reflects into the analytic properties of the amplitude, and thus into the details of the line shape of the state.
This is schematically represented in Fig.~\ref{fig:complexplaneZc}, where the pole position appear in different Riemann sheets, suggesting different interpretations.

\begin{figure}
    \includegraphics[width=.48\textwidth]{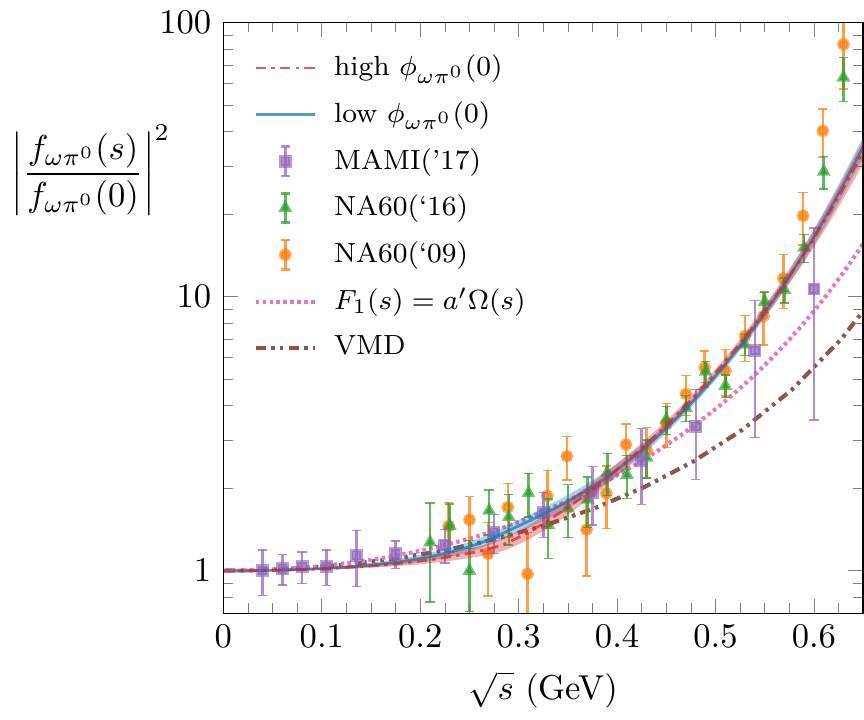}
    \raisebox{.2cm}{\includegraphics[width=.48\textwidth]{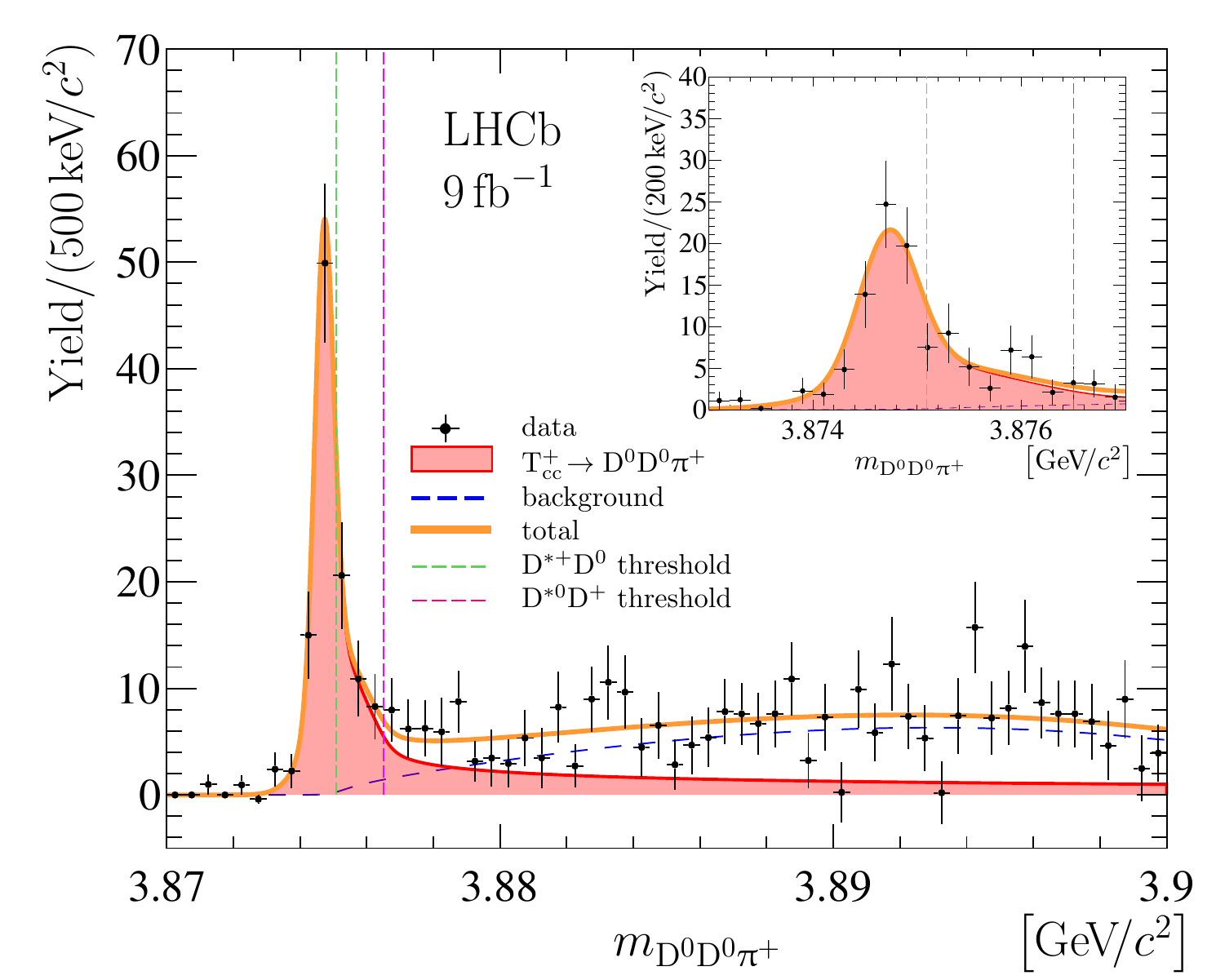}}
      \caption{Left panel: Transition form factor squared, $\left\lvert f_{\omega\pi^0}(s) \right\rvert^2/\left\lvert f_{\omega\pi^0}(0) \right\rvert^2$. Data are taken by A2 and NA60~\cite{Adlarson:2016hpp,Arnaldi:2009aa,*Arnaldi:2016pzu}. The results of Ref.~\cite{JPAC:2020umo} are shown by the red and blue bands, corresponding to the high and low $\phi_{\omega\pi^0}(0)$ phase fits.
    Figure from~\cite{JPAC:2020umo}. Right panel: $D^0 D^0 \pi^+$ spectrum from LHCb showing the prominent $T_{cc}^+$ signal. 
      Figure from~\cite{LHCb:2021auc}. }
       \label{fig:3body}
\end{figure}

For the pentaquark case, several analyses point to a bound state assignment~\cite{Du:2019pij,Du:2021fmf}. However, this can be biased by the fact that such bound states are built in the amplitude model adopted to fit the line shape. It is thus crucial to have complementary studies that keep the model assumptions to the bare minimum and investigate what data alone can tell us about the nature of the signal. This was done in~\cite{Fernandez-Ramirez:2019koa} following the bottom-up approach. The amplitude is expanded model-independently at the \SigmaD threshold. If one expands at the lowest order, the amplitude features either bound or virtual states, depending on the sign of one of the amplitude parameters. A proper statistical analysis favors the virtual state interpretation at the $2.7\sigma$ level. The same analysis was performed using Neural Network in~\cite{Ng:2021ibr}, as we will discuss in Section~\ref{sec:tools}.
Similar scenarios hold for the $Z_c(3900)$, with the addition of a triangle singularity closeby that can play a major role. Data are available for two different final states, although data quality is not as good as for the $P_c(4312)$. The analysis was done in~\cite{Albaladejo:2015lob,Pilloni:2016obd}, but statistics prevents from drawing strong conclusions. The lineshape of $T_{cc}^+$ in $D^0 D^0 \pi^+$ is also known with great detail, and will be discussed in the context of 3-body interactions in Section~\ref{sec:three}.

For the future, two complementary actions can be taken. On one hand, one has to verify the model predictions, in particular as for the existence of multiplets, or of flavor/spin partners. On the other hand, more and more precise data allow for lineshape studies as the examples given here. In this respect, $\overline{\text{P}}$ANDA will greatly improve the present limits of detector resolution, allowing us to sort the $XYZ$ puzzle out.

\section{3-body problem}\label{sec:three}

In recent years, the problem of describing multihadron scattering processes has generated significant interest. It is a known fact that most resonances couple strongly to three or more particles~\cite{pdg}. Some of these are exotics, as they do not fit the na\"ive quark model expectations, like the Roper resonance $N(1440)$, the $a_1(1420)$ seen by COMPASS~\cite{COMPASS:2015kdx,COMPASS:2020yhb}, and the hybrid candidate $\pi_1(1600)$ discussed above~\cite{COMPASS:2009xrl, COMPASS:2021ogp}. In the heavy sector, several \XYZ states have significant three-particle decay modes, most notably the $X(3872)$ and the $T_{cc}^+$~\cite{Brambilla:2019esw}. Three-body couplings might lead to non-standard line shapes and complicated structure of the amplitudes~\cite{Aitchison:1979fj,Szczepaniak:2015hya}, allowing for ambiguities in interpretations of the hadron of interest~\cite{Szczepaniak:2015eza,Nakamura:2019btl,Olsen:2017bmm,Guo:2019twa}.
The spin structure also contains theoretical subtleties that have been often overlooked in the experimental analyses~\cite{JPAC:2017vtd,*JPAC:2018dfc,Albaladejo:2019huw,JPAC:2019ufm,*Wang:2020giv}.

When dealing with three-body decays at low energies, the most rigorous formalism is the Khuri-Treiman (KT) equations~\cite{Khuri:1960zz}. 
The method has been validated 
against $\pi\pi$ scattering data~\cite{Albaladejo:2018gif}
and is extensively applied in the study of the isospin breaking $\eta\rightarrow 3\pi$~\cite{Guo:2015zqa,*Guo:2016wsi,*Colangelo:2016jmc, *Albaladejo:2017hhj}, and several other reactions~\cite{Niecknig:2012sj, *Danilkin:2014cra,*Niecknig:2015ija,*Isken:2017dkw,*Niecknig:2017ylb}. Among the various applications, the decay of $\omega,\phi \to 3\pi$ serves as one of the benchmark cases for dispersive formalisms. An example of a recent combined analysis  of $\omega\to3\pi$ and of the $\omega\pi^0$ transition form factor by JPAC is shown in Fig.~\ref{fig:3body}, which points to the need of adding more subtractions than what is required by the minimal sum rules~\cite{JPAC:2020umo}.
One of the most important applications is the precise calculation of the muon anomalous magnetic moment $(g-2)_\mu$, whose large discrepancy between theory~\cite{Aoyama:2020ynm} and experiment~\cite{Muong-2:2021ojo} is a promising signal of New Physics.
The reaction $\gamma^* \to 3\pi$ can be built similarly to the previous ones~\cite{Hoferichter:2014vra, *Hoferichter:2018kwz, *Hoferichter:2018dmo,Hoferichter:2019mqg}, and gives the second-largest individual contribution to the Hadron Vacuum Polarization~\cite{Muong-2:2021ojo}. It also enters the doubly virtual pion transition form factor~\cite{Hoferichter:2018kwz,*Hoferichter:2018dmo}, which in turn gives the leading contribution to the hadronic light-by-light~\cite{Colangelo:2014pva}. 
These techniques can be extended to 3-body heavy meson decays, which are of interest for flavor physics. In this case, the larger phase-space might require extending the formalism to include Regge-like contributions.

\begin{figure}[t]
\raisebox{-1cm}{\includegraphics[width=.45\textwidth]{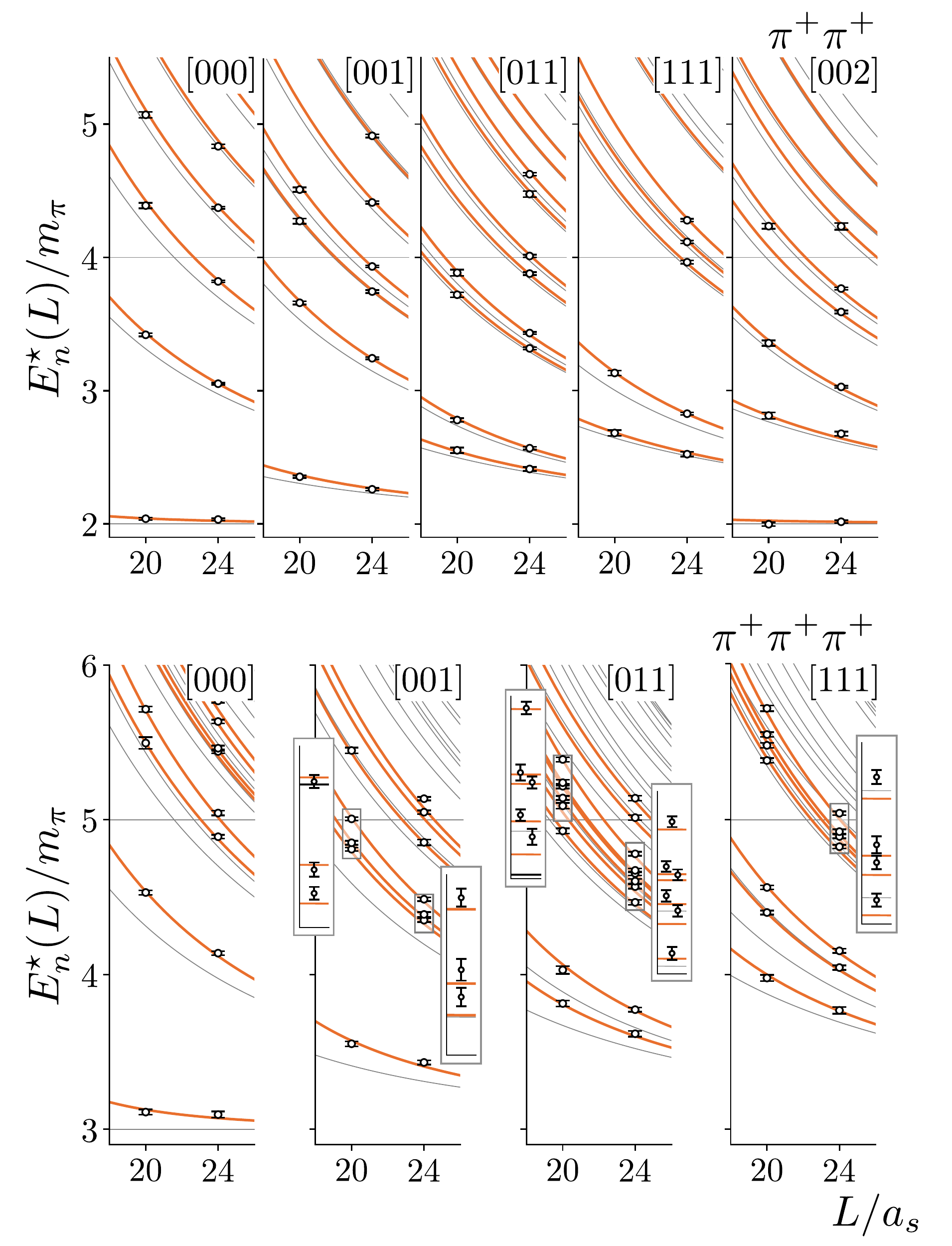}} \hspace{1cm}
\includegraphics[width=.45\textwidth]{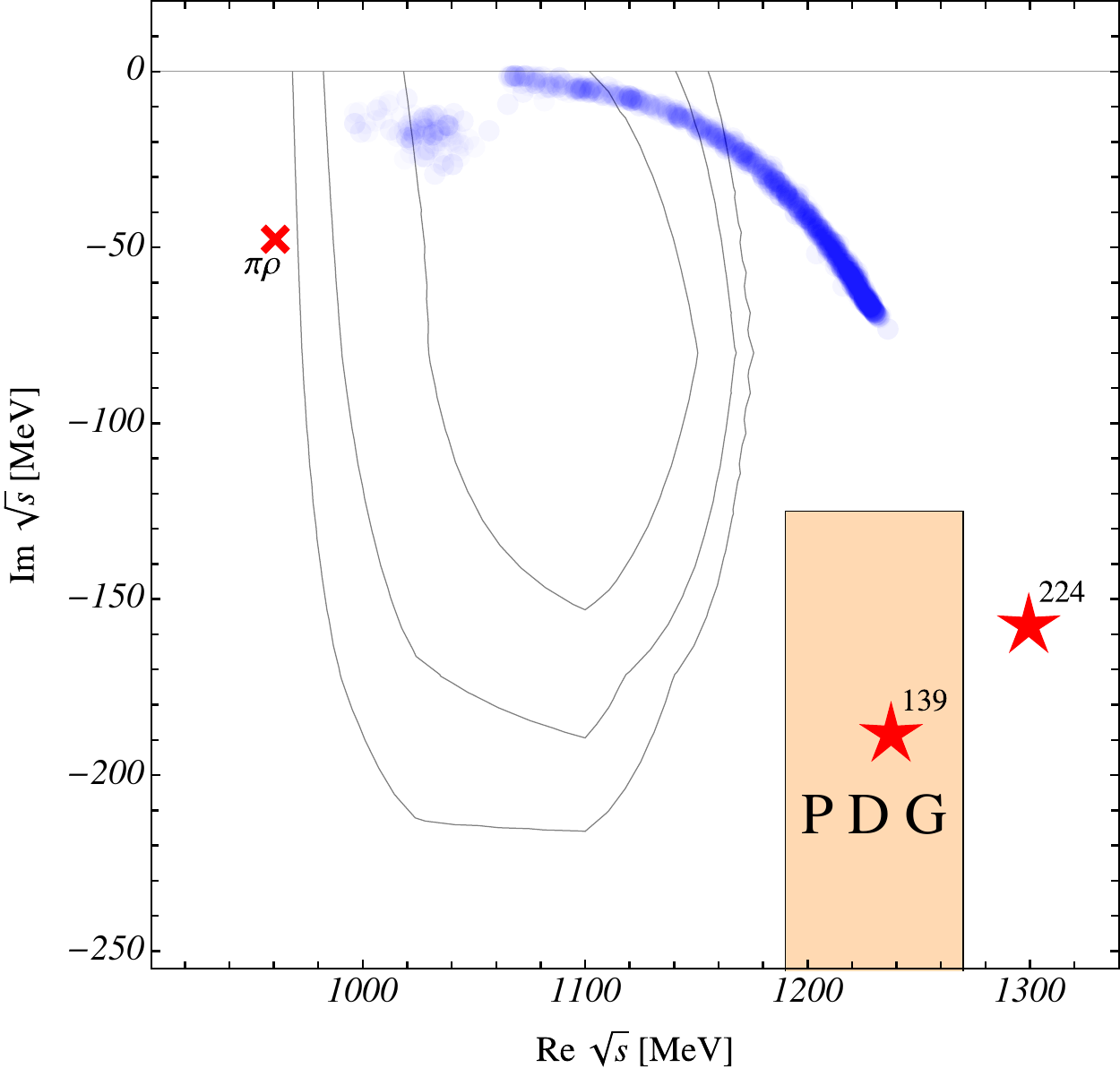}
\caption{Left panel: The $2\pi^+$ and $3\pi^+$ finite-volume spectra in the center-of-momentum frame computed by HadSpec. 
Grey curves are the ``non-interacting'' finite-volume energies, while orange curves are predictions from the finite-volume formalism based only on the two-particle scattering length. Figure from~\cite{Hansen:2020otl}. Right panel: The $a_1(1260)$ pole positions from Lattice QCD, in blue. The PDG estimation is included as the orange rectangle. Possible extrapolations to the physical pion mass are reported as red stars. Figure from~\cite{Mai:2021nul}.
}
\label{fig:3blattice}
\end{figure}
Further advancements in extracting the hadron spectrum from experimental data require the use of $3\to 3$ scattering amplitudes, for example like the ones developed in~\cite{Mai:2017vot,*Jackura:2018xnx,Mikhasenko:2019vhk}.
Since these formalisms require to solve integral equations hard to implement in data analysis, it is possible to introduce additional assumptions or simplifications, which make the equation algebraic. This was done for example in~\cite{JPAC:2018zwp} to describe the $a_1(1260)$ which dominates the $\tau \to 3\pi\nu$ decay~\cite{Davier:2013sfa, ALEPH:2005qgp},
or to describe the $\pi_2$ system in the $3\pi$ COMPASS dataset~\cite{COMPASS:2015gxz, JPAC:2021rxu}. The 3-body effects are most relevant
for $D^0D^0\pi^+$ scattering, where the prominent $T_{cc}^+$ signal reported in Fig.~\ref{fig:3body}  is seen~\cite{LHCb:2021vvq,LHCb:2021auc}, and have been studied with different levels of approximation in~\cite{Albaladejo:2021vln,LHCb:2021auc,Du:2021zzh}.

In addition to phenomenological studies, a tremendous effort has been put in calculating the resonant spectrum from first principles using Lattice QCD. Since simulations are performed for imaginary time, the $S$-matrix cannot be accessed directly. 
However, L\"uscher quantization conditions relate the scattering observables to the volume dependence of the lattice spectrum~\cite{Luscher:1986pf,Luscher:1990ux}. The $2\to2$ scattering has been generalized to any quantum numbers and masses of the four particles~\cite{Rummukainen:1995vs,*Liu:2005kr,*Kim:2005gf,Briceno:2014oea}. Matrix elements can be calculated as well~\cite{Lellouch:2000pv,Briceno:2014uqa,Briceno:2015csa}. 
Many systems of physical relevance have been studied~\cite{Briceno:2017max}. The $3\to3$ generalization of the L\"usher's idea has been developed, leading to different three-particle quantization conditions~\cite{Polejaeva:2012ut,Hansen:2015zga, Mai:2017bge, Hammer:2017uqm, Hammer:2017kms,*Jackura:2019bmu,Blanton:2020gha}, that eventually proved to be equivalent. Such reaction depends upon 8 independent Mandelstam variables, which makes the formalism way more involved than L\"uscher's, and hence the first analyses focus on non-resonant systems~\cite{Mai:2019fba,Alexandru:2020xqf,Hansen:2020otl,Brett:2021wyd}. In Fig.~\ref{fig:3blattice} we show two examples, including the first extraction of a 3-body resonance in Lattice QCD~\cite{Mai:2021nul}. All these works set the ground for future explorations in the few-body physics world.

We conclude by recalling that the quantization condition allows us to calculate the amplitude on the real axis numerically. However, it still has to be fitted with infinite volume amplitude parametrizations in order to perform the analytic continuation and extract the resonance information. This requires a parallel effort of phenomenology to provide more and more refined analytic tools.

%%%%%%%%%%%%%%%%%%%%%%%%%%%%%%%%%%%%
%	Production
%%%%%%%%%%%%%%%%%%%%%%%%%%%%%%%%%%%%
\section{Production mechanisms} \label{sec:production}

Understanding production mechanisms of (exotic) resonances offers complementary information to study their properties and eventually determine their nature.
In the light sector, lots of data come from heavy hadron decays (produced {\it e.g.} at LHC, BESIII and, in the future, at Belle~II), that can be analyzed with the multibody techniques discussed in Section~\ref{sec:three}. Other important processes are the diffractive dissociation of hadron (COMPASS, J-PARC) and lepton/photon beams (JLab and, in the future, the EIC), or the central/peripheral exclusive production at the LHC. 
These reactions at high energies are well understood in Regge theory. Processes are saturated by the exchange of a small number of towers of particles of increasing mass and spin in the cross-channel, named Reggeons. Each tower is referred to as a Regge trajectory, and its properties are inherited by the lightest particle belonging to it (Reggeon-particle duality). One can consider also trajectories dual to purely gluonic states, as the Pomeron and Odderon, that actually dominate the elastic scattering at high energies~\cite{Donnachie:2002en,TOTEM:2020zzr}.
In the dispersive language, each trajectory corresponds to a pole in the complex angular momentum plane of the cross-channel amplitude. One can introduce systematically also subleading terms, such as the exchange of heavier `daughter' trajectories, or rescattering corrections that generate cuts in the complex angular momentum plane. 
This framework allows to predict hadro- and photoproduction cross sections and polarization observables at high energies~\cite{Mathieu:2015eia,*JointPhysicsAnalysisCenter:2017del,*Mathieu:2017jjs,*Mathieu:2018xyc,*Mathieu:2020zpm,Nys:2018vck}.

\begin{figure*}[t]
\begin{center}
	\includegraphics[width=0.43\textwidth]{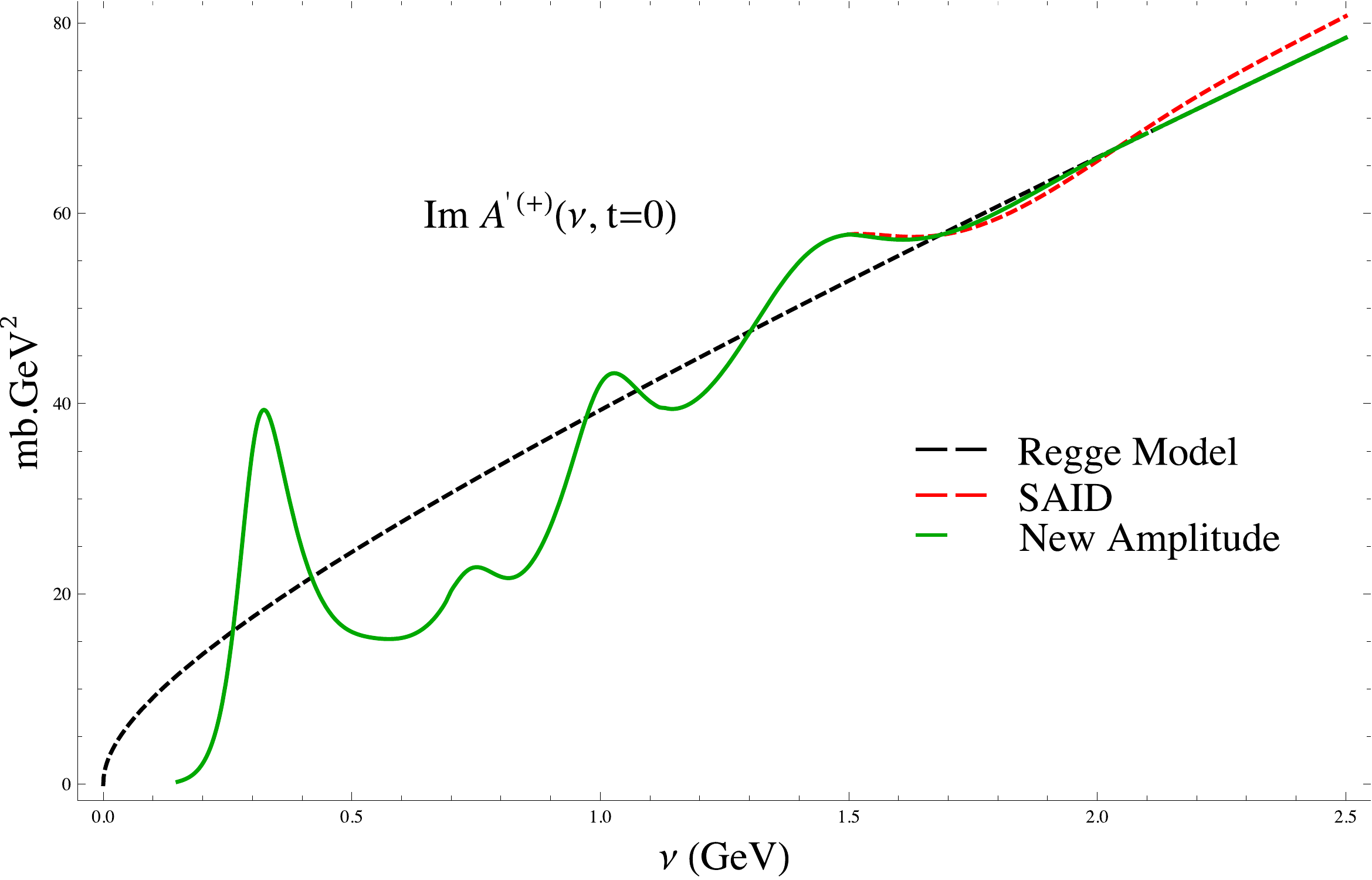} \hspace*{0.08cm}
	\includegraphics[width=0.43\textwidth]{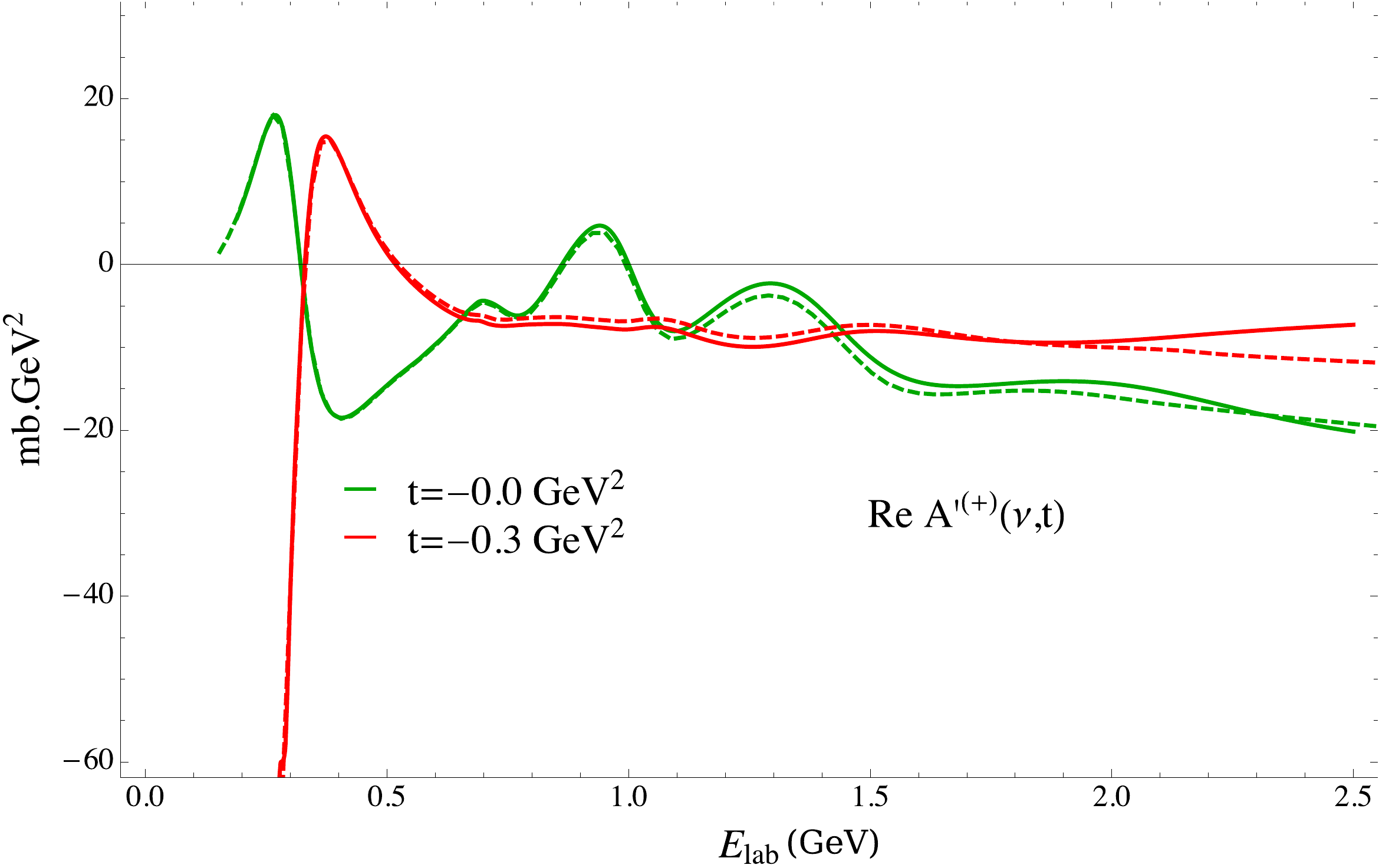}
\end{center}
\caption{(left) Illustration of FESR. The Regge parametrization is equivalent to the average of the imaginary part of the amplitude. (right) The real part of the amplitude reconstructed from the dispersion relation (dashed) matches the original real part from the partial wave analysis of SAID (solid)~\cite{Workman:2012jf}. 
Figures from~\cite{JPAC:2020umo}. \label{fig:FESR2} }
\end{figure*}

This high energy regime is smoothly connected to the resonance region, where the amplitude is saturated by a finite number of partial waves. One can write dispersion relations, that result in sum rules for partial waves that have to match the Regge expectations at high energies. 
This further constrains the available low-energy partial wave analyses, and eventually the extraction of resonance parameters~\cite{Mathieu:2015gxa,*Nys:2016vjz,*Mathieu:2017but,*Mathieu:2018mjw}. These relations are generally called finite-energy sum rules (FESR). An example of matching between low- and high-energy regions is reported in Fig.~\ref{fig:FESR2}.
The inclusion of FESRs in the analysis of the mass dependence of two-body final states will be key in reducing systematic uncertainties in the extraction of exotic candidates from the 
analysis of photoproduction experiments. While this has been studied extensively for single meson production, the extension to $2\to 3$ reactions is presently under development. Simple Regge models for the asymptotic region have been proposed already~\cite{Shi:2014nea,*Bibrzycki:2021rwh}. Further steps require a joint effort of theory and experiments, in order to access kinematic regions that are usually overlooked. 

For heavy hadrons, similar mechanisms can occur and can be studied with the same methods. Predictions of $XYZ$ and pentaquark peripheral photoproduction have been given~\cite{Albaladejo:2020tzt,HillerBlin:2016odx,*Winney:2019edt}, and are currently being used to shape the spectroscopy program at the forthcoming EIC~\cite{AbdulKhalek:2021gbh}. Additionally, one can obtain predictions for prompt inclusive production, for example with NRQCD~\cite{Bodwin:1994jh}. Extracting the long-distance matrix elements from data, and comparing them with model expectations, can offer insights into the nature of exotic states~\cite{Bignamini:2009sk,*Artoisenet:2009wk}. Finally, modeling final state interactions of $XYZ$, for example in high multiplicity enviroments~\cite{ExHIC:2017smd,*Zhang:2020dwn,*Wu:2020zbx,*Esposito:2020ywk,*Braaten:2020iqw}, also can give us yet another piece of the puzzle.

%%%%%%%%%%%%%%%%%%%%%%%%%%%%%%%%%%%%
%	Amplitude/Formalism
%%%%%%%%%%%%%%%%%%%%%%%%%%%%%%%%%%%%

\section{Tools}
\label{sec:tools}
Computing power has grown exponentially in the last decades. Heavily expensive statistical methods, based on pseudodata generation
are becoming the standard tool to have a reliable error estimation on complicated models as the ones discussed above. Similarly, Machine Learning (ML) algorithms have improved dramatically and can be applied to solve spectroscopy problems in a model-independent way.

Monte Carlo approaches 
allow for a systematic analysis of the statistical uncertainties. Most analyses at JPAC adopt statistical bootstrap as a resampling method to perform error analysis~\cite{EfroTibs93}. 
Model dependencies can be reduced with statistical learning (cross-validation, ridge methods, and stability selection)~\cite{StatisticalLearning,*Guegan:2015mea,*Landay:2016cjw}. 
Clustering methods can be used to separate the physical resonances from the artifacts of the amplitude parametrizations~\cite{JPAC:2018zyd,JPAC:2021rxu}.
All these studies allow one to give a robust determination of resonances and of their properties.
Although computationally expensive, these kind of analyses will become mandatory for the interpretation of future high-precision data.

\begin{figure}[t]
    \centering
    \includegraphics[scale=0.8,keepaspectratio]{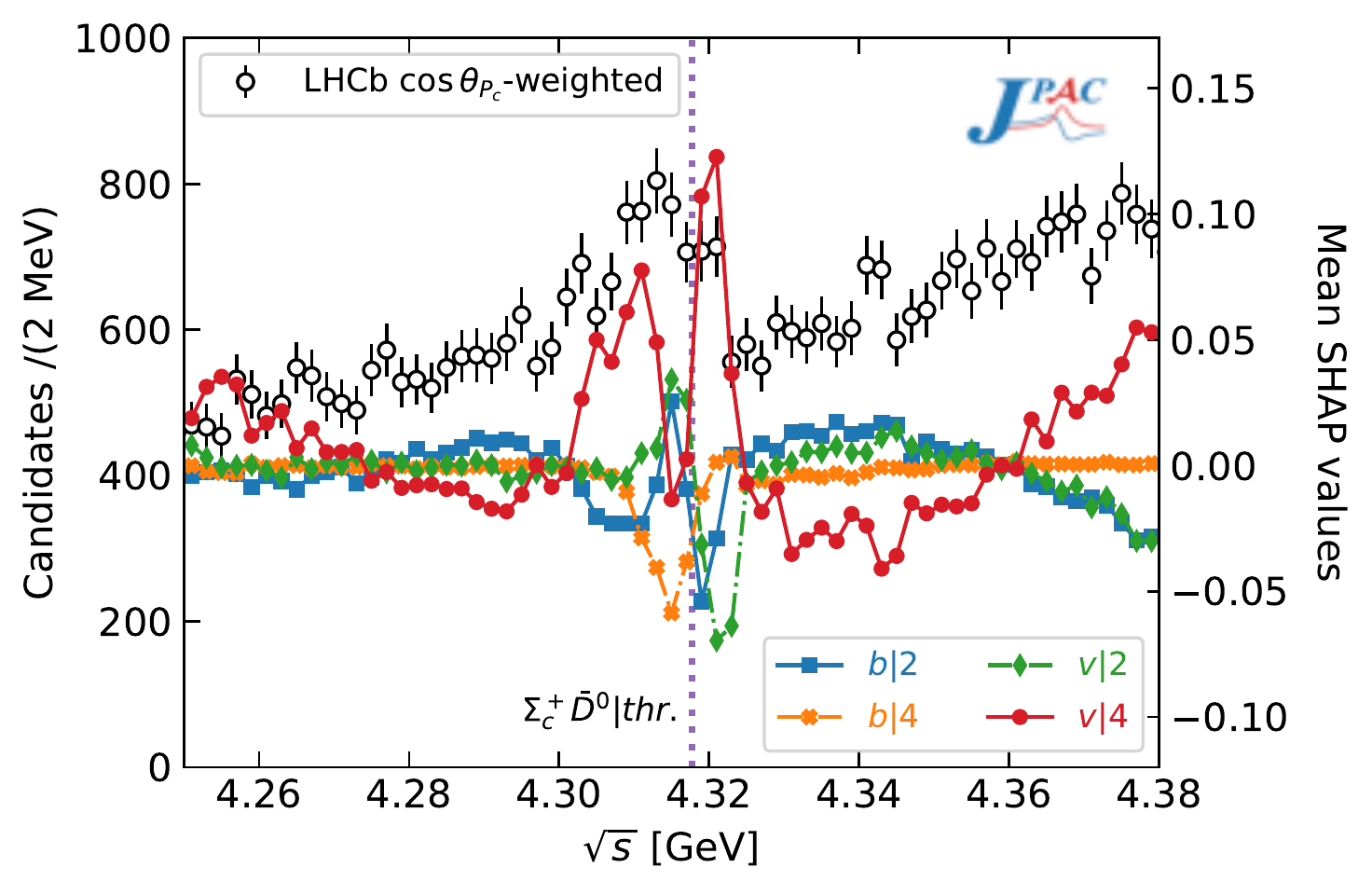}
    \caption{The $P_c(4312)$ data from LHCb~\cite{LHCb:2019kea} (left axis) and breakdown of the
    of their mean 
    SHAP values (right axis) 
    as a function of the $J/\psi \,p$ invariant mass for the four classes. The SHAP values provide how much a certain datapoint favors or disfavors a particular class. Figure from~\cite{Ng:2021ibr}.
    \label{fig:shap}}
\end{figure}

Recently, ML in the form of deep neural network (DNN)
classifiers have been explored as a tool to gain insight
on the underlying nature of the hadron states~\cite{Sombillo:2020ccg,*Sombillo:2021rxv,*Sombillo:2021yxe,Ng:2021ibr}.
The technique is very promising and complementary to the
standard procedure. 
The idea is to teach the DNN how to recognize the nature of a state, more specifically in~\cite{Ng:2021ibr} how to distinguish virtual from bound states. The DNN targets specific regions of the parameter space (which yield stable solutions) that might be difficult to reach during optimization, or might require high-resolution data. Standard $\chi^2$ fit can be indeed unstable, and a small change in the input data can induce large changes in the parameter values and therefore in the physics interpretation. 
Moreover, rather than testing a single model hypothesis as a $\chi^2$ fit would, the DNN determines the probability of each of the classes of interest, given the experimental uncertainties. The latter is possible, since the DNN learns the subtle classification boundary between the different classes. Additionally, 
using a systematic method based on SHapley Additive exPlanations (SHAP) values it is possible to break down a prediction to show how each datapoint impacts classification as done in~\cite{Ng:2021ibr} for the LHCb data on the $P_c(4312)$, as shown in Fig.~\ref{fig:shap}. 

The potential of ML for hadron spectroscopy is still largely unexplored. For example, one should study automatic methods to perform analytic continuation and hunt for resonances~\cite{Yoon:2018,*Fournier:2018}, or consider Neural Network parametrizations on which one can impose the $S$-matrix constraints numerically, on the lines of what done for parametrizing the parton distribution or spectral functions~\cite{Rojo:2004iq,*NNPDF:2021uiq}.

\section{Outlook}
The ultimate goal of Hadron Spectroscopy is to infer the pattern that underlies the dynamics of quark and gluons, and produces the excited spectrum. To do so, it is necessary to have a precise determination of the resonance properties. 
The LHC experiments and BESIII are already recording data with unprecedented statistics. More exciting results are expected from JLab and Belle-II in the near future.
Future planned facilities include Electron-Ion Colliders, a $p\bar p$ machine at the charm energies ($\overline{\text{P}}$ANDA), and COMPASS++/AMBER able to run with a variety of beam species.
In addition, in the near future Lattice QCD is expected to deliver new observables that are unaccessible experimentally.
These precise datasets require adequate analysis and theoretical methods, as the one discussed in this White Paper. Extracting the resonance parameters and understanding the different production mechanisms give a complete description of the excited states, that will allow to establish their nature with great accuracy. The use of proper statistical tools and machine learning algorithms will produce model-independent robust results.

For these analyses to be successful, we stress the importance of tight collaborative efforts
between theorists and experimentalists, in the same spirit as the ones pioneered by JPAC.

\section*{acknowledgments}
This work was supported by the U.S.~Department of Energy under Grant No.~DE-AC05-06OR23177  under which Jefferson Science Associates, LLC, manages and operates Jefferson Lab, No.~DE-FG02-87ER40365 at Indiana University, and
No.~DE-SC0018416 at the College of William \& Mary,
National Science Foundation under Grant No.~PHY-2013184,
Polish Science Center (NCN) under Grant No.~2018/29/B/ST2/02576,
Spanish Ministerio de Econom\'ia y Competitividad and  
Ministerio de Ciencia e Innovaci\'on under Grants 
No.~PID2019–106080 GB-C21, 
No.~PID2019-105439G-C22,
No.~PID2020-118758GB-I00
and No.~PID2020-112777GB-I00 (Ref.~10.13039/501100011033),
UNAM-PAPIIT under Grant No.~IN106921,
CONACYT under Grant No.~A1-S-21389,
National Natural Science Foundation of China Grant No.~12035007
and the NSFC and the Deutsche Forschungsgemeinschaft (DFG, German Research Foundation) through 
% MM
the Germany's Excellence Strategy – EXC-2094 – 390783311 as well as 
the funds provided to the Sino-German Collaborative Research Center TRR110 ``Symmetries and the Emergence of Structure in QCD" (NSFC Grant No.~12070131001, DFG Project-ID~196253076-TRR~110).
MA is supported by Generalitat Valenciana under Grant No.~CIDEGENT/2020/002.
CFR is supported by Spanish Ministerio de Educaci\'on y Formaci\'on Profesional under Grant No.~BG20/00133.
VM is a Serra Húnter fellow and acknowledges support from the Spanish national Grant No. PID2019–106080 GB-C21 and PID2020-118758GB-I00.
JASC is supported by CONACYT under Grant No.~734789.
SGS is supported by the Laboratory Directed Research and Development program of Los Alamos National Laboratory under project No.~20210944PRD2, and by the U.S. Department of Energy through the Los Alamos National Laboratory. Los Alamos National Laboratory is operated by Triad National Security, LLC, for the National Nuclear Security Administration of U.S.~Department of Energy (Contract No.~89233218CNA000001).
ANHB is supported by the Deutsche Forschungsgemeinschaft (DFG) through the Research Unit FOR
2926 (project number 40824754).

\bibliographystyle{apsrev4-2.bst}
\bibliography{quattro}

\begin{thebibliography}{196}%
\makeatletter
\providecommand \@ifxundefined [1]{%
 \@ifx{#1\undefined}
}%
\providecommand \@ifnum [1]{%
 \ifnum #1\expandafter \@firstoftwo
 \else \expandafter \@secondoftwo
 \fi
}%
\providecommand \@ifx [1]{%
 \ifx #1\expandafter \@firstoftwo
 \else \expandafter \@secondoftwo
 \fi
}%
\providecommand \natexlab [1]{#1}%
\providecommand \enquote  [1]{``#1''}%
\providecommand \bibnamefont  [1]{#1}%
\providecommand \bibfnamefont [1]{#1}%
\providecommand \citenamefont [1]{#1}%
\providecommand \href@noop [0]{\@secondoftwo}%
\providecommand \href [0]{\begingroup \@sanitize@url \@href}%
\providecommand \@href[1]{\@@startlink{#1}\@@href}%
\providecommand \@@href[1]{\endgroup#1\@@endlink}%
\providecommand \@sanitize@url [0]{\catcode `\\12\catcode `\$12\catcode
  `\&12\catcode `\#12\catcode `\^12\catcode `\_12\catcode `\%12\relax}%
\providecommand \@@startlink[1]{}%
\providecommand \@@endlink[0]{}%
\providecommand \url  [0]{\begingroup\@sanitize@url \@url }%
\providecommand \@url [1]{\endgroup\@href {#1}{\urlprefix }}%
\providecommand \urlprefix  [0]{URL }%
\providecommand \Eprint [0]{\href }%
\providecommand \doibase [0]{https://doi.org/}%
\providecommand \selectlanguage [0]{\@gobble}%
\providecommand \bibinfo  [0]{\@secondoftwo}%
\providecommand \bibfield  [0]{\@secondoftwo}%
\providecommand \translation [1]{[#1]}%
\providecommand \BibitemOpen [0]{}%
\providecommand \bibitemStop [0]{}%
\providecommand \bibitemNoStop [0]{.\EOS\space}%
\providecommand \EOS [0]{\spacefactor3000\relax}%
\providecommand \BibitemShut  [1]{\csname bibitem#1\endcsname}%
\let\auto@bib@innerbib\@empty
%</preamble>
\bibitem [{\citenamefont {Zyla}\ \emph {et~al.}(2020)\citenamefont {Zyla} \emph
  {et~al.}}]{pdg}%
  \BibitemOpen
  \bibfield  {author} {\bibinfo {author} {\bibfnamefont {P.~A.}\ \bibnamefont
  {Zyla}} \emph {et~al.} (\bibinfo {collaboration} {Particle Data Group}),\
  }\href {https://doi.org/10.1093/ptep/ptaa104} {\bibfield  {journal} {\bibinfo
   {journal} {PTEP}\ }\textbf {\bibinfo {volume} {2020}},\ \bibinfo {pages}
  {083C01} (\bibinfo {year} {2020})}\BibitemShut {NoStop}%
\bibitem [{\citenamefont {Esposito}\ \emph {et~al.}(2017)\citenamefont
  {Esposito}, \citenamefont {Pilloni},\ and\ \citenamefont
  {Polosa}}]{Esposito:2016noz}%
  \BibitemOpen
  \bibfield  {author} {\bibinfo {author} {\bibfnamefont {A.}~\bibnamefont
  {Esposito}}, \bibinfo {author} {\bibfnamefont {A.}~\bibnamefont {Pilloni}},\
  and\ \bibinfo {author} {\bibfnamefont {A.~D.}\ \bibnamefont {Polosa}},\
  }\href {https://doi.org/10.1016/j.physrep.2016.11.002} {\bibfield  {journal}
  {\bibinfo  {journal} {Phys.Rept.}\ }\textbf {\bibinfo {volume} {668}},\
  \bibinfo {pages} {1} (\bibinfo {year} {2017})},\ \Eprint
  {https://arxiv.org/abs/1611.07920} {arXiv:1611.07920 [hep-ph]} \BibitemShut
  {NoStop}%
%%CITATION = ARXIV:1611.07920;%%
\bibitem [{\citenamefont {Olsen}\ \emph {et~al.}(2018)\citenamefont {Olsen},
  \citenamefont {Skwarnicki},\ and\ \citenamefont {Zieminska}}]{Olsen:2017bmm}%
  \BibitemOpen
  \bibfield  {author} {\bibinfo {author} {\bibfnamefont {S.~L.}\ \bibnamefont
  {Olsen}}, \bibinfo {author} {\bibfnamefont {T.}~\bibnamefont {Skwarnicki}},\
  and\ \bibinfo {author} {\bibfnamefont {D.}~\bibnamefont {Zieminska}},\ }\href
  {https://doi.org/10.1103/RevModPhys.90.015003} {\bibfield  {journal}
  {\bibinfo  {journal} {Rev.Mod.Phys.}\ }\textbf {\bibinfo {volume} {90}},\
  \bibinfo {pages} {015003} (\bibinfo {year} {2018})},\ \Eprint
  {https://arxiv.org/abs/1708.04012} {arXiv:1708.04012 [hep-ph]} \BibitemShut
  {NoStop}%
%%CITATION = ARXIV:1708.04012;%%
\bibitem [{\citenamefont {Guo}\ \emph {et~al.}(2018)\citenamefont {Guo},
  \citenamefont {Hanhart}, \citenamefont {Mei{\ss}ner}, \citenamefont {Wang},
  \citenamefont {Zhao},\ and\ \citenamefont {Zou}}]{Guo:2017jvc}%
  \BibitemOpen
  \bibfield  {author} {\bibinfo {author} {\bibfnamefont {F.-K.}\ \bibnamefont
  {Guo}}, \bibinfo {author} {\bibfnamefont {C.}~\bibnamefont {Hanhart}},
  \bibinfo {author} {\bibfnamefont {U.-G.}\ \bibnamefont {Mei{\ss}ner}},
  \bibinfo {author} {\bibfnamefont {Q.}~\bibnamefont {Wang}}, \bibinfo {author}
  {\bibfnamefont {Q.}~\bibnamefont {Zhao}},\ and\ \bibinfo {author}
  {\bibfnamefont {B.-S.}\ \bibnamefont {Zou}},\ }\href
  {https://doi.org/10.1103/RevModPhys.90.015004} {\bibfield  {journal}
  {\bibinfo  {journal} {Rev.Mod.Phys.}\ }\textbf {\bibinfo {volume} {90}},\
  \bibinfo {pages} {015004} (\bibinfo {year} {2018})},\ \Eprint
  {https://arxiv.org/abs/1705.00141} {arXiv:1705.00141 [hep-ph]} \BibitemShut
  {NoStop}%
%%CITATION = ARXIV:1705.00141;%%
\bibitem [{\citenamefont {Lebed}\ \emph {et~al.}(2017)\citenamefont {Lebed},
  \citenamefont {Mitchell},\ and\ \citenamefont {Swanson}}]{Lebed:2016hpi}%
  \BibitemOpen
  \bibfield  {author} {\bibinfo {author} {\bibfnamefont {R.~F.}\ \bibnamefont
  {Lebed}}, \bibinfo {author} {\bibfnamefont {R.~E.}\ \bibnamefont
  {Mitchell}},\ and\ \bibinfo {author} {\bibfnamefont {E.~S.}\ \bibnamefont
  {Swanson}},\ }\href {https://doi.org/10.1016/j.ppnp.2016.11.003} {\bibfield
  {journal} {\bibinfo  {journal} {Prog.Part.Nucl.Phys.}\ }\textbf {\bibinfo
  {volume} {93}},\ \bibinfo {pages} {143} (\bibinfo {year} {2017})},\ \Eprint
  {https://arxiv.org/abs/1610.04528} {arXiv:1610.04528 [hep-ph]} \BibitemShut
  {NoStop}%
%%CITATION = ARXIV:1610.04528;%%
\bibitem [{\citenamefont {Karliner}\ \emph {et~al.}(2018)\citenamefont
  {Karliner}, \citenamefont {Rosner},\ and\ \citenamefont
  {Skwarnicki}}]{Karliner:2017qhf}%
  \BibitemOpen
  \bibfield  {author} {\bibinfo {author} {\bibfnamefont {M.}~\bibnamefont
  {Karliner}}, \bibinfo {author} {\bibfnamefont {J.~L.}\ \bibnamefont
  {Rosner}},\ and\ \bibinfo {author} {\bibfnamefont {T.}~\bibnamefont
  {Skwarnicki}},\ }\href {https://doi.org/10.1146/annurev-nucl-101917-020902}
  {\bibfield  {journal} {\bibinfo  {journal} {Ann.Rev.Nucl.Part.Sci}\ }\textbf
  {\bibinfo {volume} {68}},\ \bibinfo {pages} {17} (\bibinfo {year} {2018})},\
  \Eprint {https://arxiv.org/abs/1711.10626} {arXiv:1711.10626 [hep-ph]}
  \BibitemShut {NoStop}%
%%CITATION = ARXIV:1711.10626;%%
\bibitem [{\citenamefont {Guo}\ \emph {et~al.}(2020)\citenamefont {Guo},
  \citenamefont {Liu},\ and\ \citenamefont {Sakai}}]{Guo:2019twa}%
  \BibitemOpen
  \bibfield  {author} {\bibinfo {author} {\bibfnamefont {F.-K.}\ \bibnamefont
  {Guo}}, \bibinfo {author} {\bibfnamefont {X.-H.}\ \bibnamefont {Liu}},\ and\
  \bibinfo {author} {\bibfnamefont {S.}~\bibnamefont {Sakai}},\ }\href
  {https://doi.org/10.1016/j.ppnp.2020.103757} {\bibfield  {journal} {\bibinfo
  {journal} {Prog.Part.Nucl.Phys.}\ }\textbf {\bibinfo {volume} {112}},\
  \bibinfo {pages} {103757} (\bibinfo {year} {2020})},\ \Eprint
  {https://arxiv.org/abs/1912.07030} {arXiv:1912.07030 [hep-ph]} \BibitemShut
  {NoStop}%
\bibitem [{\citenamefont {Ali}\ \emph {et~al.}(2019)\citenamefont {Ali},
  \citenamefont {Maiani},\ and\ \citenamefont {Polosa}}]{ali2019multiquark}%
  \BibitemOpen
  \bibfield  {author} {\bibinfo {author} {\bibfnamefont {A.}~\bibnamefont
  {Ali}}, \bibinfo {author} {\bibfnamefont {L.}~\bibnamefont {Maiani}},\ and\
  \bibinfo {author} {\bibfnamefont {A.~D.}\ \bibnamefont {Polosa}},\ }\href
  {https://doi.org/10.1017/9781316761465} {\emph {\bibinfo {title} {Multiquark
  Hadrons}}}\ (\bibinfo  {publisher} {Cambridge University Press},\ \bibinfo
  {year} {2019})\BibitemShut {NoStop}%
\bibitem [{\citenamefont {Brambilla}\ \emph {et~al.}(2020)\citenamefont
  {Brambilla}, \citenamefont {Eidelman}, \citenamefont {Hanhart}, \citenamefont
  {Nefediev}, \citenamefont {Shen}, \citenamefont {Thomas}, \citenamefont
  {Vairo},\ and\ \citenamefont {Yuan}}]{Brambilla:2019esw}%
  \BibitemOpen
  \bibfield  {author} {\bibinfo {author} {\bibfnamefont {N.}~\bibnamefont
  {Brambilla}}, \bibinfo {author} {\bibfnamefont {S.}~\bibnamefont {Eidelman}},
  \bibinfo {author} {\bibfnamefont {C.}~\bibnamefont {Hanhart}}, \bibinfo
  {author} {\bibfnamefont {A.}~\bibnamefont {Nefediev}}, \bibinfo {author}
  {\bibfnamefont {C.-P.}\ \bibnamefont {Shen}}, \bibinfo {author}
  {\bibfnamefont {C.~E.}\ \bibnamefont {Thomas}}, \bibinfo {author}
  {\bibfnamefont {A.}~\bibnamefont {Vairo}},\ and\ \bibinfo {author}
  {\bibfnamefont {C.-Z.}\ \bibnamefont {Yuan}},\ }\href
  {https://doi.org/10.1016/j.physrep.2020.05.001} {\bibfield  {journal}
  {\bibinfo  {journal} {Phys.Rept.}\ }\textbf {\bibinfo {volume} {873}},\
  \bibinfo {pages} {1} (\bibinfo {year} {2020})},\ \Eprint
  {https://arxiv.org/abs/1907.07583} {arXiv:1907.07583 [hep-ex]} \BibitemShut
  {NoStop}%
\bibitem [{\citenamefont {Shepherd}\ \emph {et~al.}(2016)\citenamefont
  {Shepherd}, \citenamefont {Dudek},\ and\ \citenamefont
  {Mitchell}}]{Shepherd:2016dni}%
  \BibitemOpen
  \bibfield  {author} {\bibinfo {author} {\bibfnamefont {M.~R.}\ \bibnamefont
  {Shepherd}}, \bibinfo {author} {\bibfnamefont {J.~J.}\ \bibnamefont
  {Dudek}},\ and\ \bibinfo {author} {\bibfnamefont {R.~E.}\ \bibnamefont
  {Mitchell}},\ }\href {https://doi.org/10.1038/nature18011} {\bibfield
  {journal} {\bibinfo  {journal} {Nature}\ }\textbf {\bibinfo {volume} {534}},\
  \bibinfo {pages} {487} (\bibinfo {year} {2016})}\BibitemShut {NoStop}%
%%CITATION = NATUA,534,487;%%
\bibitem [{\citenamefont {Brice\~no}\ \emph {et~al.}(2018)\citenamefont
  {Brice\~no}, \citenamefont {Dudek},\ and\ \citenamefont
  {Young}}]{Briceno:2017max}%
  \BibitemOpen
  \bibfield  {author} {\bibinfo {author} {\bibfnamefont {R.~A.}\ \bibnamefont
  {Brice\~no}}, \bibinfo {author} {\bibfnamefont {J.~J.}\ \bibnamefont
  {Dudek}},\ and\ \bibinfo {author} {\bibfnamefont {R.~D.}\ \bibnamefont
  {Young}},\ }\href {https://doi.org/10.1103/RevModPhys.90.025001} {\bibfield
  {journal} {\bibinfo  {journal} {Rev.Mod.Phys.}\ }\textbf {\bibinfo {volume}
  {90}},\ \bibinfo {pages} {025001} (\bibinfo {year} {2018})},\ \Eprint
  {https://arxiv.org/abs/1706.06223} {arXiv:1706.06223 [hep-lat]} \BibitemShut
  {NoStop}%
%%CITATION = ARXIV:1706.06223;%%
\bibitem [{\citenamefont {Polonyi}(2003)}]{Polonyi:2001se}%
  \BibitemOpen
  \bibfield  {author} {\bibinfo {author} {\bibfnamefont {J.}~\bibnamefont
  {Polonyi}},\ }\href {https://doi.org/10.2478/BF02475552} {\bibfield
  {journal} {\bibinfo  {journal} {Central Eur.J.Phys.}\ }\textbf {\bibinfo
  {volume} {1}},\ \bibinfo {pages} {1} (\bibinfo {year} {2003})},\ \Eprint
  {https://arxiv.org/abs/hep-th/0110026} {arXiv:hep-th/0110026} \BibitemShut
  {NoStop}%
\bibitem [{\citenamefont {Maris}\ and\ \citenamefont
  {Roberts}(2003)}]{Maris:2003vk}%
  \BibitemOpen
  \bibfield  {author} {\bibinfo {author} {\bibfnamefont {P.}~\bibnamefont
  {Maris}}\ and\ \bibinfo {author} {\bibfnamefont {C.~D.}\ \bibnamefont
  {Roberts}},\ }\href {https://doi.org/10.1142/S0218301303001326} {\bibfield
  {journal} {\bibinfo  {journal} {Int.J.Mod.Phys.}\ }\textbf {\bibinfo {volume}
  {E12}},\ \bibinfo {pages} {297} (\bibinfo {year} {2003})},\ \Eprint
  {https://arxiv.org/abs/nucl-th/0301049} {arXiv:nucl-th/0301049} \BibitemShut
  {NoStop}%
\bibitem [{\citenamefont {Plessas}(2015)}]{Plessas:2015mpa}%
  \BibitemOpen
  \bibfield  {author} {\bibinfo {author} {\bibfnamefont {W.}~\bibnamefont
  {Plessas}},\ }\href {https://doi.org/10.1142/S0217751X15300136} {\bibfield
  {journal} {\bibinfo  {journal} {Int.J.Mod.Phys.}\ }\textbf {\bibinfo {volume}
  {A30}},\ \bibinfo {pages} {1530013} (\bibinfo {year} {2015})}\BibitemShut
  {NoStop}%
\bibitem [{\citenamefont {Brodsky}\ \emph {et~al.}(2015)\citenamefont
  {Brodsky}, \citenamefont {de~Teramond}, \citenamefont {Dosch},\ and\
  \citenamefont {Erlich}}]{Brodsky:2014yha}%
  \BibitemOpen
  \bibfield  {author} {\bibinfo {author} {\bibfnamefont {S.~J.}\ \bibnamefont
  {Brodsky}}, \bibinfo {author} {\bibfnamefont {G.~F.}\ \bibnamefont
  {de~Teramond}}, \bibinfo {author} {\bibfnamefont {H.~G.}\ \bibnamefont
  {Dosch}},\ and\ \bibinfo {author} {\bibfnamefont {J.}~\bibnamefont
  {Erlich}},\ }\href {https://doi.org/10.1016/j.physrep.2015.05.001} {\bibfield
   {journal} {\bibinfo  {journal} {Phys.Rept.}\ }\textbf {\bibinfo {volume}
  {584}},\ \bibinfo {pages} {1} (\bibinfo {year} {2015})},\ \Eprint
  {https://arxiv.org/abs/1407.8131} {arXiv:1407.8131 [hep-ph]} \BibitemShut
  {NoStop}%
\bibitem [{\citenamefont {Eden}\ \emph {et~al.}(1966)\citenamefont {Eden},
  \citenamefont {Landshoff}, \citenamefont {Olive},\ and\ \citenamefont
  {Polkinghorne}}]{Eden:1966dnq}%
  \BibitemOpen
  \bibfield  {author} {\bibinfo {author} {\bibfnamefont {R.~J.}\ \bibnamefont
  {Eden}}, \bibinfo {author} {\bibfnamefont {P.~V.}\ \bibnamefont {Landshoff}},
  \bibinfo {author} {\bibfnamefont {D.~I.}\ \bibnamefont {Olive}},\ and\
  \bibinfo {author} {\bibfnamefont {J.~C.}\ \bibnamefont {Polkinghorne}},\
  }\href {https://cds.cern.ch/record/98637} {\emph {\bibinfo {title} {{The
  analytic $S$-matrix}}}}\ (\bibinfo  {publisher} {Cambridge Univ. Press},\
  \bibinfo {address} {Cambridge},\ \bibinfo {year} {1966})\BibitemShut
  {NoStop}%
\bibitem [{\citenamefont {Martin}\ and\ \citenamefont
  {Spearman}(1970)}]{Martin:1970}%
  \BibitemOpen
  \bibfield  {author} {\bibinfo {author} {\bibfnamefont {A.}~\bibnamefont
  {Martin}}\ and\ \bibinfo {author} {\bibfnamefont {T.}~\bibnamefont
  {Spearman}},\ }\href {https://books.google.com/books?id=sxAzAAAAMAAJ} {\emph
  {\bibinfo {title} {Elementary particle theory}}}\ (\bibinfo  {publisher}
  {North-Holland Pub. Co.},\ \bibinfo {year} {1970})\BibitemShut {NoStop}%
\bibitem [{\citenamefont {Gribov}(2012)}]{Gribov:2009zz}%
  \BibitemOpen
  \bibfield  {author} {\bibinfo {author} {\bibfnamefont {V.~N.}\ \bibnamefont
  {Gribov}},\ }\href
  {http://cambridge.org/catalogue/catalogue.asp?isbn=9780521856096} {\emph
  {\bibinfo {title} {{Strong interactions of hadrons at high energies: Gribov
  lectures on Theoretical Physics}}}},\ edited by\ \bibinfo {editor}
  {\bibfnamefont {Y.~L.}\ \bibnamefont {Dokshitzer}}\ and\ \bibinfo {editor}
  {\bibfnamefont {J.}~\bibnamefont {Nyiri}}\ (\bibinfo  {publisher} {Cambridge
  University Press},\ \bibinfo {year} {2012})\BibitemShut {NoStop}%
%%CITATION = INSPIRE-833953;%%
\bibitem [{\citenamefont {Godfrey}\ and\ \citenamefont
  {Isgur}(1985)}]{Godfrey:1985xj}%
  \BibitemOpen
  \bibfield  {author} {\bibinfo {author} {\bibfnamefont {S.}~\bibnamefont
  {Godfrey}}\ and\ \bibinfo {author} {\bibfnamefont {N.}~\bibnamefont
  {Isgur}},\ }\href {https://doi.org/10.1103/PhysRevD.32.189} {\bibfield
  {journal} {\bibinfo  {journal} {Phys.Rev.}\ }\textbf {\bibinfo {volume}
  {D32}},\ \bibinfo {pages} {189} (\bibinfo {year} {1985})}\BibitemShut
  {NoStop}%
%%CITATION = PHRVA,D32,189;%%
\bibitem [{\citenamefont {Capstick}\ and\ \citenamefont
  {Isgur}(1986)}]{Capstick:1985xss}%
  \BibitemOpen
  \bibfield  {author} {\bibinfo {author} {\bibfnamefont {S.}~\bibnamefont
  {Capstick}}\ and\ \bibinfo {author} {\bibfnamefont {N.}~\bibnamefont
  {Isgur}},\ }\href {https://doi.org/10.1103/PhysRevD.34.2809} {\bibfield
  {journal} {\bibinfo  {journal} {Phys.Rev.}\ }\textbf {\bibinfo {volume}
  {D34}},\ \bibinfo {pages} {2809} (\bibinfo {year} {1986})}\BibitemShut
  {NoStop}%
\bibitem [{\citenamefont {Weinberg}(1979)}]{Weinberg:1978kz}%
  \BibitemOpen
  \bibfield  {author} {\bibinfo {author} {\bibfnamefont {S.}~\bibnamefont
  {Weinberg}},\ }\href {https://doi.org/10.1016/0378-4371(79)90223-1}
  {\bibfield  {journal} {\bibinfo  {journal} {Physica A}\ }\textbf {\bibinfo
  {volume} {96}},\ \bibinfo {pages} {327} (\bibinfo {year} {1979})}\BibitemShut
  {NoStop}%
\bibitem [{\citenamefont {Weinberg}(1990)}]{Weinberg:1990rz}%
  \BibitemOpen
  \bibfield  {author} {\bibinfo {author} {\bibfnamefont {S.}~\bibnamefont
  {Weinberg}},\ }\href {https://doi.org/10.1016/0370-2693(90)90938-3}
  {\bibfield  {journal} {\bibinfo  {journal} {Phys.Lett.}\ }\textbf {\bibinfo
  {volume} {B251}},\ \bibinfo {pages} {288} (\bibinfo {year}
  {1990})}\BibitemShut {NoStop}%
\bibitem [{\citenamefont {Jenkins}\ and\ \citenamefont
  {Manohar}(1991)}]{Jenkins:1990jv}%
  \BibitemOpen
  \bibfield  {author} {\bibinfo {author} {\bibfnamefont {E.~E.}\ \bibnamefont
  {Jenkins}}\ and\ \bibinfo {author} {\bibfnamefont {A.~V.}\ \bibnamefont
  {Manohar}},\ }\href {https://doi.org/10.1016/0370-2693(91)90266-S} {\bibfield
   {journal} {\bibinfo  {journal} {Phys.Lett.}\ }\textbf {\bibinfo {volume}
  {B255}},\ \bibinfo {pages} {558} (\bibinfo {year} {1991})}\BibitemShut
  {NoStop}%
\bibitem [{\citenamefont {Gasser}\ and\ \citenamefont
  {Leutwyler}(1984)}]{Gasser:1983yg}%
  \BibitemOpen
  \bibfield  {author} {\bibinfo {author} {\bibfnamefont {J.}~\bibnamefont
  {Gasser}}\ and\ \bibinfo {author} {\bibfnamefont {H.}~\bibnamefont
  {Leutwyler}},\ }\href {https://doi.org/10.1016/0003-4916(84)90242-2}
  {\bibfield  {journal} {\bibinfo  {journal} {Annals Phys.}\ }\textbf {\bibinfo
  {volume} {158}},\ \bibinfo {pages} {142} (\bibinfo {year}
  {1984})}\BibitemShut {NoStop}%
\bibitem [{\citenamefont {Gasser}\ and\ \citenamefont
  {Leutwyler}(1985)}]{Gasser:1984gg}%
  \BibitemOpen
  \bibfield  {author} {\bibinfo {author} {\bibfnamefont {J.}~\bibnamefont
  {Gasser}}\ and\ \bibinfo {author} {\bibfnamefont {H.}~\bibnamefont
  {Leutwyler}},\ }\href {https://doi.org/10.1016/0550-3213(85)90492-4}
  {\bibfield  {journal} {\bibinfo  {journal} {Nucl.Phys.}\ }\textbf {\bibinfo
  {volume} {B250}},\ \bibinfo {pages} {465} (\bibinfo {year}
  {1985})}\BibitemShut {NoStop}%
\bibitem [{\citenamefont {Epelbaum}\ \emph {et~al.}(2009)\citenamefont
  {Epelbaum}, \citenamefont {Hammer},\ and\ \citenamefont
  {Mei{\ss}ner}}]{Epelbaum:2008ga}%
  \BibitemOpen
  \bibfield  {author} {\bibinfo {author} {\bibfnamefont {E.}~\bibnamefont
  {Epelbaum}}, \bibinfo {author} {\bibfnamefont {H.-W.}\ \bibnamefont
  {Hammer}},\ and\ \bibinfo {author} {\bibfnamefont {U.-G.}\ \bibnamefont
  {Mei{\ss}ner}},\ }\href {https://doi.org/10.1103/RevModPhys.81.1773}
  {\bibfield  {journal} {\bibinfo  {journal} {Rev.Mod.Phys.}\ }\textbf
  {\bibinfo {volume} {81}},\ \bibinfo {pages} {1773} (\bibinfo {year}
  {2009})},\ \Eprint {https://arxiv.org/abs/0811.1338} {arXiv:0811.1338
  [nucl-th]} \BibitemShut {NoStop}%
\bibitem [{\citenamefont {Oller}(2020)}]{Oller:2020guq}%
  \BibitemOpen
  \bibfield  {author} {\bibinfo {author} {\bibfnamefont {J.~A.}\ \bibnamefont
  {Oller}},\ }\href {https://doi.org/10.3390/sym12071114} {\bibfield  {journal}
  {\bibinfo  {journal} {Symmetry}\ }\textbf {\bibinfo {volume} {12}},\ \bibinfo
  {pages} {1114} (\bibinfo {year} {2020})},\ \Eprint
  {https://arxiv.org/abs/2005.14417} {arXiv:2005.14417 [hep-ph]} \BibitemShut
  {NoStop}%
\bibitem [{\citenamefont {Roy}(1971)}]{Roy:1971tc}%
  \BibitemOpen
  \bibfield  {author} {\bibinfo {author} {\bibfnamefont {S.~M.}\ \bibnamefont
  {Roy}},\ }\href {https://doi.org/10.1016/0370-2693(71)90724-6} {\bibfield
  {journal} {\bibinfo  {journal} {Phys.Lett.}\ }\textbf {\bibinfo {volume}
  {36B}},\ \bibinfo {pages} {353} (\bibinfo {year} {1971})}\BibitemShut
  {NoStop}%
%%CITATION = PHLTA,36B,353;%%
\bibitem [{\citenamefont {Hite}\ and\ \citenamefont
  {Steiner}(1973)}]{Hite:1973pm}%
  \BibitemOpen
  \bibfield  {author} {\bibinfo {author} {\bibfnamefont {G.~E.}\ \bibnamefont
  {Hite}}\ and\ \bibinfo {author} {\bibfnamefont {F.}~\bibnamefont {Steiner}},\
  }\href {https://doi.org/10.1007/BF02722827} {\bibfield  {journal} {\bibinfo
  {journal} {Nuovo Cim. A}\ }\textbf {\bibinfo {volume} {18}},\ \bibinfo
  {pages} {237} (\bibinfo {year} {1973})}\BibitemShut {NoStop}%
\bibitem [{\citenamefont {Pel\'aez}(2016)}]{Pelaez:2015qba}%
  \BibitemOpen
  \bibfield  {author} {\bibinfo {author} {\bibfnamefont {J.~R.}\ \bibnamefont
  {Pel\'aez}},\ }\href {https://doi.org/10.1016/j.physrep.2016.09.001}
  {\bibfield  {journal} {\bibinfo  {journal} {Phys.Rept.}\ }\textbf {\bibinfo
  {volume} {658}},\ \bibinfo {pages} {1} (\bibinfo {year} {2016})},\ \Eprint
  {https://arxiv.org/abs/1510.00653} {arXiv:1510.00653 [hep-ph]} \BibitemShut
  {NoStop}%
%%CITATION = ARXIV:1510.00653;%%
\bibitem [{\citenamefont {Caprini}\ \emph {et~al.}(2006)\citenamefont
  {Caprini}, \citenamefont {Colangelo},\ and\ \citenamefont
  {Leutwyler}}]{Caprini:2005zr}%
  \BibitemOpen
  \bibfield  {author} {\bibinfo {author} {\bibfnamefont {I.}~\bibnamefont
  {Caprini}}, \bibinfo {author} {\bibfnamefont {G.}~\bibnamefont {Colangelo}},\
  and\ \bibinfo {author} {\bibfnamefont {H.}~\bibnamefont {Leutwyler}},\ }\href
  {https://doi.org/10.1103/PhysRevLett.96.132001} {\bibfield  {journal}
  {\bibinfo  {journal} {Phys.Rev.Lett.}\ }\textbf {\bibinfo {volume} {96}},\
  \bibinfo {pages} {132001} (\bibinfo {year} {2006})},\ \Eprint
  {https://arxiv.org/abs/hep-ph/0512364} {arXiv:hep-ph/0512364 [hep-ph]}
  \BibitemShut {NoStop}%
%%CITATION = HEP-PH/0512364;%%
\bibitem [{\citenamefont {Garc\'ia-Mart\'in}\ \emph {et~al.}(2011)\citenamefont
  {Garc\'ia-Mart\'in}, \citenamefont {Kaminski}, \citenamefont {Pel\'aez},\
  and\ \citenamefont {Ruiz~de Elvira}}]{GarciaMartin:2011nna}%
  \BibitemOpen
  \bibfield  {author} {\bibinfo {author} {\bibfnamefont {R.}~\bibnamefont
  {Garc\'ia-Mart\'in}}, \bibinfo {author} {\bibfnamefont {R.}~\bibnamefont
  {Kaminski}}, \bibinfo {author} {\bibfnamefont {J.~R.}\ \bibnamefont
  {Pel\'aez}},\ and\ \bibinfo {author} {\bibfnamefont {J.}~\bibnamefont
  {Ruiz~de Elvira}},\ }\href {https://doi.org/10.1103/PhysRevLett.107.072001}
  {\bibfield  {journal} {\bibinfo  {journal} {Phys.Rev.Lett.}\ }\textbf
  {\bibinfo {volume} {107}},\ \bibinfo {pages} {072001} (\bibinfo {year}
  {2011})},\ \Eprint {https://arxiv.org/abs/1107.1635} {arXiv:1107.1635
  [hep-ph]} \BibitemShut {NoStop}%
%%CITATION = ARXIV:1107.1635;%%
\bibitem [{\citenamefont {Moussallam}(2011)}]{Moussallam:2011zg}%
  \BibitemOpen
  \bibfield  {author} {\bibinfo {author} {\bibfnamefont {B.}~\bibnamefont
  {Moussallam}},\ }\href {https://doi.org/10.1140/epjc/s10052-011-1814-z}
  {\bibfield  {journal} {\bibinfo  {journal} {Eur.Phys.J.}\ }\textbf {\bibinfo
  {volume} {C71}},\ \bibinfo {pages} {1814} (\bibinfo {year} {2011})},\ \Eprint
  {https://arxiv.org/abs/1110.6074} {arXiv:1110.6074 [hep-ph]} \BibitemShut
  {NoStop}%
%%CITATION = ARXIV:1110.6074;%%
\bibitem [{\citenamefont {Pel\'aez}\ and\ \citenamefont
  {Rodas}(2020)}]{Pelaez:2020uiw}%
  \BibitemOpen
  \bibfield  {author} {\bibinfo {author} {\bibfnamefont {J.}~\bibnamefont
  {Pel\'aez}}\ and\ \bibinfo {author} {\bibfnamefont {A.}~\bibnamefont
  {Rodas}},\ }\href {https://doi.org/10.1103/PhysRevLett.124.172001} {\bibfield
   {journal} {\bibinfo  {journal} {Phys.Rev.Lett.}\ }\textbf {\bibinfo {volume}
  {124}},\ \bibinfo {pages} {172001} (\bibinfo {year} {2020})},\ \Eprint
  {https://arxiv.org/abs/2001.08153} {arXiv:2001.08153 [hep-ph]} \BibitemShut
  {NoStop}%
\bibitem [{\citenamefont {Rodas}\ \emph {et~al.}(2022)\citenamefont {Rodas},
  \citenamefont {Pilloni}, \citenamefont {Albaladejo}, \citenamefont
  {Fernandez-Ramirez}, \citenamefont {Mathieu},\ and\ \citenamefont
  {Szczepaniak}}]{Rodas:2021tyb}%
  \BibitemOpen
  \bibfield  {author} {\bibinfo {author} {\bibfnamefont {A.}~\bibnamefont
  {Rodas}}, \bibinfo {author} {\bibfnamefont {A.}~\bibnamefont {Pilloni}},
  \bibinfo {author} {\bibfnamefont {M.}~\bibnamefont {Albaladejo}}, \bibinfo
  {author} {\bibfnamefont {C.}~\bibnamefont {Fernandez-Ramirez}}, \bibinfo
  {author} {\bibfnamefont {V.}~\bibnamefont {Mathieu}},\ and\ \bibinfo {author}
  {\bibfnamefont {A.~P.}\ \bibnamefont {Szczepaniak}} (\bibinfo {collaboration}
  {Joint Physics Analysis Center}),\ }\href
  {https://doi.org/10.1140/epjc/s10052-022-10014-8} {\bibfield  {journal}
  {\bibinfo  {journal} {Eur.Phys.J.}\ }\textbf {\bibinfo {volume} {C82}},\
  \bibinfo {pages} {80} (\bibinfo {year} {2022})},\ \Eprint
  {https://arxiv.org/abs/2110.00027} {arXiv:2110.00027 [hep-ph]} \BibitemShut
  {NoStop}%
\bibitem [{\citenamefont {Mathieu}\ \emph {et~al.}(2009)\citenamefont
  {Mathieu}, \citenamefont {Kochelev},\ and\ \citenamefont
  {Vento}}]{Mathieu:2008me}%
  \BibitemOpen
  \bibfield  {author} {\bibinfo {author} {\bibfnamefont {V.}~\bibnamefont
  {Mathieu}}, \bibinfo {author} {\bibfnamefont {N.}~\bibnamefont {Kochelev}},\
  and\ \bibinfo {author} {\bibfnamefont {V.}~\bibnamefont {Vento}},\ }\href
  {https://doi.org/10.1142/S0218301309012124} {\bibfield  {journal} {\bibinfo
  {journal} {Int.J.Mod.Phys.}\ }\textbf {\bibinfo {volume} {E18}},\ \bibinfo
  {pages} {1} (\bibinfo {year} {2009})},\ \Eprint
  {https://arxiv.org/abs/0810.4453} {arXiv:0810.4453 [hep-ph]} \BibitemShut
  {NoStop}%
%%CITATION = ARXIV:0810.4453;%%
\bibitem [{\citenamefont {Llanes-Estrada}(2021)}]{Llanes-Estrada:2021evz}%
  \BibitemOpen
  \bibfield  {author} {\bibinfo {author} {\bibfnamefont {F.~J.}\ \bibnamefont
  {Llanes-Estrada}},\ }\bibfield  {journal} {\bibinfo  {journal}
  {Eur.Phys.J.ST}\ }\href {https://doi.org/10.1140/epjs/s11734-021-00143-8}
  {10.1140/epjs/s11734-021-00143-8} (\bibinfo {year} {2021}),\ \Eprint
  {https://arxiv.org/abs/2101.05366} {arXiv:2101.05366 [hep-ph]} \BibitemShut
  {NoStop}%
\bibitem [{\citenamefont {Meyer}\ and\ \citenamefont
  {Swanson}(2015)}]{Meyer:2015eta}%
  \BibitemOpen
  \bibfield  {author} {\bibinfo {author} {\bibfnamefont {C.~A.}\ \bibnamefont
  {Meyer}}\ and\ \bibinfo {author} {\bibfnamefont {E.~S.}\ \bibnamefont
  {Swanson}},\ }\href {https://doi.org/10.1016/j.ppnp.2015.03.001} {\bibfield
  {journal} {\bibinfo  {journal} {Prog.Part.Nucl.Phys.}\ }\textbf {\bibinfo
  {volume} {82}},\ \bibinfo {pages} {21} (\bibinfo {year} {2015})},\ \Eprint
  {https://arxiv.org/abs/1502.07276} {arXiv:1502.07276 [hep-ph]} \BibitemShut
  {NoStop}%
%%CITATION = ARXIV:1502.07276;%%
\bibitem [{\citenamefont {Morningstar}\ and\ \citenamefont
  {Peardon}(1999)}]{Morningstar:1999rf}%
  \BibitemOpen
  \bibfield  {author} {\bibinfo {author} {\bibfnamefont {C.~J.}\ \bibnamefont
  {Morningstar}}\ and\ \bibinfo {author} {\bibfnamefont {M.~J.}\ \bibnamefont
  {Peardon}},\ }\href {https://doi.org/10.1103/PhysRevD.60.034509} {\bibfield
  {journal} {\bibinfo  {journal} {Phys.Rev.}\ }\textbf {\bibinfo {volume}
  {D60}},\ \bibinfo {pages} {034509} (\bibinfo {year} {1999})},\ \Eprint
  {https://arxiv.org/abs/hep-lat/9901004} {arXiv:hep-lat/9901004 [hep-lat]}
  \BibitemShut {NoStop}%
%%CITATION = HEP-LAT/9901004;%%
\bibitem [{\citenamefont {Szczepaniak}\ and\ \citenamefont
  {Swanson}(2003)}]{Szczepaniak:2003mr}%
  \BibitemOpen
  \bibfield  {author} {\bibinfo {author} {\bibfnamefont {A.~P.}\ \bibnamefont
  {Szczepaniak}}\ and\ \bibinfo {author} {\bibfnamefont {E.~S.}\ \bibnamefont
  {Swanson}},\ }\href {https://doi.org/10.1016/j.physletb.2003.10.008}
  {\bibfield  {journal} {\bibinfo  {journal} {Phys.Lett.}\ }\textbf {\bibinfo
  {volume} {B577}},\ \bibinfo {pages} {61} (\bibinfo {year} {2003})},\ \Eprint
  {https://arxiv.org/abs/hep-ph/0308268} {arXiv:hep-ph/0308268} \BibitemShut
  {NoStop}%
\bibitem [{\citenamefont {Athenodorou}\ and\ \citenamefont
  {Teper}(2020)}]{Athenodorou:2020ani}%
  \BibitemOpen
  \bibfield  {author} {\bibinfo {author} {\bibfnamefont {A.}~\bibnamefont
  {Athenodorou}}\ and\ \bibinfo {author} {\bibfnamefont {M.}~\bibnamefont
  {Teper}},\ }\href {https://doi.org/10.1007/JHEP11(2020)172} {\bibfield
  {journal} {\bibinfo  {journal} {JHEP}\ }\textbf {\bibinfo {volume} {11}},\
  \bibinfo {pages} {172}},\ \Eprint {https://arxiv.org/abs/2007.06422}
  {arXiv:2007.06422 [hep-lat]} \BibitemShut {NoStop}%
\bibitem [{\citenamefont {Ablikim}\ \emph {et~al.}(2015)\citenamefont {Ablikim}
  \emph {et~al.}}]{BESIII:2015rug}%
  \BibitemOpen
  \bibfield  {author} {\bibinfo {author} {\bibfnamefont {M.}~\bibnamefont
  {Ablikim}} \emph {et~al.} (\bibinfo {collaboration} {BESIII}),\ }\href
  {https://doi.org/10.1103/PhysRevD.92.052003} {\bibfield  {journal} {\bibinfo
  {journal} {Phys.Rev.}\ }\textbf {\bibinfo {volume} {D92}},\ \bibinfo {pages}
  {052003} (\bibinfo {year} {2015})},\ \bibinfo {note} {[Erratum:
  Phys.Rev.D93,no.3,039906(2016)]},\ \Eprint {https://arxiv.org/abs/1506.00546}
  {arXiv:1506.00546 [hep-ex]} \BibitemShut {NoStop}%
%%CITATION = ARXIV:1506.00546;%%
\bibitem [{\citenamefont {Ablikim}\ \emph {et~al.}(2018)\citenamefont {Ablikim}
  \emph {et~al.}}]{BESIII:2018ubj}%
  \BibitemOpen
  \bibfield  {author} {\bibinfo {author} {\bibfnamefont {M.}~\bibnamefont
  {Ablikim}} \emph {et~al.} (\bibinfo {collaboration} {BESIII}),\ }\href
  {https://doi.org/10.1103/PhysRevD.98.072003} {\bibfield  {journal} {\bibinfo
  {journal} {Phys.Rev.}\ }\textbf {\bibinfo {volume} {D98}},\ \bibinfo {pages}
  {072003} (\bibinfo {year} {2018})},\ \Eprint
  {https://arxiv.org/abs/1808.06946} {arXiv:1808.06946 [hep-ex]} \BibitemShut
  {NoStop}%
\bibitem [{\citenamefont {Alexeev}\ \emph {et~al.}(2022)\citenamefont {Alexeev}
  \emph {et~al.}}]{COMPASS:2021ogp}%
  \BibitemOpen
  \bibfield  {author} {\bibinfo {author} {\bibfnamefont {G.~D.}\ \bibnamefont
  {Alexeev}} \emph {et~al.} (\bibinfo {collaboration} {COMPASS}),\ }\href
  {https://doi.org/10.1103/PhysRevD.105.012005} {\bibfield  {journal} {\bibinfo
   {journal} {Phys.Rev.}\ }\textbf {\bibinfo {volume} {D105}},\ \bibinfo
  {pages} {012005} (\bibinfo {year} {2022})},\ \Eprint
  {https://arxiv.org/abs/2108.01744} {arXiv:2108.01744 [hep-ex]} \BibitemShut
  {NoStop}%
\bibitem [{\citenamefont {Ablikim}\ \emph {et~al.}(2022)\citenamefont {Ablikim}
  \emph {et~al.}}]{BESIII:2022riz}%
  \BibitemOpen
  \bibfield  {author} {\bibinfo {author} {\bibfnamefont {M.}~\bibnamefont
  {Ablikim}} \emph {et~al.} (\bibinfo {collaboration} {BESIII}),\ }\Eprint
  {https://arxiv.org/abs/2202.00621} {arXiv:2202.00621 [hep-ex]}  (\bibinfo
  {year} {2022})\BibitemShut {NoStop}%
\bibitem [{\citenamefont {Close}\ and\ \citenamefont
  {Lipkin}(1987)}]{Close:1987aw}%
  \BibitemOpen
  \bibfield  {author} {\bibinfo {author} {\bibfnamefont {F.~E.}\ \bibnamefont
  {Close}}\ and\ \bibinfo {author} {\bibfnamefont {H.~J.}\ \bibnamefont
  {Lipkin}},\ }\href {https://doi.org/10.1016/0370-2693(87)90613-7} {\bibfield
  {journal} {\bibinfo  {journal} {Phys.Lett.}\ }\textbf {\bibinfo {volume}
  {B196}},\ \bibinfo {pages} {245} (\bibinfo {year} {1987})}\BibitemShut
  {NoStop}%
%%CITATION = PHLTA,B196,245;%%
\bibitem [{\citenamefont {Adolph}\ \emph
  {et~al.}(2015{\natexlab{a}})\citenamefont {Adolph} \emph
  {et~al.}}]{COMPASS:2014vkj}%
  \BibitemOpen
  \bibfield  {author} {\bibinfo {author} {\bibfnamefont {C.}~\bibnamefont
  {Adolph}} \emph {et~al.} (\bibinfo {collaboration} {COMPASS}),\ }\href
  {https://doi.org/10.1016/j.physletb.2014.11.058} {\bibfield  {journal}
  {\bibinfo  {journal} {Phys.Lett.}\ }\textbf {\bibinfo {volume} {B740}},\
  \bibinfo {pages} {303} (\bibinfo {year} {2015}{\natexlab{a}})},\ \bibinfo
  {note} {[Corrigendum: Phys.Lett.B 811, 135913 (2020)]},\ \Eprint
  {https://arxiv.org/abs/1408.4286} {arXiv:1408.4286 [hep-ex]} \BibitemShut
  {NoStop}%
%%CITATION = ARXIV:1408.4286;%%
\bibitem [{\citenamefont {Jackura}\ \emph {et~al.}(2018)\citenamefont {Jackura}
  \emph {et~al.}}]{JPAC:2017dbi}%
  \BibitemOpen
  \bibfield  {author} {\bibinfo {author} {\bibfnamefont {A.}~\bibnamefont
  {Jackura}} \emph {et~al.} (\bibinfo {collaboration} {COMPASS and JPAC}),\
  }\href {https://doi.org/10.1016/j.physletb.2018.01.017} {\bibfield  {journal}
  {\bibinfo  {journal} {Phys.Lett.}\ }\textbf {\bibinfo {volume} {B779}},\
  \bibinfo {pages} {464–472} (\bibinfo {year} {2018})},\ \Eprint
  {https://arxiv.org/abs/1707.02848} {arXiv:1707.02848 [hep-ph]} \BibitemShut
  {NoStop}%
%%CITATION = ARXIV:1707.02848;%%
\bibitem [{\citenamefont {Rodas}\ \emph {et~al.}(2019)\citenamefont {Rodas}
  \emph {et~al.}}]{JPAC:2018zyd}%
  \BibitemOpen
  \bibfield  {author} {\bibinfo {author} {\bibfnamefont {A.}~\bibnamefont
  {Rodas}} \emph {et~al.} (\bibinfo {collaboration} {JPAC}),\ }\href
  {https://doi.org/10.1103/PhysRevLett.122.042002} {\bibfield  {journal}
  {\bibinfo  {journal} {Phys.Rev.Lett.}\ }\textbf {\bibinfo {volume} {122}},\
  \bibinfo {pages} {042002} (\bibinfo {year} {2019})},\ \Eprint
  {https://arxiv.org/abs/1810.04171} {arXiv:1810.04171 [hep-ph]} \BibitemShut
  {NoStop}%
%%CITATION = ARXIV:1810.04171;%%
\bibitem [{\citenamefont {Chew}\ and\ \citenamefont
  {Mandelstam}(1960)}]{Chew:1960iv}%
  \BibitemOpen
  \bibfield  {author} {\bibinfo {author} {\bibfnamefont {G.~F.}\ \bibnamefont
  {Chew}}\ and\ \bibinfo {author} {\bibfnamefont {S.}~\bibnamefont
  {Mandelstam}},\ }\href {https://doi.org/10.1103/PhysRev.119.467} {\bibfield
  {journal} {\bibinfo  {journal} {Phys.Rev.}\ }\textbf {\bibinfo {volume}
  {119}},\ \bibinfo {pages} {467} (\bibinfo {year} {1960})}\BibitemShut
  {NoStop}%
%%CITATION = PHRVA,119,467;%%
\bibitem [{\citenamefont {Woss}\ \emph {et~al.}(2021)\citenamefont {Woss},
  \citenamefont {Dudek}, \citenamefont {Edwards}, \citenamefont {Thomas},\ and\
  \citenamefont {Wilson}}]{Woss:2020ayi}%
  \BibitemOpen
  \bibfield  {author} {\bibinfo {author} {\bibfnamefont {A.~J.}\ \bibnamefont
  {Woss}}, \bibinfo {author} {\bibfnamefont {J.~J.}\ \bibnamefont {Dudek}},
  \bibinfo {author} {\bibfnamefont {R.~G.}\ \bibnamefont {Edwards}}, \bibinfo
  {author} {\bibfnamefont {C.~E.}\ \bibnamefont {Thomas}},\ and\ \bibinfo
  {author} {\bibfnamefont {D.~J.}\ \bibnamefont {Wilson}} (\bibinfo
  {collaboration} {Hadron Spectrum}),\ }\href
  {https://doi.org/10.1103/PhysRevD.103.054502} {\bibfield  {journal} {\bibinfo
   {journal} {Phys.Rev.}\ }\textbf {\bibinfo {volume} {D103}},\ \bibinfo
  {pages} {054502} (\bibinfo {year} {2021})},\ \Eprint
  {https://arxiv.org/abs/2009.10034} {arXiv:2009.10034 [hep-lat]} \BibitemShut
  {NoStop}%
\bibitem [{\citenamefont {Kopf}\ \emph {et~al.}(2021)\citenamefont {Kopf},
  \citenamefont {Albrecht}, \citenamefont {Koch}, \citenamefont {K\"u\ss{}ner},
  \citenamefont {Pychy}, \citenamefont {Qin},\ and\ \citenamefont
  {Wiedner}}]{Kopf:2020yoa}%
  \BibitemOpen
  \bibfield  {author} {\bibinfo {author} {\bibfnamefont {B.}~\bibnamefont
  {Kopf}}, \bibinfo {author} {\bibfnamefont {M.}~\bibnamefont {Albrecht}},
  \bibinfo {author} {\bibfnamefont {H.}~\bibnamefont {Koch}}, \bibinfo {author}
  {\bibfnamefont {M.}~\bibnamefont {K\"u\ss{}ner}}, \bibinfo {author}
  {\bibfnamefont {J.}~\bibnamefont {Pychy}}, \bibinfo {author} {\bibfnamefont
  {X.}~\bibnamefont {Qin}},\ and\ \bibinfo {author} {\bibfnamefont
  {U.}~\bibnamefont {Wiedner}},\ }\href
  {https://doi.org/10.1140/epjc/s10052-021-09821-2} {\bibfield  {journal}
  {\bibinfo  {journal} {Eur.Phys.J.}\ }\textbf {\bibinfo {volume} {C81}},\
  \bibinfo {pages} {1056} (\bibinfo {year} {2021})},\ \Eprint
  {https://arxiv.org/abs/2008.11566} {arXiv:2008.11566 [hep-ph]} \BibitemShut
  {NoStop}%
\bibitem [{\citenamefont {Aghasyan}\ \emph {et~al.}(2018)\citenamefont
  {Aghasyan} \emph {et~al.}}]{COMPASS:2018uzl}%
  \BibitemOpen
  \bibfield  {author} {\bibinfo {author} {\bibfnamefont {M.}~\bibnamefont
  {Aghasyan}} \emph {et~al.} (\bibinfo {collaboration} {COMPASS}),\ }\href
  {https://doi.org/10.1103/PhysRevD.98.092003} {\bibfield  {journal} {\bibinfo
  {journal} {Phys.Rev.}\ }\textbf {\bibinfo {volume} {D98}},\ \bibinfo {pages}
  {092003} (\bibinfo {year} {2018})},\ \Eprint
  {https://arxiv.org/abs/1802.05913} {arXiv:1802.05913 [hep-ex]} \BibitemShut
  {NoStop}%
%%CITATION = ARXIV:1802.05913;%%
\bibitem [{\citenamefont {Choi}\ \emph {et~al.}(2003)\citenamefont {Choi} \emph
  {et~al.}}]{Belle:2003nnu}%
  \BibitemOpen
  \bibfield  {author} {\bibinfo {author} {\bibfnamefont {S.-K.}\ \bibnamefont
  {Choi}} \emph {et~al.} (\bibinfo {collaboration} {\belle}),\ }\href
  {https://doi.org/10.1103/PhysRevLett.91.262001} {\bibfield  {journal}
  {\bibinfo  {journal} {Phys.Rev.Lett.}\ }\textbf {\bibinfo {volume} {91}},\
  \bibinfo {pages} {262001} (\bibinfo {year} {2003})},\ \Eprint
  {https://arxiv.org/abs/hep-ex/0309032} {arXiv:hep-ex/0309032} \BibitemShut
  {NoStop}%
%%CITATION = HEP-EX/0309032;%%
\bibitem [{\citenamefont {Aaij}\ \emph
  {et~al.}(2021{\natexlab{a}})\citenamefont {Aaij} \emph
  {et~al.}}]{LHCb:2021vvq}%
  \BibitemOpen
  \bibfield  {author} {\bibinfo {author} {\bibfnamefont {R.}~\bibnamefont
  {Aaij}} \emph {et~al.} (\bibinfo {collaboration} {LHCb}),\ }\Eprint
  {https://arxiv.org/abs/2109.01038} {arXiv:2109.01038 [hep-ex]}  (\bibinfo
  {year} {2021}{\natexlab{a}})\BibitemShut {NoStop}%
\bibitem [{\citenamefont {Aaij}\ \emph
  {et~al.}(2021{\natexlab{b}})\citenamefont {Aaij} \emph
  {et~al.}}]{LHCb:2021auc}%
  \BibitemOpen
  \bibfield  {author} {\bibinfo {author} {\bibfnamefont {R.}~\bibnamefont
  {Aaij}} \emph {et~al.} (\bibinfo {collaboration} {LHCb}),\ }\Eprint
  {https://arxiv.org/abs/2109.01056} {arXiv:2109.01056 [hep-ex]}  (\bibinfo
  {year} {2021}{\natexlab{b}})\BibitemShut {NoStop}%
\bibitem [{\citenamefont {Aaij}\ \emph
  {et~al.}(2020{\natexlab{a}})\citenamefont {Aaij} \emph
  {et~al.}}]{LHCb:2020bwg}%
  \BibitemOpen
  \bibfield  {author} {\bibinfo {author} {\bibfnamefont {R.}~\bibnamefont
  {Aaij}} \emph {et~al.} (\bibinfo {collaboration} {LHCb}),\ }\href
  {https://doi.org/10.1016/j.scib.2020.08.032} {\bibfield  {journal} {\bibinfo
  {journal} {Sci.Bull.}\ }\textbf {\bibinfo {volume} {65}},\ \bibinfo {pages}
  {1983} (\bibinfo {year} {2020}{\natexlab{a}})},\ \Eprint
  {https://arxiv.org/abs/2006.16957} {arXiv:2006.16957 [hep-ex]} \BibitemShut
  {NoStop}%
\bibitem [{\citenamefont {Aaij}\ \emph
  {et~al.}(2020{\natexlab{b}})\citenamefont {Aaij} \emph
  {et~al.}}]{LHCb:2020xds}%
  \BibitemOpen
  \bibfield  {author} {\bibinfo {author} {\bibfnamefont {R.}~\bibnamefont
  {Aaij}} \emph {et~al.} (\bibinfo {collaboration} {LHCb}),\ }\href
  {https://doi.org/10.1103/PhysRevD.102.092005} {\bibfield  {journal} {\bibinfo
   {journal} {Phys.Rev.}\ }\textbf {\bibinfo {volume} {D102}},\ \bibinfo
  {pages} {092005} (\bibinfo {year} {2020}{\natexlab{b}})},\ \Eprint
  {https://arxiv.org/abs/2005.13419} {arXiv:2005.13419 [hep-ex]} \BibitemShut
  {NoStop}%
\bibitem [{\citenamefont {Esposito}\ \emph {et~al.}(2022)\citenamefont
  {Esposito}, \citenamefont {Maiani}, \citenamefont {Pilloni}, \citenamefont
  {Polosa},\ and\ \citenamefont {Riquer}}]{Esposito:2021vhu}%
  \BibitemOpen
  \bibfield  {author} {\bibinfo {author} {\bibfnamefont {A.}~\bibnamefont
  {Esposito}}, \bibinfo {author} {\bibfnamefont {L.}~\bibnamefont {Maiani}},
  \bibinfo {author} {\bibfnamefont {A.}~\bibnamefont {Pilloni}}, \bibinfo
  {author} {\bibfnamefont {A.~D.}\ \bibnamefont {Polosa}},\ and\ \bibinfo
  {author} {\bibfnamefont {V.}~\bibnamefont {Riquer}},\ }\href
  {https://doi.org/10.1103/PhysRevD.105.L031503} {\bibfield  {journal}
  {\bibinfo  {journal} {Phys.Rev.}\ }\textbf {\bibinfo {volume} {D105}},\
  \bibinfo {pages} {L031503} (\bibinfo {year} {2022})},\ \Eprint
  {https://arxiv.org/abs/2108.11413} {arXiv:2108.11413 [hep-ph]} \BibitemShut
  {NoStop}%
\bibitem [{\citenamefont {Baru}\ \emph {et~al.}(2021)\citenamefont {Baru},
  \citenamefont {Dong}, \citenamefont {Du}, \citenamefont {Filin},
  \citenamefont {Guo}, \citenamefont {Hanhart}, \citenamefont {Nefediev},
  \citenamefont {Nieves},\ and\ \citenamefont {Wang}}]{Baru:2021ldu}%
  \BibitemOpen
  \bibfield  {author} {\bibinfo {author} {\bibfnamefont {V.}~\bibnamefont
  {Baru}}, \bibinfo {author} {\bibfnamefont {X.-K.}\ \bibnamefont {Dong}},
  \bibinfo {author} {\bibfnamefont {M.-L.}\ \bibnamefont {Du}}, \bibinfo
  {author} {\bibfnamefont {A.}~\bibnamefont {Filin}}, \bibinfo {author}
  {\bibfnamefont {F.-K.}\ \bibnamefont {Guo}}, \bibinfo {author} {\bibfnamefont
  {C.}~\bibnamefont {Hanhart}}, \bibinfo {author} {\bibfnamefont
  {A.}~\bibnamefont {Nefediev}}, \bibinfo {author} {\bibfnamefont
  {J.}~\bibnamefont {Nieves}},\ and\ \bibinfo {author} {\bibfnamefont
  {Q.}~\bibnamefont {Wang}},\ }\Eprint {https://arxiv.org/abs/2110.07484}
  {arXiv:2110.07484 [hep-ph]}  (\bibinfo {year} {2021})\BibitemShut {NoStop}%
\bibitem [{\citenamefont {Tornqvist}(1994)}]{Tornqvist:1993ng}%
  \BibitemOpen
  \bibfield  {author} {\bibinfo {author} {\bibfnamefont {N.~A.}\ \bibnamefont
  {Tornqvist}},\ }\href {https://doi.org/10.1007/BF01413192} {\bibfield
  {journal} {\bibinfo  {journal} {Z.Phys.}\ }\textbf {\bibinfo {volume}
  {C61}},\ \bibinfo {pages} {525} (\bibinfo {year} {1994})},\ \Eprint
  {https://arxiv.org/abs/hep-ph/9310247} {arXiv:hep-ph/9310247 [hep-ph]}
  \BibitemShut {NoStop}%
%%CITATION = HEP-PH/9310247;%%
\bibitem [{\citenamefont {Braaten}\ and\ \citenamefont
  {Kusunoki}(2004)}]{Braaten:2003he}%
  \BibitemOpen
  \bibfield  {author} {\bibinfo {author} {\bibfnamefont {E.}~\bibnamefont
  {Braaten}}\ and\ \bibinfo {author} {\bibfnamefont {M.}~\bibnamefont
  {Kusunoki}},\ }\href {https://doi.org/10.1103/PhysRevD.69.074005} {\bibfield
  {journal} {\bibinfo  {journal} {Phys.Rev.}\ }\textbf {\bibinfo {volume}
  {D69}},\ \bibinfo {pages} {074005} (\bibinfo {year} {2004})},\ \Eprint
  {https://arxiv.org/abs/hep-ph/0311147} {arXiv:hep-ph/0311147 [hep-ph]}
  \BibitemShut {NoStop}%
\bibitem [{\citenamefont {Voloshin}(2004)}]{Voloshin:2003nt}%
  \BibitemOpen
  \bibfield  {author} {\bibinfo {author} {\bibfnamefont {M.}~\bibnamefont
  {Voloshin}},\ }\href {https://doi.org/10.1016/j.physletb.2003.11.014}
  {\bibfield  {journal} {\bibinfo  {journal} {Phys.Lett.}\ }\textbf {\bibinfo
  {volume} {B579}},\ \bibinfo {pages} {316} (\bibinfo {year} {2004})},\ \Eprint
  {https://arxiv.org/abs/hep-ph/0309307} {arXiv:hep-ph/0309307 [hep-ph]}
  \BibitemShut {NoStop}%
%%CITATION = HEP-PH/0309307;%%
\bibitem [{\citenamefont {Close}\ and\ \citenamefont
  {Page}(2004)}]{Close:2003sg}%
  \BibitemOpen
  \bibfield  {author} {\bibinfo {author} {\bibfnamefont {F.~E.}\ \bibnamefont
  {Close}}\ and\ \bibinfo {author} {\bibfnamefont {P.~R.}\ \bibnamefont
  {Page}},\ }\href {https://doi.org/10.1016/j.physletb.2003.10.032} {\bibfield
  {journal} {\bibinfo  {journal} {Phys.Lett.}\ }\textbf {\bibinfo {volume}
  {B578}},\ \bibinfo {pages} {119} (\bibinfo {year} {2004})},\ \Eprint
  {https://arxiv.org/abs/hep-ph/0309253} {arXiv:hep-ph/0309253 [hep-ph]}
  \BibitemShut {NoStop}%
\bibitem [{\citenamefont {Swanson}(2006)}]{Swanson:2006st}%
  \BibitemOpen
  \bibfield  {author} {\bibinfo {author} {\bibfnamefont {E.~S.}\ \bibnamefont
  {Swanson}},\ }\href {https://doi.org/10.1016/j.physrep.2006.04.003}
  {\bibfield  {journal} {\bibinfo  {journal} {Phys.Rept.}\ }\textbf {\bibinfo
  {volume} {429}},\ \bibinfo {pages} {243} (\bibinfo {year} {2006})},\ \Eprint
  {https://arxiv.org/abs/hep-ph/0601110} {arXiv:hep-ph/0601110 [hep-ph]}
  \BibitemShut {NoStop}%
\bibitem [{\citenamefont {Maiani}\ \emph {et~al.}(2005)\citenamefont {Maiani},
  \citenamefont {Piccinini}, \citenamefont {Polosa},\ and\ \citenamefont
  {Riquer}}]{Maiani:2004vq}%
  \BibitemOpen
  \bibfield  {author} {\bibinfo {author} {\bibfnamefont {L.}~\bibnamefont
  {Maiani}}, \bibinfo {author} {\bibfnamefont {F.}~\bibnamefont {Piccinini}},
  \bibinfo {author} {\bibfnamefont {A.~D.}\ \bibnamefont {Polosa}},\ and\
  \bibinfo {author} {\bibfnamefont {V.}~\bibnamefont {Riquer}},\ }\href
  {https://doi.org/10.1103/PhysRevD.71.014028} {\bibfield  {journal} {\bibinfo
  {journal} {Phys.Rev.}\ }\textbf {\bibinfo {volume} {D71}},\ \bibinfo {pages}
  {014028} (\bibinfo {year} {2005})},\ \Eprint
  {https://arxiv.org/abs/hep-ph/0412098} {arXiv:hep-ph/0412098 [hep-ph]}
  \BibitemShut {NoStop}%
%%CITATION = HEP-PH/0412098;%%
\bibitem [{\citenamefont {Ali}\ \emph {et~al.}(2012)\citenamefont {Ali},
  \citenamefont {Hambrock},\ and\ \citenamefont {Wang}}]{Ali:2011ug}%
  \BibitemOpen
  \bibfield  {author} {\bibinfo {author} {\bibfnamefont {A.}~\bibnamefont
  {Ali}}, \bibinfo {author} {\bibfnamefont {C.}~\bibnamefont {Hambrock}},\ and\
  \bibinfo {author} {\bibfnamefont {W.}~\bibnamefont {Wang}},\ }\href
  {https://doi.org/10.1103/PhysRevD.85.054011} {\bibfield  {journal} {\bibinfo
  {journal} {Phys.Rev.}\ }\textbf {\bibinfo {volume} {D85}},\ \bibinfo {pages}
  {054011} (\bibinfo {year} {2012})},\ \Eprint
  {https://arxiv.org/abs/1110.1333} {arXiv:1110.1333 [hep-ph]} \BibitemShut
  {NoStop}%
%%CITATION = ARXIV:1110.1333;%%
\bibitem [{\citenamefont {Ali}\ \emph {et~al.}(2015)\citenamefont {Ali},
  \citenamefont {Maiani}, \citenamefont {Polosa},\ and\ \citenamefont
  {Riquer}}]{Ali:2014dva}%
  \BibitemOpen
  \bibfield  {author} {\bibinfo {author} {\bibfnamefont {A.}~\bibnamefont
  {Ali}}, \bibinfo {author} {\bibfnamefont {L.}~\bibnamefont {Maiani}},
  \bibinfo {author} {\bibfnamefont {A.}~\bibnamefont {Polosa}},\ and\ \bibinfo
  {author} {\bibfnamefont {V.}~\bibnamefont {Riquer}},\ }\href
  {https://doi.org/10.1103/PhysRevD.91.017502} {\bibfield  {journal} {\bibinfo
  {journal} {Phys.Rev.}\ }\textbf {\bibinfo {volume} {D91}},\ \bibinfo {pages}
  {017502} (\bibinfo {year} {2015})},\ \Eprint
  {https://arxiv.org/abs/1412.2049} {arXiv:1412.2049 [hep-ph]} \BibitemShut
  {NoStop}%
%%CITATION = ARXIV:1412.2049;%%
\bibitem [{\citenamefont {Aaij}\ \emph {et~al.}(2015)\citenamefont {Aaij} \emph
  {et~al.}}]{Aaij:2015tga}%
  \BibitemOpen
  \bibfield  {author} {\bibinfo {author} {\bibfnamefont {R.}~\bibnamefont
  {Aaij}} \emph {et~al.} (\bibinfo {collaboration} {LHCb}),\ }\href
  {https://doi.org/10.1103/PhysRevLett.115.072001} {\bibfield  {journal}
  {\bibinfo  {journal} {Phys.Rev.Lett.}\ }\textbf {\bibinfo {volume} {115}},\
  \bibinfo {pages} {072001} (\bibinfo {year} {2015})},\ \Eprint
  {https://arxiv.org/abs/1507.03414} {arXiv:1507.03414 [hep-ex]} \BibitemShut
  {NoStop}%
%%CITATION = ARXIV:1507.03414;%%
\bibitem [{\citenamefont {Aaij}\ \emph {et~al.}(2019)\citenamefont {Aaij} \emph
  {et~al.}}]{LHCb:2019kea}%
  \BibitemOpen
  \bibfield  {author} {\bibinfo {author} {\bibfnamefont {R.}~\bibnamefont
  {Aaij}} \emph {et~al.} (\bibinfo {collaboration} {LHCb}),\ }\href
  {https://doi.org/10.1103/PhysRevLett.122.222001} {\bibfield  {journal}
  {\bibinfo  {journal} {Phys.Rev.Lett}\ }\textbf {\bibinfo {volume} {122}},\
  \bibinfo {pages} {222001} (\bibinfo {year} {2019})},\ \Eprint
  {https://arxiv.org/abs/1904.03947} {arXiv:1904.03947 [hep-ex]} \BibitemShut
  {NoStop}%
%%CITATION = ARXIV:1904.03947;%%
\bibitem [{\citenamefont {Maiani}\ \emph {et~al.}(2015)\citenamefont {Maiani},
  \citenamefont {Polosa},\ and\ \citenamefont {Riquer}}]{Maiani:2015vwa}%
  \BibitemOpen
  \bibfield  {author} {\bibinfo {author} {\bibfnamefont {L.}~\bibnamefont
  {Maiani}}, \bibinfo {author} {\bibfnamefont {A.~D.}\ \bibnamefont {Polosa}},\
  and\ \bibinfo {author} {\bibfnamefont {V.}~\bibnamefont {Riquer}},\ }\href
  {https://doi.org/10.1016/j.physletb.2015.08.008} {\bibfield  {journal}
  {\bibinfo  {journal} {Phys.Lett.}\ }\textbf {\bibinfo {volume} {B749}},\
  \bibinfo {pages} {289} (\bibinfo {year} {2015})},\ \Eprint
  {https://arxiv.org/abs/1507.04980} {arXiv:1507.04980 [hep-ph]} \BibitemShut
  {NoStop}%
%%CITATION = ARXIV:1507.04980;%%
\bibitem [{\citenamefont {Lebed}(2015)}]{Lebed:2015tna}%
  \BibitemOpen
  \bibfield  {author} {\bibinfo {author} {\bibfnamefont {R.~F.}\ \bibnamefont
  {Lebed}},\ }\href {https://doi.org/10.1016/j.physletb.2015.08.032} {\bibfield
   {journal} {\bibinfo  {journal} {Phys.Lett.}\ }\textbf {\bibinfo {volume}
  {B749}},\ \bibinfo {pages} {454} (\bibinfo {year} {2015})},\ \Eprint
  {https://arxiv.org/abs/1507.05867} {arXiv:1507.05867 [hep-ph]} \BibitemShut
  {NoStop}%
%%CITATION = ARXIV:1507.05867;%%
\bibitem [{\citenamefont {Anisovich}\ \emph {et~al.}(2015)\citenamefont
  {Anisovich}, \citenamefont {Matveev}, \citenamefont {Nyiri}, \citenamefont
  {Sarantsev},\ and\ \citenamefont {Semenova}}]{Anisovich:2015cia}%
  \BibitemOpen
  \bibfield  {author} {\bibinfo {author} {\bibfnamefont {V.~V.}\ \bibnamefont
  {Anisovich}}, \bibinfo {author} {\bibfnamefont {M.~A.}\ \bibnamefont
  {Matveev}}, \bibinfo {author} {\bibfnamefont {J.}~\bibnamefont {Nyiri}},
  \bibinfo {author} {\bibfnamefont {A.~V.}\ \bibnamefont {Sarantsev}},\ and\
  \bibinfo {author} {\bibfnamefont {A.~N.}\ \bibnamefont {Semenova}},\ }\Eprint
  {https://arxiv.org/abs/1507.07652} {arXiv:1507.07652 [hep-ph]}  (\bibinfo
  {year} {2015})\BibitemShut {NoStop}%
%%CITATION = ARXIV:1507.07652;%%
\bibitem [{\citenamefont {Ali}\ and\ \citenamefont
  {Parkhomenko}(2019)}]{Ali:2019npk}%
  \BibitemOpen
  \bibfield  {author} {\bibinfo {author} {\bibfnamefont {A.}~\bibnamefont
  {Ali}}\ and\ \bibinfo {author} {\bibfnamefont {A.~{\relax Ya}.}\ \bibnamefont
  {Parkhomenko}},\ }\href {https://doi.org/10.1016/j.physletb.2019.05.002}
  {\bibfield  {journal} {\bibinfo  {journal} {Phys.Lett.}\ }\textbf {\bibinfo
  {volume} {B793}},\ \bibinfo {pages} {365} (\bibinfo {year} {2019})},\ \Eprint
  {https://arxiv.org/abs/1904.00446} {arXiv:1904.00446 [hep-ph]} \BibitemShut
  {NoStop}%
%%CITATION = ARXIV:1904.00446;%%
\bibitem [{\citenamefont {Chen}\ \emph
  {et~al.}(2015{\natexlab{a}})\citenamefont {Chen}, \citenamefont {Liu},
  \citenamefont {Li},\ and\ \citenamefont {Zhu}}]{Chen:2015loa}%
  \BibitemOpen
  \bibfield  {author} {\bibinfo {author} {\bibfnamefont {R.}~\bibnamefont
  {Chen}}, \bibinfo {author} {\bibfnamefont {X.}~\bibnamefont {Liu}}, \bibinfo
  {author} {\bibfnamefont {X.-Q.}\ \bibnamefont {Li}},\ and\ \bibinfo {author}
  {\bibfnamefont {S.-L.}\ \bibnamefont {Zhu}},\ }\href
  {https://doi.org/10.1103/PhysRevLett.115.132002} {\bibfield  {journal}
  {\bibinfo  {journal} {Phys.Rev.Lett.}\ }\textbf {\bibinfo {volume} {115}},\
  \bibinfo {pages} {132002} (\bibinfo {year} {2015}{\natexlab{a}})},\ \Eprint
  {https://arxiv.org/abs/1507.03704} {arXiv:1507.03704 [hep-ph]} \BibitemShut
  {NoStop}%
%%CITATION = ARXIV:1507.03704;%%
\bibitem [{\citenamefont {Chen}\ \emph
  {et~al.}(2015{\natexlab{b}})\citenamefont {Chen}, \citenamefont {Chen},
  \citenamefont {Liu}, \citenamefont {Steele},\ and\ \citenamefont
  {Zhu}}]{Chen:2015moa}%
  \BibitemOpen
  \bibfield  {author} {\bibinfo {author} {\bibfnamefont {H.-X.}\ \bibnamefont
  {Chen}}, \bibinfo {author} {\bibfnamefont {W.}~\bibnamefont {Chen}}, \bibinfo
  {author} {\bibfnamefont {X.}~\bibnamefont {Liu}}, \bibinfo {author}
  {\bibfnamefont {T.~G.}\ \bibnamefont {Steele}},\ and\ \bibinfo {author}
  {\bibfnamefont {S.-L.}\ \bibnamefont {Zhu}},\ }\href
  {https://doi.org/10.1103/PhysRevLett.115.172001} {\bibfield  {journal}
  {\bibinfo  {journal} {Phys.Rev.Lett.}\ }\textbf {\bibinfo {volume} {115}},\
  \bibinfo {pages} {172001} (\bibinfo {year} {2015}{\natexlab{b}})},\ \Eprint
  {https://arxiv.org/abs/1507.03717} {arXiv:1507.03717 [hep-ph]} \BibitemShut
  {NoStop}%
%%CITATION = ARXIV:1507.03717;%%
\bibitem [{\citenamefont {Roca}\ \emph {et~al.}(2015)\citenamefont {Roca},
  \citenamefont {Nieves},\ and\ \citenamefont {Oset}}]{Roca:2015dva}%
  \BibitemOpen
  \bibfield  {author} {\bibinfo {author} {\bibfnamefont {L.}~\bibnamefont
  {Roca}}, \bibinfo {author} {\bibfnamefont {J.}~\bibnamefont {Nieves}},\ and\
  \bibinfo {author} {\bibfnamefont {E.}~\bibnamefont {Oset}},\ }\href
  {https://doi.org/10.1103/PhysRevD.92.094003} {\bibfield  {journal} {\bibinfo
  {journal} {Phys.Rev.}\ }\textbf {\bibinfo {volume} {D92}},\ \bibinfo {pages}
  {094003} (\bibinfo {year} {2015})},\ \Eprint
  {https://arxiv.org/abs/1507.04249} {arXiv:1507.04249 [hep-ph]} \BibitemShut
  {NoStop}%
%%CITATION = ARXIV:1507.04249;%%
\bibitem [{\citenamefont {Guo}\ \emph {et~al.}(2019)\citenamefont {Guo},
  \citenamefont {Jing}, \citenamefont {Mei{\ss}ner},\ and\ \citenamefont
  {Sakai}}]{Guo:2019fdo}%
  \BibitemOpen
  \bibfield  {author} {\bibinfo {author} {\bibfnamefont {F.-K.}\ \bibnamefont
  {Guo}}, \bibinfo {author} {\bibfnamefont {H.-J.}\ \bibnamefont {Jing}},
  \bibinfo {author} {\bibfnamefont {U.-G.}\ \bibnamefont {Mei{\ss}ner}},\ and\
  \bibinfo {author} {\bibfnamefont {S.}~\bibnamefont {Sakai}},\ }\href
  {https://doi.org/10.1103/PhysRevD.99.091501} {\bibfield  {journal} {\bibinfo
  {journal} {Phys.Rev.}\ }\textbf {\bibinfo {volume} {D99}},\ \bibinfo {pages}
  {091501} (\bibinfo {year} {2019})},\ \Eprint
  {https://arxiv.org/abs/1903.11503} {arXiv:1903.11503 [hep-ph]} \BibitemShut
  {NoStop}%
%%CITATION = ARXIV:1903.11503;%%
\bibitem [{\citenamefont {Guo}\ and\ \citenamefont
  {Oller}(2019)}]{Guo:2019kdc}%
  \BibitemOpen
  \bibfield  {author} {\bibinfo {author} {\bibfnamefont {Z.-H.}\ \bibnamefont
  {Guo}}\ and\ \bibinfo {author} {\bibfnamefont {J.~A.}\ \bibnamefont
  {Oller}},\ }\href {https://doi.org/10.1016/j.physletb.2019.04.053} {\bibfield
   {journal} {\bibinfo  {journal} {Phys.Lett.}\ }\textbf {\bibinfo {volume}
  {B793}},\ \bibinfo {pages} {144} (\bibinfo {year} {2019})},\ \Eprint
  {https://arxiv.org/abs/1904.00851} {arXiv:1904.00851 [hep-ph]} \BibitemShut
  {NoStop}%
%%CITATION = ARXIV:1904.00851;%%
\bibitem [{\citenamefont {Liu}\ \emph {et~al.}(2019)\citenamefont {Liu},
  \citenamefont {Pan}, \citenamefont {Peng}, \citenamefont {Sanchez~Sanchez},
  \citenamefont {Geng}, \citenamefont {Hosaka},\ and\ \citenamefont
  {Pavon~Valderrama}}]{Liu:2019tjn}%
  \BibitemOpen
  \bibfield  {author} {\bibinfo {author} {\bibfnamefont {M.-Z.}\ \bibnamefont
  {Liu}}, \bibinfo {author} {\bibfnamefont {Y.-W.}\ \bibnamefont {Pan}},
  \bibinfo {author} {\bibfnamefont {F.-Z.}\ \bibnamefont {Peng}}, \bibinfo
  {author} {\bibfnamefont {M.}~\bibnamefont {Sanchez~Sanchez}}, \bibinfo
  {author} {\bibfnamefont {L.-S.}\ \bibnamefont {Geng}}, \bibinfo {author}
  {\bibfnamefont {A.}~\bibnamefont {Hosaka}},\ and\ \bibinfo {author}
  {\bibfnamefont {M.}~\bibnamefont {Pavon~Valderrama}},\ }\href
  {https://doi.org/10.1103/PhysRevLett.122.242001} {\bibfield  {journal}
  {\bibinfo  {journal} {Phys.Rev.Lett.}\ }\textbf {\bibinfo {volume} {122}},\
  \bibinfo {pages} {242001} (\bibinfo {year} {2019})},\ \Eprint
  {https://arxiv.org/abs/1903.11560} {arXiv:1903.11560 [hep-ph]} \BibitemShut
  {NoStop}%
%%CITATION = ARXIV:1903.11560;%%
\bibitem [{\citenamefont {Szczepaniak}(2016)}]{Szczepaniak:2015hya}%
  \BibitemOpen
  \bibfield  {author} {\bibinfo {author} {\bibfnamefont {A.~P.}\ \bibnamefont
  {Szczepaniak}},\ }\href {https://doi.org/10.1016/j.physletb.2016.03.064}
  {\bibfield  {journal} {\bibinfo  {journal} {Phys.Lett.}\ }\textbf {\bibinfo
  {volume} {B757}},\ \bibinfo {pages} {61} (\bibinfo {year} {2016})},\ \Eprint
  {https://arxiv.org/abs/1510.01789} {arXiv:1510.01789 [hep-ph]} \BibitemShut
  {NoStop}%
%%CITATION = ARXIV:1510.01789;%%
\bibitem [{\citenamefont {Mei{\ss}ner}\ and\ \citenamefont
  {Oller}(2015)}]{Meissner:2015mza}%
  \BibitemOpen
  \bibfield  {author} {\bibinfo {author} {\bibfnamefont {U.-G.}\ \bibnamefont
  {Mei{\ss}ner}}\ and\ \bibinfo {author} {\bibfnamefont {J.~A.}\ \bibnamefont
  {Oller}},\ }\href {https://doi.org/10.1016/j.physletb.2015.10.015} {\bibfield
   {journal} {\bibinfo  {journal} {Phys.Lett.}\ }\textbf {\bibinfo {volume}
  {B751}},\ \bibinfo {pages} {59} (\bibinfo {year} {2015})},\ \Eprint
  {https://arxiv.org/abs/1507.07478} {arXiv:1507.07478 [hep-ph]} \BibitemShut
  {NoStop}%
%%CITATION = ARXIV:1507.07478;%%
\bibitem [{\citenamefont {Mikhasenko}(2015)}]{Mikhasenko:2015vca}%
  \BibitemOpen
  \bibfield  {author} {\bibinfo {author} {\bibfnamefont {M.}~\bibnamefont
  {Mikhasenko}},\ }\Eprint {https://arxiv.org/abs/1507.06552} {arXiv:1507.06552
  [hep-ph]}  (\bibinfo {year} {2015})\BibitemShut {NoStop}%
%%CITATION = ARXIV:1507.06552;%%
\bibitem [{\citenamefont {Guo}\ \emph {et~al.}(2016)\citenamefont {Guo},
  \citenamefont {Mei{\ss}ner}, \citenamefont {Nieves},\ and\ \citenamefont
  {Yang}}]{Guo:2016bkl}%
  \BibitemOpen
  \bibfield  {author} {\bibinfo {author} {\bibfnamefont {F.-K.}\ \bibnamefont
  {Guo}}, \bibinfo {author} {\bibfnamefont {U.-G.}\ \bibnamefont
  {Mei{\ss}ner}}, \bibinfo {author} {\bibfnamefont {J.}~\bibnamefont
  {Nieves}},\ and\ \bibinfo {author} {\bibfnamefont {Z.}~\bibnamefont {Yang}},\
  }\href {https://doi.org/10.1140/epja/i2016-16318-4} {\bibfield  {journal}
  {\bibinfo  {journal} {Eur.Phys.J.}\ }\textbf {\bibinfo {volume} {A52}},\
  \bibinfo {pages} {318} (\bibinfo {year} {2016})},\ \Eprint
  {https://arxiv.org/abs/1605.05113} {arXiv:1605.05113 [hep-ph]} \BibitemShut
  {NoStop}%
%%CITATION = ARXIV:1605.05113;%%
\bibitem [{\citenamefont {Bayar}\ \emph {et~al.}(2016)\citenamefont {Bayar},
  \citenamefont {Aceti}, \citenamefont {Guo},\ and\ \citenamefont
  {Oset}}]{Bayar:2016ftu}%
  \BibitemOpen
  \bibfield  {author} {\bibinfo {author} {\bibfnamefont {M.}~\bibnamefont
  {Bayar}}, \bibinfo {author} {\bibfnamefont {F.}~\bibnamefont {Aceti}},
  \bibinfo {author} {\bibfnamefont {F.-K.}\ \bibnamefont {Guo}},\ and\ \bibinfo
  {author} {\bibfnamefont {E.}~\bibnamefont {Oset}},\ }\href
  {https://doi.org/10.1103/PhysRevD.94.074039} {\bibfield  {journal} {\bibinfo
  {journal} {Phys.Rev.}\ }\textbf {\bibinfo {volume} {D94}},\ \bibinfo {pages}
  {074039} (\bibinfo {year} {2016})},\ \Eprint
  {https://arxiv.org/abs/1609.04133} {arXiv:1609.04133 [hep-ph]} \BibitemShut
  {NoStop}%
\bibitem [{\citenamefont {Nakamura}(2021)}]{Nakamura:2021qvy}%
  \BibitemOpen
  \bibfield  {author} {\bibinfo {author} {\bibfnamefont {S.~X.}\ \bibnamefont
  {Nakamura}},\ }\href {https://doi.org/10.1103/PhysRevD.103.L111503}
  {\bibfield  {journal} {\bibinfo  {journal} {Phys.Rev.}\ }\textbf {\bibinfo
  {volume} {D103}},\ \bibinfo {pages} {111503} (\bibinfo {year} {2021})},\
  \Eprint {https://arxiv.org/abs/2103.06817} {arXiv:2103.06817 [hep-ph]}
  \BibitemShut {NoStop}%
\bibitem [{\citenamefont {Eden}\ and\ \citenamefont
  {Taylor}(1964)}]{Eden:1964zz}%
  \BibitemOpen
  \bibfield  {author} {\bibinfo {author} {\bibfnamefont {R.~J.}\ \bibnamefont
  {Eden}}\ and\ \bibinfo {author} {\bibfnamefont {J.~R.}\ \bibnamefont
  {Taylor}},\ }\href {https://doi.org/10.1103/PhysRev.133.B1575} {\bibfield
  {journal} {\bibinfo  {journal} {Phys.Rev.}\ }\textbf {\bibinfo {volume}
  {133}},\ \bibinfo {pages} {B1575} (\bibinfo {year} {1964})}\BibitemShut
  {NoStop}%
%%CITATION = PHRVA,133,B1575;%%
\bibitem [{\citenamefont {Hammer}\ and\ \citenamefont
  {K{\"o}nig}(2014)}]{Hammer:2014rba}%
  \BibitemOpen
  \bibfield  {author} {\bibinfo {author} {\bibfnamefont {H.~W.}\ \bibnamefont
  {Hammer}}\ and\ \bibinfo {author} {\bibfnamefont {S.}~\bibnamefont
  {K{\"o}nig}},\ }\href {https://doi.org/10.1016/j.physletb.2014.07.015}
  {\bibfield  {journal} {\bibinfo  {journal} {Phys.Lett.}\ }\textbf {\bibinfo
  {volume} {B736}},\ \bibinfo {pages} {208} (\bibinfo {year} {2014})},\ \Eprint
  {https://arxiv.org/abs/1406.1359} {arXiv:1406.1359 [nucl-th]} \BibitemShut
  {NoStop}%
%%CITATION = ARXIV:1406.1359;%%
\bibitem [{\citenamefont {Albaladejo}\ \emph {et~al.}(2021)\citenamefont
  {Albaladejo} \emph {et~al.}}]{JPAC:2021rxu}%
  \BibitemOpen
  \bibfield  {author} {\bibinfo {author} {\bibfnamefont {M.}~\bibnamefont
  {Albaladejo}} \emph {et~al.} (\bibinfo {collaboration} {JPAC}),\ }\Eprint
  {https://arxiv.org/abs/2112.13436} {arXiv:2112.13436 [hep-ph]}  (\bibinfo
  {year} {2021})\BibitemShut {NoStop}%
\bibitem [{\citenamefont {Fern\'andez-Ram\'irez}\ \emph
  {et~al.}(2019)\citenamefont {Fern\'andez-Ram\'irez}, \citenamefont {Pilloni},
  \citenamefont {Albaladejo}, \citenamefont {Jackura}, \citenamefont {Mathieu},
  \citenamefont {Mikhasenko}, \citenamefont {Silva-Castro},\ and\ \citenamefont
  {Szczepaniak}}]{Fernandez-Ramirez:2019koa}%
  \BibitemOpen
  \bibfield  {author} {\bibinfo {author} {\bibfnamefont {C.}~\bibnamefont
  {Fern\'andez-Ram\'irez}}, \bibinfo {author} {\bibfnamefont {A.}~\bibnamefont
  {Pilloni}}, \bibinfo {author} {\bibfnamefont {M.}~\bibnamefont {Albaladejo}},
  \bibinfo {author} {\bibfnamefont {A.}~\bibnamefont {Jackura}}, \bibinfo
  {author} {\bibfnamefont {V.}~\bibnamefont {Mathieu}}, \bibinfo {author}
  {\bibfnamefont {M.}~\bibnamefont {Mikhasenko}}, \bibinfo {author}
  {\bibfnamefont {J.~A.}\ \bibnamefont {Silva-Castro}},\ and\ \bibinfo {author}
  {\bibfnamefont {A.~P.}\ \bibnamefont {Szczepaniak}} (\bibinfo {collaboration}
  {JPAC}),\ }\href {https://doi.org/10.1103/PhysRevLett.123.092001} {\bibfield
  {journal} {\bibinfo  {journal} {Phys.Rev.Lett.}\ }\textbf {\bibinfo {volume}
  {123}},\ \bibinfo {pages} {092001} (\bibinfo {year} {2019})},\ \Eprint
  {https://arxiv.org/abs/1904.10021} {arXiv:1904.10021 [hep-ph]} \BibitemShut
  {NoStop}%
%%CITATION = ARXIV:1904.10021;%%
\bibitem [{\citenamefont {Adlarson}\ \emph {et~al.}(2017)\citenamefont
  {Adlarson} \emph {et~al.}}]{Adlarson:2016hpp}%
  \BibitemOpen
  \bibfield  {author} {\bibinfo {author} {\bibfnamefont {P.}~\bibnamefont
  {Adlarson}} \emph {et~al.},\ }\href
  {https://doi.org/10.1103/PhysRevC.95.035208} {\bibfield  {journal} {\bibinfo
  {journal} {Phys.Rev.}\ }\textbf {\bibinfo {volume} {C95}},\ \bibinfo {pages}
  {035208} (\bibinfo {year} {2017})},\ \Eprint
  {https://arxiv.org/abs/1609.04503} {arXiv:1609.04503 [hep-ex]} \BibitemShut
  {NoStop}%
\bibitem [{\citenamefont {Arnaldi}\ \emph {et~al.}(2009)\citenamefont {Arnaldi}
  \emph {et~al.}}]{Arnaldi:2009aa}%
  \BibitemOpen
  \bibfield  {author} {\bibinfo {author} {\bibfnamefont {R.}~\bibnamefont
  {Arnaldi}} \emph {et~al.} (\bibinfo {collaboration} {NA60}),\ }\href
  {https://doi.org/10.1016/j.physletb.2009.05.029} {\bibfield  {journal}
  {\bibinfo  {journal} {Phys.Lett.}\ }\textbf {\bibinfo {volume} {B677}},\
  \bibinfo {pages} {260} (\bibinfo {year} {2009})},\ \Eprint
  {https://arxiv.org/abs/0902.2547} {arXiv:0902.2547 [hep-ph]} \BibitemShut
  {NoStop}%
\bibitem [{\citenamefont {Arnaldi}\ \emph {et~al.}(2016)\citenamefont {Arnaldi}
  \emph {et~al.}}]{Arnaldi:2016pzu}%
  \BibitemOpen
  \bibfield  {author} {\bibinfo {author} {\bibfnamefont {R.}~\bibnamefont
  {Arnaldi}} \emph {et~al.} (\bibinfo {collaboration} {NA60}),\ }\href
  {https://doi.org/10.1016/j.physletb.2016.04.013} {\bibfield  {journal}
  {\bibinfo  {journal} {Phys.Lett.}\ }\textbf {\bibinfo {volume} {B757}},\
  \bibinfo {pages} {437} (\bibinfo {year} {2016})},\ \Eprint
  {https://arxiv.org/abs/1608.07898} {arXiv:1608.07898 [hep-ex]} \BibitemShut
  {NoStop}%
\bibitem [{\citenamefont {Albaladejo}\ \emph
  {et~al.}(2020{\natexlab{a}})\citenamefont {Albaladejo}, \citenamefont
  {Danilkin}, \citenamefont {Gonz\`alez-Sol\'is}, \citenamefont {Winney},
  \citenamefont {Fern\'andez-Ram\'irez}, \citenamefont {Hiller~Blin},
  \citenamefont {Mathieu}, \citenamefont {Mikhasenko}, \citenamefont
  {Pilloni},\ and\ \citenamefont {Szczepaniak}}]{JPAC:2020umo}%
  \BibitemOpen
  \bibfield  {author} {\bibinfo {author} {\bibfnamefont {M.}~\bibnamefont
  {Albaladejo}}, \bibinfo {author} {\bibfnamefont {I.}~\bibnamefont
  {Danilkin}}, \bibinfo {author} {\bibfnamefont {S.}~\bibnamefont
  {Gonz\`alez-Sol\'is}}, \bibinfo {author} {\bibfnamefont {D.}~\bibnamefont
  {Winney}}, \bibinfo {author} {\bibfnamefont {C.}~\bibnamefont
  {Fern\'andez-Ram\'irez}}, \bibinfo {author} {\bibfnamefont {A.}~\bibnamefont
  {Hiller~Blin}}, \bibinfo {author} {\bibfnamefont {V.}~\bibnamefont
  {Mathieu}}, \bibinfo {author} {\bibfnamefont {M.}~\bibnamefont {Mikhasenko}},
  \bibinfo {author} {\bibfnamefont {A.}~\bibnamefont {Pilloni}},\ and\ \bibinfo
  {author} {\bibfnamefont {A.}~\bibnamefont {Szczepaniak}},\ }\href
  {https://doi.org/10.1140/epjc/s10052-020-08576-6} {\bibfield  {journal}
  {\bibinfo  {journal} {Eur.Phys.J.}\ }\textbf {\bibinfo {volume} {C80}},\
  \bibinfo {pages} {1107} (\bibinfo {year} {2020}{\natexlab{a}})},\ \Eprint
  {https://arxiv.org/abs/2006.01058} {arXiv:2006.01058 [hep-ph]} \BibitemShut
  {NoStop}%
\bibitem [{\citenamefont {Du}\ \emph {et~al.}(2020)\citenamefont {Du},
  \citenamefont {Baru}, \citenamefont {Guo}, \citenamefont {Hanhart},
  \citenamefont {Mei{\ss}ner}, \citenamefont {Oller},\ and\ \citenamefont
  {Wang}}]{Du:2019pij}%
  \BibitemOpen
  \bibfield  {author} {\bibinfo {author} {\bibfnamefont {M.-L.}\ \bibnamefont
  {Du}}, \bibinfo {author} {\bibfnamefont {V.}~\bibnamefont {Baru}}, \bibinfo
  {author} {\bibfnamefont {F.-K.}\ \bibnamefont {Guo}}, \bibinfo {author}
  {\bibfnamefont {C.}~\bibnamefont {Hanhart}}, \bibinfo {author} {\bibfnamefont
  {U.-G.}\ \bibnamefont {Mei{\ss}ner}}, \bibinfo {author} {\bibfnamefont
  {J.~A.}\ \bibnamefont {Oller}},\ and\ \bibinfo {author} {\bibfnamefont
  {Q.}~\bibnamefont {Wang}},\ }\href
  {https://doi.org/10.1103/PhysRevLett.124.072001} {\bibfield  {journal}
  {\bibinfo  {journal} {Phys.Rev.Lett.}\ }\textbf {\bibinfo {volume} {124}},\
  \bibinfo {pages} {072001} (\bibinfo {year} {2020})},\ \Eprint
  {https://arxiv.org/abs/1910.11846} {arXiv:1910.11846 [hep-ph]} \BibitemShut
  {NoStop}%
%%CITATION = ARXIV:1910.11846;%%
\bibitem [{\citenamefont {Du}\ \emph {et~al.}(2021)\citenamefont {Du},
  \citenamefont {Baru}, \citenamefont {Guo}, \citenamefont {Hanhart},
  \citenamefont {Mei\ss{}ner}, \citenamefont {Oller},\ and\ \citenamefont
  {Wang}}]{Du:2021fmf}%
  \BibitemOpen
  \bibfield  {author} {\bibinfo {author} {\bibfnamefont {M.-L.}\ \bibnamefont
  {Du}}, \bibinfo {author} {\bibfnamefont {V.}~\bibnamefont {Baru}}, \bibinfo
  {author} {\bibfnamefont {F.-K.}\ \bibnamefont {Guo}}, \bibinfo {author}
  {\bibfnamefont {C.}~\bibnamefont {Hanhart}}, \bibinfo {author} {\bibfnamefont
  {U.-G.}\ \bibnamefont {Mei\ss{}ner}}, \bibinfo {author} {\bibfnamefont
  {J.~A.}\ \bibnamefont {Oller}},\ and\ \bibinfo {author} {\bibfnamefont
  {Q.}~\bibnamefont {Wang}},\ }\href {https://doi.org/10.1007/JHEP08(2021)157}
  {\bibfield  {journal} {\bibinfo  {journal} {JHEP}\ }\textbf {\bibinfo
  {volume} {08}},\ \bibinfo {pages} {157}},\ \Eprint
  {https://arxiv.org/abs/2102.07159} {arXiv:2102.07159 [hep-ph]} \BibitemShut
  {NoStop}%
\bibitem [{\citenamefont {Ng}\ \emph {et~al.}(2021)\citenamefont {Ng},
  \citenamefont {Bibrzycki}, \citenamefont {Nys}, \citenamefont
  {Fernandez-Ramirez}, \citenamefont {Pilloni}, \citenamefont {Mathieu},
  \citenamefont {Rasmusson},\ and\ \citenamefont {Szczepaniak}}]{Ng:2021ibr}%
  \BibitemOpen
  \bibfield  {author} {\bibinfo {author} {\bibfnamefont {L.}~\bibnamefont
  {Ng}}, \bibinfo {author} {\bibfnamefont {L.}~\bibnamefont {Bibrzycki}},
  \bibinfo {author} {\bibfnamefont {J.}~\bibnamefont {Nys}}, \bibinfo {author}
  {\bibfnamefont {C.}~\bibnamefont {Fernandez-Ramirez}}, \bibinfo {author}
  {\bibfnamefont {A.}~\bibnamefont {Pilloni}}, \bibinfo {author} {\bibfnamefont
  {V.}~\bibnamefont {Mathieu}}, \bibinfo {author} {\bibfnamefont {A.~J.}\
  \bibnamefont {Rasmusson}},\ and\ \bibinfo {author} {\bibfnamefont {A.~P.}\
  \bibnamefont {Szczepaniak}} (\bibinfo {collaboration} {JPAC}),\ }\Eprint
  {https://arxiv.org/abs/2110.13742} {arXiv:2110.13742 [hep-ph]}  (\bibinfo
  {year} {2021})\BibitemShut {NoStop}%
\bibitem [{\citenamefont {Albaladejo}\ \emph {et~al.}(2016)\citenamefont
  {Albaladejo}, \citenamefont {Guo}, \citenamefont {Hidalgo-Duque},\ and\
  \citenamefont {Nieves}}]{Albaladejo:2015lob}%
  \BibitemOpen
  \bibfield  {author} {\bibinfo {author} {\bibfnamefont {M.}~\bibnamefont
  {Albaladejo}}, \bibinfo {author} {\bibfnamefont {F.-K.}\ \bibnamefont {Guo}},
  \bibinfo {author} {\bibfnamefont {C.}~\bibnamefont {Hidalgo-Duque}},\ and\
  \bibinfo {author} {\bibfnamefont {J.}~\bibnamefont {Nieves}},\ }\href
  {https://doi.org/10.1016/j.physletb.2016.02.025} {\bibfield  {journal}
  {\bibinfo  {journal} {Phys.Lett.}\ }\textbf {\bibinfo {volume} {B755}},\
  \bibinfo {pages} {337} (\bibinfo {year} {2016})},\ \Eprint
  {https://arxiv.org/abs/1512.03638} {arXiv:1512.03638 [hep-ph]} \BibitemShut
  {NoStop}%
%%CITATION = ARXIV:1512.03638;%%
\bibitem [{\citenamefont {Pilloni}\ \emph {et~al.}(2017)\citenamefont
  {Pilloni}, \citenamefont {Fern\'andez-Ram\'irez}, \citenamefont {Jackura},
  \citenamefont {Mathieu}, \citenamefont {Mikhasenko}, \citenamefont {Nys},\
  and\ \citenamefont {Szczepaniak}}]{Pilloni:2016obd}%
  \BibitemOpen
  \bibfield  {author} {\bibinfo {author} {\bibfnamefont {A.}~\bibnamefont
  {Pilloni}}, \bibinfo {author} {\bibfnamefont {C.}~\bibnamefont
  {Fern\'andez-Ram\'irez}}, \bibinfo {author} {\bibfnamefont {A.}~\bibnamefont
  {Jackura}}, \bibinfo {author} {\bibfnamefont {V.}~\bibnamefont {Mathieu}},
  \bibinfo {author} {\bibfnamefont {M.}~\bibnamefont {Mikhasenko}}, \bibinfo
  {author} {\bibfnamefont {J.}~\bibnamefont {Nys}},\ and\ \bibinfo {author}
  {\bibfnamefont {A.~P.}\ \bibnamefont {Szczepaniak}} (\bibinfo {collaboration}
  {JPAC}),\ }\href {https://doi.org/10.1016/j.physletb.2017.06.030} {\bibfield
  {journal} {\bibinfo  {journal} {Phys.Lett.}\ }\textbf {\bibinfo {volume}
  {B772}},\ \bibinfo {pages} {200} (\bibinfo {year} {2017})},\ \Eprint
  {https://arxiv.org/abs/1612.06490} {arXiv:1612.06490 [hep-ph]} \BibitemShut
  {NoStop}%
%%CITATION = ARXIV:1612.06490;%%
\bibitem [{\citenamefont {Adolph}\ \emph
  {et~al.}(2015{\natexlab{b}})\citenamefont {Adolph} \emph
  {et~al.}}]{COMPASS:2015kdx}%
  \BibitemOpen
  \bibfield  {author} {\bibinfo {author} {\bibfnamefont {C.}~\bibnamefont
  {Adolph}} \emph {et~al.} (\bibinfo {collaboration} {COMPASS}),\ }\href
  {https://doi.org/10.1103/PhysRevLett.115.082001} {\bibfield  {journal}
  {\bibinfo  {journal} {Phys.Rev.Lett.}\ }\textbf {\bibinfo {volume} {115}},\
  \bibinfo {pages} {082001} (\bibinfo {year} {2015}{\natexlab{b}})},\ \Eprint
  {https://arxiv.org/abs/1501.05732} {arXiv:1501.05732 [hep-ex]} \BibitemShut
  {NoStop}%
%%CITATION = ARXIV:1501.05732;%%
\bibitem [{\citenamefont {Alexeev}\ \emph {et~al.}(2021)\citenamefont {Alexeev}
  \emph {et~al.}}]{COMPASS:2020yhb}%
  \BibitemOpen
  \bibfield  {author} {\bibinfo {author} {\bibfnamefont {G.~D.}\ \bibnamefont
  {Alexeev}} \emph {et~al.} (\bibinfo {collaboration} {COMPASS}),\ }\href
  {https://doi.org/10.1103/PhysRevLett.127.082501} {\bibfield  {journal}
  {\bibinfo  {journal} {Phys.Rev.Lett.}\ }\textbf {\bibinfo {volume} {127}},\
  \bibinfo {pages} {082501} (\bibinfo {year} {2021})},\ \Eprint
  {https://arxiv.org/abs/2006.05342} {arXiv:2006.05342 [hep-ph]} \BibitemShut
  {NoStop}%
\bibitem [{\citenamefont {Alekseev}\ \emph {et~al.}(2010)\citenamefont
  {Alekseev} \emph {et~al.}}]{COMPASS:2009xrl}%
  \BibitemOpen
  \bibfield  {author} {\bibinfo {author} {\bibfnamefont {M.}~\bibnamefont
  {Alekseev}} \emph {et~al.} (\bibinfo {collaboration} {COMPASS}),\ }\href
  {https://doi.org/10.1103/PhysRevLett.104.241803} {\bibfield  {journal}
  {\bibinfo  {journal} {Phys.Rev.Lett.}\ }\textbf {\bibinfo {volume} {104}},\
  \bibinfo {pages} {241803} (\bibinfo {year} {2010})},\ \Eprint
  {https://arxiv.org/abs/0910.5842} {arXiv:0910.5842 [hep-ex]} \BibitemShut
  {NoStop}%
%%CITATION = ARXIV:0910.5842;%%
\bibitem [{\citenamefont {Aitchison}\ and\ \citenamefont
  {Brehm}(1979)}]{Aitchison:1979fj}%
  \BibitemOpen
  \bibfield  {author} {\bibinfo {author} {\bibfnamefont {I.~J.~R.}\
  \bibnamefont {Aitchison}}\ and\ \bibinfo {author} {\bibfnamefont {J.~J.}\
  \bibnamefont {Brehm}},\ }\href {https://doi.org/10.1016/0370-2693(79)90056-X}
  {\bibfield  {journal} {\bibinfo  {journal} {Phys.Lett.}\ }\textbf {\bibinfo
  {volume} {84B}},\ \bibinfo {pages} {349} (\bibinfo {year}
  {1979})}\BibitemShut {NoStop}%
%%CITATION = PHLTA,84B,349;%%
\bibitem [{\citenamefont {Szczepaniak}(2015)}]{Szczepaniak:2015eza}%
  \BibitemOpen
  \bibfield  {author} {\bibinfo {author} {\bibfnamefont {A.~P.}\ \bibnamefont
  {Szczepaniak}},\ }\href {https://doi.org/10.1016/j.physletb.2015.06.029}
  {\bibfield  {journal} {\bibinfo  {journal} {Phys.Lett.}\ }\textbf {\bibinfo
  {volume} {B747}},\ \bibinfo {pages} {410} (\bibinfo {year} {2015})},\ \Eprint
  {https://arxiv.org/abs/1501.01691} {arXiv:1501.01691 [hep-ph]} \BibitemShut
  {NoStop}%
%%CITATION = ARXIV:1501.01691;%%
\bibitem [{\citenamefont {Nakamura}\ and\ \citenamefont
  {Tsushima}(2019)}]{Nakamura:2019btl}%
  \BibitemOpen
  \bibfield  {author} {\bibinfo {author} {\bibfnamefont {S.}~\bibnamefont
  {Nakamura}}\ and\ \bibinfo {author} {\bibfnamefont {K.}~\bibnamefont
  {Tsushima}},\ }\href {https://doi.org/10.1103/PhysRevD.100.051502} {\bibfield
   {journal} {\bibinfo  {journal} {Phys.Rev.}\ }\textbf {\bibinfo {volume}
  {D100}},\ \bibinfo {pages} {051502} (\bibinfo {year} {2019})},\ \Eprint
  {https://arxiv.org/abs/1901.07385} {arXiv:1901.07385 [hep-ph]} \BibitemShut
  {NoStop}%
\bibitem [{\citenamefont {Mikhasenko}\ \emph
  {et~al.}(2018{\natexlab{a}})\citenamefont {Mikhasenko}, \citenamefont
  {Pilloni}, \citenamefont {Nys}, \citenamefont {Albaladejo}, \citenamefont
  {Fern\'andez-Ram\'irez}, \citenamefont {Jackura}, \citenamefont {Mathieu},
  \citenamefont {Sherrill}, \citenamefont {Skwarnicki},\ and\ \citenamefont
  {Szczepaniak}}]{JPAC:2017vtd}%
  \BibitemOpen
  \bibfield  {author} {\bibinfo {author} {\bibfnamefont {M.}~\bibnamefont
  {Mikhasenko}}, \bibinfo {author} {\bibfnamefont {A.}~\bibnamefont {Pilloni}},
  \bibinfo {author} {\bibfnamefont {J.}~\bibnamefont {Nys}}, \bibinfo {author}
  {\bibfnamefont {M.}~\bibnamefont {Albaladejo}}, \bibinfo {author}
  {\bibfnamefont {C.}~\bibnamefont {Fern\'andez-Ram\'irez}}, \bibinfo {author}
  {\bibfnamefont {A.}~\bibnamefont {Jackura}}, \bibinfo {author} {\bibfnamefont
  {V.}~\bibnamefont {Mathieu}}, \bibinfo {author} {\bibfnamefont
  {N.}~\bibnamefont {Sherrill}}, \bibinfo {author} {\bibfnamefont
  {T.}~\bibnamefont {Skwarnicki}},\ and\ \bibinfo {author} {\bibfnamefont
  {A.~P.}\ \bibnamefont {Szczepaniak}} (\bibinfo {collaboration} {JPAC}),\
  }\href {https://doi.org/10.1140/epjc/s10052-018-5670-y} {\bibfield  {journal}
  {\bibinfo  {journal} {Eur.Phys.J.}\ }\textbf {\bibinfo {volume} {C78}},\
  \bibinfo {pages} {229} (\bibinfo {year} {2018}{\natexlab{a}})},\ \Eprint
  {https://arxiv.org/abs/1712.02815} {arXiv:1712.02815 [hep-ph]} \BibitemShut
  {NoStop}%
%%CITATION = ARXIV:1712.02815;%%
\bibitem [{\citenamefont {Pilloni}\ \emph {et~al.}(2018)\citenamefont
  {Pilloni}, \citenamefont {Nys}, \citenamefont {Mikhasenko}, \citenamefont
  {Albaladejo}, \citenamefont {Fern\'andez-Ram\'irez}, \citenamefont {Jackura},
  \citenamefont {Mathieu}, \citenamefont {Sherrill}, \citenamefont
  {Skwarnicki},\ and\ \citenamefont {Szczepaniak}}]{JPAC:2018dfc}%
  \BibitemOpen
  \bibfield  {author} {\bibinfo {author} {\bibfnamefont {A.}~\bibnamefont
  {Pilloni}}, \bibinfo {author} {\bibfnamefont {J.}~\bibnamefont {Nys}},
  \bibinfo {author} {\bibfnamefont {M.}~\bibnamefont {Mikhasenko}}, \bibinfo
  {author} {\bibfnamefont {M.}~\bibnamefont {Albaladejo}}, \bibinfo {author}
  {\bibfnamefont {C.}~\bibnamefont {Fern\'andez-Ram\'irez}}, \bibinfo {author}
  {\bibfnamefont {A.}~\bibnamefont {Jackura}}, \bibinfo {author} {\bibfnamefont
  {V.}~\bibnamefont {Mathieu}}, \bibinfo {author} {\bibfnamefont
  {N.}~\bibnamefont {Sherrill}}, \bibinfo {author} {\bibfnamefont
  {T.}~\bibnamefont {Skwarnicki}},\ and\ \bibinfo {author} {\bibfnamefont
  {A.~P.}\ \bibnamefont {Szczepaniak}},\ }\href
  {https://doi.org/10.1140/epjc/s10052-018-6177-2} {\bibfield  {journal}
  {\bibinfo  {journal} {Eur.Phys.J.}\ }\textbf {\bibinfo {volume} {C78}},\
  \bibinfo {pages} {727} (\bibinfo {year} {2018})},\ \Eprint
  {https://arxiv.org/abs/1805.02113} {arXiv:1805.02113 [hep-ph]} \BibitemShut
  {NoStop}%
%%CITATION = ARXIV:1805.02113;%%
\bibitem [{\citenamefont {Albaladejo}\ \emph
  {et~al.}(2020{\natexlab{b}})\citenamefont {Albaladejo}, \citenamefont
  {Winney}, \citenamefont {Danilkin}, \citenamefont {Fernández-Ramírez},
  \citenamefont {Mathieu}, \citenamefont {Mikhasenko}, \citenamefont {Pilloni},
  \citenamefont {Silva-Castro},\ and\ \citenamefont
  {Szczepaniak}}]{Albaladejo:2019huw}%
  \BibitemOpen
  \bibfield  {author} {\bibinfo {author} {\bibfnamefont {M.}~\bibnamefont
  {Albaladejo}}, \bibinfo {author} {\bibfnamefont {D.}~\bibnamefont {Winney}},
  \bibinfo {author} {\bibfnamefont {I.}~\bibnamefont {Danilkin}}, \bibinfo
  {author} {\bibfnamefont {C.}~\bibnamefont {Fernández-Ramírez}}, \bibinfo
  {author} {\bibfnamefont {V.}~\bibnamefont {Mathieu}}, \bibinfo {author}
  {\bibfnamefont {M.}~\bibnamefont {Mikhasenko}}, \bibinfo {author}
  {\bibfnamefont {A.}~\bibnamefont {Pilloni}}, \bibinfo {author} {\bibfnamefont
  {J.}~\bibnamefont {Silva-Castro}},\ and\ \bibinfo {author} {\bibfnamefont
  {A.}~\bibnamefont {Szczepaniak}} (\bibinfo {collaboration} {JPAC}),\ }\href
  {https://doi.org/10.1103/PhysRevD.101.054018} {\bibfield  {journal} {\bibinfo
   {journal} {Phys.Rev.}\ }\textbf {\bibinfo {volume} {D101}},\ \bibinfo
  {pages} {054018} (\bibinfo {year} {2020}{\natexlab{b}})},\ \Eprint
  {https://arxiv.org/abs/1910.03107} {arXiv:1910.03107 [hep-ph]} \BibitemShut
  {NoStop}%
\bibitem [{\citenamefont {Mikhasenko}\ \emph {et~al.}(2020)\citenamefont
  {Mikhasenko}, \citenamefont {Albaladejo}, \citenamefont {Bibrzycki},
  \citenamefont {Fern\'andez-Ram\'irez}, \citenamefont {Mathieu}, \citenamefont
  {Mitchell}, \citenamefont {Pappagallo}, \citenamefont {Pilloni},
  \citenamefont {Winney}, \citenamefont {Skwarnicki},\ and\ \citenamefont
  {Szczepaniak}}]{JPAC:2019ufm}%
  \BibitemOpen
  \bibfield  {author} {\bibinfo {author} {\bibfnamefont {M.}~\bibnamefont
  {Mikhasenko}}, \bibinfo {author} {\bibfnamefont {M.}~\bibnamefont
  {Albaladejo}}, \bibinfo {author} {\bibfnamefont {L.}~\bibnamefont
  {Bibrzycki}}, \bibinfo {author} {\bibfnamefont {C.}~\bibnamefont
  {Fern\'andez-Ram\'irez}}, \bibinfo {author} {\bibfnamefont {V.}~\bibnamefont
  {Mathieu}}, \bibinfo {author} {\bibfnamefont {S.}~\bibnamefont {Mitchell}},
  \bibinfo {author} {\bibfnamefont {M.}~\bibnamefont {Pappagallo}}, \bibinfo
  {author} {\bibfnamefont {A.}~\bibnamefont {Pilloni}}, \bibinfo {author}
  {\bibfnamefont {D.}~\bibnamefont {Winney}}, \bibinfo {author} {\bibfnamefont
  {T.}~\bibnamefont {Skwarnicki}},\ and\ \bibinfo {author} {\bibfnamefont
  {A.}~\bibnamefont {Szczepaniak}} (\bibinfo {collaboration} {JPAC}),\ }\href
  {https://doi.org/10.1103/PhysRevD.101.034033} {\bibfield  {journal} {\bibinfo
   {journal} {Phys.Rev.}\ }\textbf {\bibinfo {volume} {D101}},\ \bibinfo
  {pages} {034033} (\bibinfo {year} {2020})},\ \Eprint
  {https://arxiv.org/abs/1910.04566} {arXiv:1910.04566 [hep-ph]} \BibitemShut
  {NoStop}%
\bibitem [{\citenamefont {Wang}\ \emph {et~al.}(2021)\citenamefont {Wang},
  \citenamefont {Jiang}, \citenamefont {Liu}, \citenamefont {Qian},
  \citenamefont {Lyu},\ and\ \citenamefont {Zhang}}]{Wang:2020giv}%
  \BibitemOpen
  \bibfield  {author} {\bibinfo {author} {\bibfnamefont {M.}~\bibnamefont
  {Wang}}, \bibinfo {author} {\bibfnamefont {Y.}~\bibnamefont {Jiang}},
  \bibinfo {author} {\bibfnamefont {Y.}~\bibnamefont {Liu}}, \bibinfo {author}
  {\bibfnamefont {W.}~\bibnamefont {Qian}}, \bibinfo {author} {\bibfnamefont
  {X.}~\bibnamefont {Lyu}},\ and\ \bibinfo {author} {\bibfnamefont
  {L.}~\bibnamefont {Zhang}},\ }\href
  {https://doi.org/10.1088/1674-1137/abf139} {\bibfield  {journal} {\bibinfo
  {journal} {Chin.Phys.}\ }\textbf {\bibinfo {volume} {C45}},\ \bibinfo {pages}
  {063103} (\bibinfo {year} {2021})},\ \Eprint
  {https://arxiv.org/abs/2012.03699} {arXiv:2012.03699 [hep-ex]} \BibitemShut
  {NoStop}%
\bibitem [{\citenamefont {Khuri}\ and\ \citenamefont
  {Treiman}(1960)}]{Khuri:1960zz}%
  \BibitemOpen
  \bibfield  {author} {\bibinfo {author} {\bibfnamefont {N.~N.}\ \bibnamefont
  {Khuri}}\ and\ \bibinfo {author} {\bibfnamefont {S.~B.}\ \bibnamefont
  {Treiman}},\ }\href {https://doi.org/10.1103/PhysRev.119.1115} {\bibfield
  {journal} {\bibinfo  {journal} {Phys.Rev.}\ }\textbf {\bibinfo {volume}
  {119}},\ \bibinfo {pages} {1115} (\bibinfo {year} {1960})}\BibitemShut
  {NoStop}%
%%CITATION = PHRVA,119,1115;%%
\bibitem [{\citenamefont {Albaladejo}\ \emph {et~al.}(2018)\citenamefont
  {Albaladejo}, \citenamefont {Sherrill}, \citenamefont
  {Fern\'andez-Ram\'irez}, \citenamefont {Jackura}, \citenamefont {Mathieu},
  \citenamefont {Mikhasenko}, \citenamefont {Nys}, \citenamefont {Pilloni},\
  and\ \citenamefont {Szczepaniak}}]{Albaladejo:2018gif}%
  \BibitemOpen
  \bibfield  {author} {\bibinfo {author} {\bibfnamefont {M.}~\bibnamefont
  {Albaladejo}}, \bibinfo {author} {\bibfnamefont {N.}~\bibnamefont
  {Sherrill}}, \bibinfo {author} {\bibfnamefont {C.}~\bibnamefont
  {Fern\'andez-Ram\'irez}}, \bibinfo {author} {\bibfnamefont {A.}~\bibnamefont
  {Jackura}}, \bibinfo {author} {\bibfnamefont {V.}~\bibnamefont {Mathieu}},
  \bibinfo {author} {\bibfnamefont {M.}~\bibnamefont {Mikhasenko}}, \bibinfo
  {author} {\bibfnamefont {J.}~\bibnamefont {Nys}}, \bibinfo {author}
  {\bibfnamefont {A.}~\bibnamefont {Pilloni}},\ and\ \bibinfo {author}
  {\bibfnamefont {A.~P.}\ \bibnamefont {Szczepaniak}},\ }\href
  {https://doi.org/10.1140/epjc/s10052-018-6045-0} {\bibfield  {journal}
  {\bibinfo  {journal} {Eur.Phys.J.}\ }\textbf {\bibinfo {volume} {C78}},\
  \bibinfo {pages} {574} (\bibinfo {year} {2018})},\ \Eprint
  {https://arxiv.org/abs/1803.06027} {arXiv:1803.06027 [hep-ph]} \BibitemShut
  {NoStop}%
%%CITATION = ARXIV:1803.06027;%%
\bibitem [{\citenamefont {Guo}\ \emph {et~al.}(2015)\citenamefont {Guo},
  \citenamefont {Danilkin}, \citenamefont {Schott}, \citenamefont
  {Fern\'andez-Ram\'irez}, \citenamefont {Mathieu},\ and\ \citenamefont
  {Szczepaniak}}]{Guo:2015zqa}%
  \BibitemOpen
  \bibfield  {author} {\bibinfo {author} {\bibfnamefont {P.}~\bibnamefont
  {Guo}}, \bibinfo {author} {\bibfnamefont {I.~V.}\ \bibnamefont {Danilkin}},
  \bibinfo {author} {\bibfnamefont {D.}~\bibnamefont {Schott}}, \bibinfo
  {author} {\bibfnamefont {C.}~\bibnamefont {Fern\'andez-Ram\'irez}}, \bibinfo
  {author} {\bibfnamefont {V.}~\bibnamefont {Mathieu}},\ and\ \bibinfo {author}
  {\bibfnamefont {A.~P.}\ \bibnamefont {Szczepaniak}},\ }\href
  {https://doi.org/10.1103/PhysRevD.92.054016} {\bibfield  {journal} {\bibinfo
  {journal} {Phys.Rev.}\ }\textbf {\bibinfo {volume} {D92}},\ \bibinfo {pages}
  {054016} (\bibinfo {year} {2015})},\ \Eprint
  {https://arxiv.org/abs/1505.01715} {arXiv:1505.01715 [hep-ph]} \BibitemShut
  {NoStop}%
%%CITATION = ARXIV:1505.01715;%%
\bibitem [{\citenamefont {Guo}\ \emph {et~al.}(2017)\citenamefont {Guo},
  \citenamefont {Danilkin}, \citenamefont {Fern\'andez-Ram\'irez},
  \citenamefont {Mathieu},\ and\ \citenamefont {Szczepaniak}}]{Guo:2016wsi}%
  \BibitemOpen
  \bibfield  {author} {\bibinfo {author} {\bibfnamefont {P.}~\bibnamefont
  {Guo}}, \bibinfo {author} {\bibfnamefont {I.~V.}\ \bibnamefont {Danilkin}},
  \bibinfo {author} {\bibfnamefont {C.}~\bibnamefont {Fern\'andez-Ram\'irez}},
  \bibinfo {author} {\bibfnamefont {V.}~\bibnamefont {Mathieu}},\ and\ \bibinfo
  {author} {\bibfnamefont {A.~P.}\ \bibnamefont {Szczepaniak}},\ }\href
  {https://doi.org/10.1016/j.physletb.2017.05.092} {\bibfield  {journal}
  {\bibinfo  {journal} {Phys.Lett.}\ }\textbf {\bibinfo {volume} {B771}},\
  \bibinfo {pages} {497} (\bibinfo {year} {2017})},\ \Eprint
  {https://arxiv.org/abs/1608.01447} {arXiv:1608.01447 [hep-ph]} \BibitemShut
  {NoStop}%
%%CITATION = ARXIV:1608.01447;%%
\bibitem [{\citenamefont {Colangelo}\ \emph {et~al.}(2017)\citenamefont
  {Colangelo}, \citenamefont {Lanz}, \citenamefont {Leutwyler},\ and\
  \citenamefont {Passemar}}]{Colangelo:2016jmc}%
  \BibitemOpen
  \bibfield  {author} {\bibinfo {author} {\bibfnamefont {G.}~\bibnamefont
  {Colangelo}}, \bibinfo {author} {\bibfnamefont {S.}~\bibnamefont {Lanz}},
  \bibinfo {author} {\bibfnamefont {H.}~\bibnamefont {Leutwyler}},\ and\
  \bibinfo {author} {\bibfnamefont {E.}~\bibnamefont {Passemar}},\ }\href
  {https://doi.org/10.1103/PhysRevLett.118.022001} {\bibfield  {journal}
  {\bibinfo  {journal} {Phys.Rev.Lett.}\ }\textbf {\bibinfo {volume} {118}},\
  \bibinfo {pages} {022001} (\bibinfo {year} {2017})},\ \Eprint
  {https://arxiv.org/abs/1610.03494} {arXiv:1610.03494 [hep-ph]} \BibitemShut
  {NoStop}%
%%CITATION = ARXIV:1610.03494;%%
\bibitem [{\citenamefont {Albaladejo}\ and\ \citenamefont
  {Moussallam}(2017)}]{Albaladejo:2017hhj}%
  \BibitemOpen
  \bibfield  {author} {\bibinfo {author} {\bibfnamefont {M.}~\bibnamefont
  {Albaladejo}}\ and\ \bibinfo {author} {\bibfnamefont {B.}~\bibnamefont
  {Moussallam}},\ }\href {https://doi.org/10.1140/epjc/s10052-017-5052-x}
  {\bibfield  {journal} {\bibinfo  {journal} {Eur.Phys.J.}\ }\textbf {\bibinfo
  {volume} {C77}},\ \bibinfo {pages} {508} (\bibinfo {year} {2017})},\ \Eprint
  {https://arxiv.org/abs/1702.04931} {arXiv:1702.04931 [hep-ph]} \BibitemShut
  {NoStop}%
%%CITATION = ARXIV:1702.04931;%%
\bibitem [{\citenamefont {Niecknig}\ \emph {et~al.}(2012)\citenamefont
  {Niecknig}, \citenamefont {Kubis},\ and\ \citenamefont
  {Schneider}}]{Niecknig:2012sj}%
  \BibitemOpen
  \bibfield  {author} {\bibinfo {author} {\bibfnamefont {F.}~\bibnamefont
  {Niecknig}}, \bibinfo {author} {\bibfnamefont {B.}~\bibnamefont {Kubis}},\
  and\ \bibinfo {author} {\bibfnamefont {S.~P.}\ \bibnamefont {Schneider}},\
  }\href {https://doi.org/10.1140/epjc/s10052-012-2014-1} {\bibfield  {journal}
  {\bibinfo  {journal} {Eur.Phys.J.}\ }\textbf {\bibinfo {volume} {C72}},\
  \bibinfo {pages} {2014} (\bibinfo {year} {2012})},\ \Eprint
  {https://arxiv.org/abs/1203.2501} {arXiv:1203.2501 [hep-ph]} \BibitemShut
  {NoStop}%
%%CITATION = ARXIV:1203.2501;%%
\bibitem [{\citenamefont {Danilkin}\ \emph {et~al.}(2015)\citenamefont
  {Danilkin}, \citenamefont {Fern\'andez-Ram\'irez}, \citenamefont {Guo},
  \citenamefont {Mathieu}, \citenamefont {Schott}, \citenamefont {Shi},\ and\
  \citenamefont {Szczepaniak}}]{Danilkin:2014cra}%
  \BibitemOpen
  \bibfield  {author} {\bibinfo {author} {\bibfnamefont {I.~V.}\ \bibnamefont
  {Danilkin}}, \bibinfo {author} {\bibfnamefont {C.}~\bibnamefont
  {Fern\'andez-Ram\'irez}}, \bibinfo {author} {\bibfnamefont {P.}~\bibnamefont
  {Guo}}, \bibinfo {author} {\bibfnamefont {V.}~\bibnamefont {Mathieu}},
  \bibinfo {author} {\bibfnamefont {D.}~\bibnamefont {Schott}}, \bibinfo
  {author} {\bibfnamefont {M.}~\bibnamefont {Shi}},\ and\ \bibinfo {author}
  {\bibfnamefont {A.~P.}\ \bibnamefont {Szczepaniak}},\ }\href
  {https://doi.org/10.1103/PhysRevD.91.094029} {\bibfield  {journal} {\bibinfo
  {journal} {Phys.Rev.}\ }\textbf {\bibinfo {volume} {D91}},\ \bibinfo {pages}
  {094029} (\bibinfo {year} {2015})},\ \Eprint
  {https://arxiv.org/abs/1409.7708} {arXiv:1409.7708 [hep-ph]} \BibitemShut
  {NoStop}%
%%CITATION = ARXIV:1409.7708;%%
\bibitem [{\citenamefont {Niecknig}\ and\ \citenamefont
  {Kubis}(2015)}]{Niecknig:2015ija}%
  \BibitemOpen
  \bibfield  {author} {\bibinfo {author} {\bibfnamefont {F.}~\bibnamefont
  {Niecknig}}\ and\ \bibinfo {author} {\bibfnamefont {B.}~\bibnamefont
  {Kubis}},\ }\href {https://doi.org/10.1007/JHEP10(2015)142} {\bibfield
  {journal} {\bibinfo  {journal} {JHEP}\ }\textbf {\bibinfo {volume} {10}},\
  \bibinfo {pages} {142}},\ \Eprint {https://arxiv.org/abs/1509.03188}
  {arXiv:1509.03188 [hep-ph]} \BibitemShut {NoStop}%
%%CITATION = ARXIV:1509.03188;%%
\bibitem [{\citenamefont {Isken}\ \emph {et~al.}(2017)\citenamefont {Isken},
  \citenamefont {Kubis}, \citenamefont {Schneider},\ and\ \citenamefont
  {Stoffer}}]{Isken:2017dkw}%
  \BibitemOpen
  \bibfield  {author} {\bibinfo {author} {\bibfnamefont {T.}~\bibnamefont
  {Isken}}, \bibinfo {author} {\bibfnamefont {B.}~\bibnamefont {Kubis}},
  \bibinfo {author} {\bibfnamefont {S.~P.}\ \bibnamefont {Schneider}},\ and\
  \bibinfo {author} {\bibfnamefont {P.}~\bibnamefont {Stoffer}},\ }\href
  {https://doi.org/10.1140/epjc/s10052-017-5024-1} {\bibfield  {journal}
  {\bibinfo  {journal} {Eur.Phys.J.}\ }\textbf {\bibinfo {volume} {C77}},\
  \bibinfo {pages} {489} (\bibinfo {year} {2017})},\ \Eprint
  {https://arxiv.org/abs/1705.04339} {arXiv:1705.04339 [hep-ph]} \BibitemShut
  {NoStop}%
%%CITATION = ARXIV:1705.04339;%%
\bibitem [{\citenamefont {Niecknig}\ and\ \citenamefont
  {Kubis}(2017)}]{Niecknig:2017ylb}%
  \BibitemOpen
  \bibfield  {author} {\bibinfo {author} {\bibfnamefont {F.}~\bibnamefont
  {Niecknig}}\ and\ \bibinfo {author} {\bibfnamefont {B.}~\bibnamefont
  {Kubis}},\ }\href {https://doi.org/10.1016/j.physletb.2018.03.048} {\bibfield
   {journal} {\bibinfo  {journal} {Phys.Lett.}\ }\textbf {\bibinfo {volume}
  {B780}},\ \bibinfo {pages} {471} (\bibinfo {year} {2017})},\ \Eprint
  {https://arxiv.org/abs/1708.00446} {arXiv:1708.00446 [hep-ph]} \BibitemShut
  {NoStop}%
%%CITATION = ARXIV:1708.00446;%%
\bibitem [{\citenamefont {Aoyama}\ \emph {et~al.}(2020)\citenamefont {Aoyama}
  \emph {et~al.}}]{Aoyama:2020ynm}%
  \BibitemOpen
  \bibfield  {author} {\bibinfo {author} {\bibfnamefont {T.}~\bibnamefont
  {Aoyama}} \emph {et~al.},\ }\href
  {https://doi.org/10.1016/j.physrep.2020.07.006} {\bibfield  {journal}
  {\bibinfo  {journal} {Phys.Rept.}\ }\textbf {\bibinfo {volume} {887}},\
  \bibinfo {pages} {1} (\bibinfo {year} {2020})},\ \Eprint
  {https://arxiv.org/abs/2006.04822} {arXiv:2006.04822 [hep-ph]} \BibitemShut
  {NoStop}%
\bibitem [{\citenamefont {Abi}\ \emph {et~al.}(2021)\citenamefont {Abi} \emph
  {et~al.}}]{Muong-2:2021ojo}%
  \BibitemOpen
  \bibfield  {author} {\bibinfo {author} {\bibfnamefont {B.}~\bibnamefont
  {Abi}} \emph {et~al.} (\bibinfo {collaboration} {Muon g-2}),\ }\href
  {https://doi.org/10.1103/PhysRevLett.126.141801} {\bibfield  {journal}
  {\bibinfo  {journal} {Phys.Rev.Lett.}\ }\textbf {\bibinfo {volume} {126}},\
  \bibinfo {pages} {141801} (\bibinfo {year} {2021})},\ \Eprint
  {https://arxiv.org/abs/2104.03281} {arXiv:2104.03281 [hep-ex]} \BibitemShut
  {NoStop}%
\bibitem [{\citenamefont {Hoferichter}\ \emph {et~al.}(2014)\citenamefont
  {Hoferichter}, \citenamefont {Kubis}, \citenamefont {Leupold}, \citenamefont
  {Niecknig},\ and\ \citenamefont {Schneider}}]{Hoferichter:2014vra}%
  \BibitemOpen
  \bibfield  {author} {\bibinfo {author} {\bibfnamefont {M.}~\bibnamefont
  {Hoferichter}}, \bibinfo {author} {\bibfnamefont {B.}~\bibnamefont {Kubis}},
  \bibinfo {author} {\bibfnamefont {S.}~\bibnamefont {Leupold}}, \bibinfo
  {author} {\bibfnamefont {F.}~\bibnamefont {Niecknig}},\ and\ \bibinfo
  {author} {\bibfnamefont {S.~P.}\ \bibnamefont {Schneider}},\ }\href
  {https://doi.org/10.1140/epjc/s10052-014-3180-0} {\bibfield  {journal}
  {\bibinfo  {journal} {Eur.Phys.J.}\ }\textbf {\bibinfo {volume} {C74}},\
  \bibinfo {pages} {3180} (\bibinfo {year} {2014})},\ \Eprint
  {https://arxiv.org/abs/1410.4691} {arXiv:1410.4691 [hep-ph]} \BibitemShut
  {NoStop}%
%%CITATION = ARXIV:1410.4691;%%
\bibitem [{\citenamefont {Hoferichter}\ \emph
  {et~al.}(2018{\natexlab{a}})\citenamefont {Hoferichter}, \citenamefont
  {Hoid}, \citenamefont {Kubis}, \citenamefont {Leupold},\ and\ \citenamefont
  {Schneider}}]{Hoferichter:2018kwz}%
  \BibitemOpen
  \bibfield  {author} {\bibinfo {author} {\bibfnamefont {M.}~\bibnamefont
  {Hoferichter}}, \bibinfo {author} {\bibfnamefont {B.-L.}\ \bibnamefont
  {Hoid}}, \bibinfo {author} {\bibfnamefont {B.}~\bibnamefont {Kubis}},
  \bibinfo {author} {\bibfnamefont {S.}~\bibnamefont {Leupold}},\ and\ \bibinfo
  {author} {\bibfnamefont {S.~P.}\ \bibnamefont {Schneider}},\ }\href
  {https://doi.org/10.1007/JHEP10(2018)141} {\bibfield  {journal} {\bibinfo
  {journal} {JHEP}\ }\textbf {\bibinfo {volume} {10}},\ \bibinfo {pages}
  {141}},\ \Eprint {https://arxiv.org/abs/1808.04823} {arXiv:1808.04823
  [hep-ph]} \BibitemShut {NoStop}%
\bibitem [{\citenamefont {Hoferichter}\ \emph
  {et~al.}(2018{\natexlab{b}})\citenamefont {Hoferichter}, \citenamefont
  {Hoid}, \citenamefont {Kubis}, \citenamefont {Leupold},\ and\ \citenamefont
  {Schneider}}]{Hoferichter:2018dmo}%
  \BibitemOpen
  \bibfield  {author} {\bibinfo {author} {\bibfnamefont {M.}~\bibnamefont
  {Hoferichter}}, \bibinfo {author} {\bibfnamefont {B.-L.}\ \bibnamefont
  {Hoid}}, \bibinfo {author} {\bibfnamefont {B.}~\bibnamefont {Kubis}},
  \bibinfo {author} {\bibfnamefont {S.}~\bibnamefont {Leupold}},\ and\ \bibinfo
  {author} {\bibfnamefont {S.~P.}\ \bibnamefont {Schneider}},\ }\href
  {https://doi.org/10.1103/PhysRevLett.121.112002} {\bibfield  {journal}
  {\bibinfo  {journal} {Phys.Rev.Lett.}\ }\textbf {\bibinfo {volume} {121}},\
  \bibinfo {pages} {112002} (\bibinfo {year} {2018}{\natexlab{b}})},\ \Eprint
  {https://arxiv.org/abs/1805.01471} {arXiv:1805.01471 [hep-ph]} \BibitemShut
  {NoStop}%
\bibitem [{\citenamefont {Hoferichter}\ \emph {et~al.}(2019)\citenamefont
  {Hoferichter}, \citenamefont {Hoid},\ and\ \citenamefont
  {Kubis}}]{Hoferichter:2019mqg}%
  \BibitemOpen
  \bibfield  {author} {\bibinfo {author} {\bibfnamefont {M.}~\bibnamefont
  {Hoferichter}}, \bibinfo {author} {\bibfnamefont {B.-L.}\ \bibnamefont
  {Hoid}},\ and\ \bibinfo {author} {\bibfnamefont {B.}~\bibnamefont {Kubis}},\
  }\href {https://doi.org/10.1007/JHEP08(2019)137} {\bibfield  {journal}
  {\bibinfo  {journal} {JHEP}\ }\textbf {\bibinfo {volume} {08}},\ \bibinfo
  {pages} {137}},\ \Eprint {https://arxiv.org/abs/1907.01556} {arXiv:1907.01556
  [hep-ph]} \BibitemShut {NoStop}%
\bibitem [{\citenamefont {Colangelo}\ \emph {et~al.}(2014)\citenamefont
  {Colangelo}, \citenamefont {Hoferichter}, \citenamefont {Kubis},
  \citenamefont {Procura},\ and\ \citenamefont {Stoffer}}]{Colangelo:2014pva}%
  \BibitemOpen
  \bibfield  {author} {\bibinfo {author} {\bibfnamefont {G.}~\bibnamefont
  {Colangelo}}, \bibinfo {author} {\bibfnamefont {M.}~\bibnamefont
  {Hoferichter}}, \bibinfo {author} {\bibfnamefont {B.}~\bibnamefont {Kubis}},
  \bibinfo {author} {\bibfnamefont {M.}~\bibnamefont {Procura}},\ and\ \bibinfo
  {author} {\bibfnamefont {P.}~\bibnamefont {Stoffer}},\ }\href
  {https://doi.org/10.1016/j.physletb.2014.09.021} {\bibfield  {journal}
  {\bibinfo  {journal} {Phys.Lett.}\ }\textbf {\bibinfo {volume} {B738}},\
  \bibinfo {pages} {6} (\bibinfo {year} {2014})},\ \Eprint
  {https://arxiv.org/abs/1408.2517} {arXiv:1408.2517 [hep-ph]} \BibitemShut
  {NoStop}%
\bibitem [{\citenamefont {Hansen}\ \emph {et~al.}(2021)\citenamefont {Hansen},
  \citenamefont {Brice\~no}, \citenamefont {Edwards}, \citenamefont {Thomas},\
  and\ \citenamefont {Wilson}}]{Hansen:2020otl}%
  \BibitemOpen
  \bibfield  {author} {\bibinfo {author} {\bibfnamefont {M.~T.}\ \bibnamefont
  {Hansen}}, \bibinfo {author} {\bibfnamefont {R.~A.}\ \bibnamefont
  {Brice\~no}}, \bibinfo {author} {\bibfnamefont {R.~G.}\ \bibnamefont
  {Edwards}}, \bibinfo {author} {\bibfnamefont {C.~E.}\ \bibnamefont
  {Thomas}},\ and\ \bibinfo {author} {\bibfnamefont {D.~J.}\ \bibnamefont
  {Wilson}} (\bibinfo {collaboration} {Hadron Spectrum}),\ }\href
  {https://doi.org/10.1103/PhysRevLett.126.012001} {\bibfield  {journal}
  {\bibinfo  {journal} {Phys.Rev.Lett.}\ }\textbf {\bibinfo {volume} {126}},\
  \bibinfo {pages} {012001} (\bibinfo {year} {2021})},\ \Eprint
  {https://arxiv.org/abs/2009.04931} {arXiv:2009.04931 [hep-lat]} \BibitemShut
  {NoStop}%
\bibitem [{\citenamefont {Mai}\ \emph {et~al.}(2021)\citenamefont {Mai},
  \citenamefont {Alexandru}, \citenamefont {Brett}, \citenamefont {Culver},
  \citenamefont {D\"oring}, \citenamefont {Lee},\ and\ \citenamefont
  {Sadasivan}}]{Mai:2021nul}%
  \BibitemOpen
  \bibfield  {author} {\bibinfo {author} {\bibfnamefont {M.}~\bibnamefont
  {Mai}}, \bibinfo {author} {\bibfnamefont {A.}~\bibnamefont {Alexandru}},
  \bibinfo {author} {\bibfnamefont {R.}~\bibnamefont {Brett}}, \bibinfo
  {author} {\bibfnamefont {C.}~\bibnamefont {Culver}}, \bibinfo {author}
  {\bibfnamefont {M.}~\bibnamefont {D\"oring}}, \bibinfo {author}
  {\bibfnamefont {F.~X.}\ \bibnamefont {Lee}},\ and\ \bibinfo {author}
  {\bibfnamefont {D.}~\bibnamefont {Sadasivan}} (\bibinfo {collaboration}
  {GWQCD}),\ }\href {https://doi.org/10.1103/PhysRevLett.127.222001} {\bibfield
   {journal} {\bibinfo  {journal} {Phys.Rev.Lett.}\ }\textbf {\bibinfo {volume}
  {127}},\ \bibinfo {pages} {222001} (\bibinfo {year} {2021})},\ \Eprint
  {https://arxiv.org/abs/2107.03973} {arXiv:2107.03973 [hep-lat]} \BibitemShut
  {NoStop}%
\bibitem [{\citenamefont {Mai}\ \emph {et~al.}(2017)\citenamefont {Mai},
  \citenamefont {Hu}, \citenamefont {D\"oring}, \citenamefont {Pilloni},\ and\
  \citenamefont {Szczepaniak}}]{Mai:2017vot}%
  \BibitemOpen
  \bibfield  {author} {\bibinfo {author} {\bibfnamefont {M.}~\bibnamefont
  {Mai}}, \bibinfo {author} {\bibfnamefont {B.}~\bibnamefont {Hu}}, \bibinfo
  {author} {\bibfnamefont {M.}~\bibnamefont {D\"oring}}, \bibinfo {author}
  {\bibfnamefont {A.}~\bibnamefont {Pilloni}},\ and\ \bibinfo {author}
  {\bibfnamefont {A.}~\bibnamefont {Szczepaniak}},\ }\href
  {https://doi.org/10.1140/epja/i2017-12368-4} {\bibfield  {journal} {\bibinfo
  {journal} {Eur.Phys.J.}\ }\textbf {\bibinfo {volume} {A53}},\ \bibinfo
  {pages} {177} (\bibinfo {year} {2017})},\ \Eprint
  {https://arxiv.org/abs/1706.06118} {arXiv:1706.06118 [nucl-th]} \BibitemShut
  {NoStop}%
%%CITATION = ARXIV:1706.06118;%%
\bibitem [{\citenamefont {Jackura}\ \emph
  {et~al.}(2019{\natexlab{a}})\citenamefont {Jackura}, \citenamefont
  {Fern\'andez-Ram\'irez}, \citenamefont {Mathieu}, \citenamefont {Mikhasenko},
  \citenamefont {Nys}, \citenamefont {Pilloni}, \citenamefont {Salda\~na},
  \citenamefont {Sherrill},\ and\ \citenamefont
  {Szczepaniak}}]{Jackura:2018xnx}%
  \BibitemOpen
  \bibfield  {author} {\bibinfo {author} {\bibfnamefont {A.}~\bibnamefont
  {Jackura}}, \bibinfo {author} {\bibfnamefont {C.}~\bibnamefont
  {Fern\'andez-Ram\'irez}}, \bibinfo {author} {\bibfnamefont {V.}~\bibnamefont
  {Mathieu}}, \bibinfo {author} {\bibfnamefont {M.}~\bibnamefont {Mikhasenko}},
  \bibinfo {author} {\bibfnamefont {J.}~\bibnamefont {Nys}}, \bibinfo {author}
  {\bibfnamefont {A.}~\bibnamefont {Pilloni}}, \bibinfo {author} {\bibfnamefont
  {K.}~\bibnamefont {Salda\~na}}, \bibinfo {author} {\bibfnamefont
  {N.}~\bibnamefont {Sherrill}},\ and\ \bibinfo {author} {\bibfnamefont
  {A.~P.}\ \bibnamefont {Szczepaniak}} (\bibinfo {collaboration} {JPAC}),\
  }\href {https://doi.org/10.1140/epjc/s10052-019-6566-1} {\bibfield  {journal}
  {\bibinfo  {journal} {Eur.Phys.J.}\ }\textbf {\bibinfo {volume} {C79}},\
  \bibinfo {pages} {56} (\bibinfo {year} {2019}{\natexlab{a}})},\ \Eprint
  {https://arxiv.org/abs/1809.10523} {arXiv:1809.10523 [hep-ph]} \BibitemShut
  {NoStop}%
%%CITATION = ARXIV:1809.10523;%%
\bibitem [{\citenamefont {Mikhasenko}\ \emph {et~al.}(2019)\citenamefont
  {Mikhasenko}, \citenamefont {Wunderlich}, \citenamefont {Jackura},
  \citenamefont {Mathieu}, \citenamefont {Pilloni}, \citenamefont {Ketzer},\
  and\ \citenamefont {Szczepaniak}}]{Mikhasenko:2019vhk}%
  \BibitemOpen
  \bibfield  {author} {\bibinfo {author} {\bibfnamefont {M.}~\bibnamefont
  {Mikhasenko}}, \bibinfo {author} {\bibfnamefont {Y.}~\bibnamefont
  {Wunderlich}}, \bibinfo {author} {\bibfnamefont {A.}~\bibnamefont {Jackura}},
  \bibinfo {author} {\bibfnamefont {V.}~\bibnamefont {Mathieu}}, \bibinfo
  {author} {\bibfnamefont {A.}~\bibnamefont {Pilloni}}, \bibinfo {author}
  {\bibfnamefont {B.}~\bibnamefont {Ketzer}},\ and\ \bibinfo {author}
  {\bibfnamefont {A.}~\bibnamefont {Szczepaniak}},\ }\href
  {https://doi.org/10.1007/JHEP08(2019)080} {\bibfield  {journal} {\bibinfo
  {journal} {JHEP}\ }\textbf {\bibinfo {volume} {08}},\ \bibinfo {pages}
  {080}},\ \Eprint {https://arxiv.org/abs/1904.11894} {arXiv:1904.11894
  [hep-ph]} \BibitemShut {NoStop}%
\bibitem [{\citenamefont {Mikhasenko}\ \emph
  {et~al.}(2018{\natexlab{b}})\citenamefont {Mikhasenko}, \citenamefont
  {Pilloni}, \citenamefont {Albaladejo}, \citenamefont
  {Fern\'andez-Ram\'\i{}rez}, \citenamefont {Jackura}, \citenamefont {Mathieu},
  \citenamefont {Nys}, \citenamefont {Rodas}, \citenamefont {Ketzer},\ and\
  \citenamefont {Szczepaniak}}]{JPAC:2018zwp}%
  \BibitemOpen
  \bibfield  {author} {\bibinfo {author} {\bibfnamefont {M.}~\bibnamefont
  {Mikhasenko}}, \bibinfo {author} {\bibfnamefont {A.}~\bibnamefont {Pilloni}},
  \bibinfo {author} {\bibfnamefont {M.}~\bibnamefont {Albaladejo}}, \bibinfo
  {author} {\bibfnamefont {C.}~\bibnamefont {Fern\'andez-Ram\'\i{}rez}},
  \bibinfo {author} {\bibfnamefont {A.}~\bibnamefont {Jackura}}, \bibinfo
  {author} {\bibfnamefont {V.}~\bibnamefont {Mathieu}}, \bibinfo {author}
  {\bibfnamefont {J.}~\bibnamefont {Nys}}, \bibinfo {author} {\bibfnamefont
  {A.}~\bibnamefont {Rodas}}, \bibinfo {author} {\bibfnamefont
  {B.}~\bibnamefont {Ketzer}},\ and\ \bibinfo {author} {\bibfnamefont {A.~P.}\
  \bibnamefont {Szczepaniak}} (\bibinfo {collaboration} {JPAC}),\ }\href
  {https://doi.org/10.1103/PhysRevD.98.096021} {\bibfield  {journal} {\bibinfo
  {journal} {Phys.Rev.}\ }\textbf {\bibinfo {volume} {D98}},\ \bibinfo {pages}
  {096021} (\bibinfo {year} {2018}{\natexlab{b}})},\ \Eprint
  {https://arxiv.org/abs/1810.00016} {arXiv:1810.00016 [hep-ph]} \BibitemShut
  {NoStop}%
\bibitem [{\citenamefont {Davier}\ \emph {et~al.}(2014)\citenamefont {Davier},
  \citenamefont {H\"ocker}, \citenamefont {Malaescu}, \citenamefont {Yuan},\
  and\ \citenamefont {Zhang}}]{Davier:2013sfa}%
  \BibitemOpen
  \bibfield  {author} {\bibinfo {author} {\bibfnamefont {M.}~\bibnamefont
  {Davier}}, \bibinfo {author} {\bibfnamefont {A.}~\bibnamefont {H\"ocker}},
  \bibinfo {author} {\bibfnamefont {B.}~\bibnamefont {Malaescu}}, \bibinfo
  {author} {\bibfnamefont {C.-Z.}\ \bibnamefont {Yuan}},\ and\ \bibinfo
  {author} {\bibfnamefont {Z.}~\bibnamefont {Zhang}},\ }\href
  {https://doi.org/10.1140/epjc/s10052-014-2803-9} {\bibfield  {journal}
  {\bibinfo  {journal} {Eur.Phys.J.}\ }\textbf {\bibinfo {volume} {C74}},\
  \bibinfo {pages} {2803} (\bibinfo {year} {2014})},\ \Eprint
  {https://arxiv.org/abs/1312.1501} {arXiv:1312.1501 [hep-ex]} \BibitemShut
  {NoStop}%
\bibitem [{\citenamefont {Schael}\ \emph {et~al.}(2005)\citenamefont {Schael}
  \emph {et~al.}}]{ALEPH:2005qgp}%
  \BibitemOpen
  \bibfield  {author} {\bibinfo {author} {\bibfnamefont {S.}~\bibnamefont
  {Schael}} \emph {et~al.} (\bibinfo {collaboration} {ALEPH}),\ }\href
  {https://doi.org/10.1016/j.physrep.2005.06.007} {\bibfield  {journal}
  {\bibinfo  {journal} {Phys.Rept.}\ }\textbf {\bibinfo {volume} {421}},\
  \bibinfo {pages} {191} (\bibinfo {year} {2005})},\ \Eprint
  {https://arxiv.org/abs/hep-ex/0506072} {arXiv:hep-ex/0506072} \BibitemShut
  {NoStop}%
\bibitem [{\citenamefont {Adolph}\ \emph {et~al.}(2017)\citenamefont {Adolph}
  \emph {et~al.}}]{COMPASS:2015gxz}%
  \BibitemOpen
  \bibfield  {author} {\bibinfo {author} {\bibfnamefont {C.}~\bibnamefont
  {Adolph}} \emph {et~al.} (\bibinfo {collaboration} {COMPASS}),\ }\href
  {https://doi.org/10.1103/PhysRevD.95.032004} {\bibfield  {journal} {\bibinfo
  {journal} {Phys.Rev.}\ }\textbf {\bibinfo {volume} {D95}},\ \bibinfo {pages}
  {032004} (\bibinfo {year} {2017})},\ \Eprint
  {https://arxiv.org/abs/1509.00992} {arXiv:1509.00992 [hep-ex]} \BibitemShut
  {NoStop}%
%%CITATION = ARXIV:1509.00992;%%
\bibitem [{\citenamefont {Albaladejo}(2021)}]{Albaladejo:2021vln}%
  \BibitemOpen
  \bibfield  {author} {\bibinfo {author} {\bibfnamefont {M.}~\bibnamefont
  {Albaladejo}},\ }\Eprint {https://arxiv.org/abs/2110.02944} {arXiv:2110.02944
  [hep-ph]}  (\bibinfo {year} {2021})\BibitemShut {NoStop}%
\bibitem [{\citenamefont {Du}\ \emph {et~al.}(2022)\citenamefont {Du},
  \citenamefont {Baru}, \citenamefont {Dong}, \citenamefont {Filin},
  \citenamefont {Guo}, \citenamefont {Hanhart}, \citenamefont {Nefediev},
  \citenamefont {Nieves},\ and\ \citenamefont {Wang}}]{Du:2021zzh}%
  \BibitemOpen
  \bibfield  {author} {\bibinfo {author} {\bibfnamefont {M.-L.}\ \bibnamefont
  {Du}}, \bibinfo {author} {\bibfnamefont {V.}~\bibnamefont {Baru}}, \bibinfo
  {author} {\bibfnamefont {X.-K.}\ \bibnamefont {Dong}}, \bibinfo {author}
  {\bibfnamefont {A.}~\bibnamefont {Filin}}, \bibinfo {author} {\bibfnamefont
  {F.-K.}\ \bibnamefont {Guo}}, \bibinfo {author} {\bibfnamefont
  {C.}~\bibnamefont {Hanhart}}, \bibinfo {author} {\bibfnamefont
  {A.}~\bibnamefont {Nefediev}}, \bibinfo {author} {\bibfnamefont
  {J.}~\bibnamefont {Nieves}},\ and\ \bibinfo {author} {\bibfnamefont
  {Q.}~\bibnamefont {Wang}},\ }\href
  {https://doi.org/10.1103/PhysRevD.105.014024} {\bibfield  {journal} {\bibinfo
   {journal} {Phys.Rev.}\ }\textbf {\bibinfo {volume} {D105}},\ \bibinfo
  {pages} {014024} (\bibinfo {year} {2022})},\ \Eprint
  {https://arxiv.org/abs/2110.13765} {arXiv:2110.13765 [hep-ph]} \BibitemShut
  {NoStop}%
\bibitem [{\citenamefont {Luscher}(1986)}]{Luscher:1986pf}%
  \BibitemOpen
  \bibfield  {author} {\bibinfo {author} {\bibfnamefont {M.}~\bibnamefont
  {Luscher}},\ }\href {https://doi.org/10.1007/BF01211097} {\bibfield
  {journal} {\bibinfo  {journal} {Commun.Math.Phys.}\ }\textbf {\bibinfo
  {volume} {105}},\ \bibinfo {pages} {153} (\bibinfo {year}
  {1986})}\BibitemShut {NoStop}%
%%CITATION = CMPHA,105,153;%%
\bibitem [{\citenamefont {L\"uscher}(1991)}]{Luscher:1990ux}%
  \BibitemOpen
  \bibfield  {author} {\bibinfo {author} {\bibfnamefont {M.}~\bibnamefont
  {L\"uscher}},\ }\href {https://doi.org/10.1016/0550-3213(91)90366-6}
  {\bibfield  {journal} {\bibinfo  {journal} {Nucl.Phys.}\ }\textbf {\bibinfo
  {volume} {B354}},\ \bibinfo {pages} {531} (\bibinfo {year}
  {1991})}\BibitemShut {NoStop}%
%%CITATION = NUPHA,B354,531;%%
\bibitem [{\citenamefont {Rummukainen}\ and\ \citenamefont
  {Gottlieb}(1995)}]{Rummukainen:1995vs}%
  \BibitemOpen
  \bibfield  {author} {\bibinfo {author} {\bibfnamefont {K.}~\bibnamefont
  {Rummukainen}}\ and\ \bibinfo {author} {\bibfnamefont {S.~A.}\ \bibnamefont
  {Gottlieb}},\ }\href {https://doi.org/10.1016/0550-3213(95)00313-H}
  {\bibfield  {journal} {\bibinfo  {journal} {Nucl.Phys.}\ }\textbf {\bibinfo
  {volume} {B450}},\ \bibinfo {pages} {397} (\bibinfo {year} {1995})},\ \Eprint
  {https://arxiv.org/abs/hep-lat/9503028} {arXiv:hep-lat/9503028} \BibitemShut
  {NoStop}%
\bibitem [{\citenamefont {Liu}\ \emph {et~al.}(2006)\citenamefont {Liu},
  \citenamefont {Feng},\ and\ \citenamefont {He}}]{Liu:2005kr}%
  \BibitemOpen
  \bibfield  {author} {\bibinfo {author} {\bibfnamefont {C.}~\bibnamefont
  {Liu}}, \bibinfo {author} {\bibfnamefont {X.}~\bibnamefont {Feng}},\ and\
  \bibinfo {author} {\bibfnamefont {S.}~\bibnamefont {He}},\ }\href
  {https://doi.org/10.1142/S0217751X06032150} {\bibfield  {journal} {\bibinfo
  {journal} {Int.J.Mod.Phys.}\ }\textbf {\bibinfo {volume} {A21}},\ \bibinfo
  {pages} {847} (\bibinfo {year} {2006})},\ \Eprint
  {https://arxiv.org/abs/hep-lat/0508022} {arXiv:hep-lat/0508022} \BibitemShut
  {NoStop}%
\bibitem [{\citenamefont {Kim}\ \emph {et~al.}(2005)\citenamefont {Kim},
  \citenamefont {Sachrajda},\ and\ \citenamefont {Sharpe}}]{Kim:2005gf}%
  \BibitemOpen
  \bibfield  {author} {\bibinfo {author} {\bibfnamefont {C.~h.}\ \bibnamefont
  {Kim}}, \bibinfo {author} {\bibfnamefont {C.~T.}\ \bibnamefont {Sachrajda}},\
  and\ \bibinfo {author} {\bibfnamefont {S.~R.}\ \bibnamefont {Sharpe}},\
  }\href {https://doi.org/10.1016/j.nuclphysb.2005.08.029} {\bibfield
  {journal} {\bibinfo  {journal} {Nucl.Phys.}\ }\textbf {\bibinfo {volume}
  {B727}},\ \bibinfo {pages} {218} (\bibinfo {year} {2005})},\ \Eprint
  {https://arxiv.org/abs/hep-lat/0507006} {arXiv:hep-lat/0507006} \BibitemShut
  {NoStop}%
\bibitem [{\citenamefont {Brice\~{n}o}(2014)}]{Briceno:2014oea}%
  \BibitemOpen
  \bibfield  {author} {\bibinfo {author} {\bibfnamefont {R.~A.}\ \bibnamefont
  {Brice\~{n}o}},\ }\href {https://doi.org/10.1103/PhysRevD.89.074507}
  {\bibfield  {journal} {\bibinfo  {journal} {Phys.Rev.}\ }\textbf {\bibinfo
  {volume} {D89}},\ \bibinfo {pages} {074507} (\bibinfo {year} {2014})},\
  \Eprint {https://arxiv.org/abs/1401.3312} {arXiv:1401.3312 [hep-lat]}
  \BibitemShut {NoStop}%
%%CITATION = ARXIV:1401.3312;%%
\bibitem [{\citenamefont {Lellouch}\ and\ \citenamefont
  {Luscher}(2001)}]{Lellouch:2000pv}%
  \BibitemOpen
  \bibfield  {author} {\bibinfo {author} {\bibfnamefont {L.}~\bibnamefont
  {Lellouch}}\ and\ \bibinfo {author} {\bibfnamefont {M.}~\bibnamefont
  {Luscher}},\ }\href {https://doi.org/10.1007/s002200100410} {\bibfield
  {journal} {\bibinfo  {journal} {Commun.Math.Phys.}\ }\textbf {\bibinfo
  {volume} {219}},\ \bibinfo {pages} {31} (\bibinfo {year} {2001})},\ \Eprint
  {https://arxiv.org/abs/hep-lat/0003023} {arXiv:hep-lat/0003023} \BibitemShut
  {NoStop}%
\bibitem [{\citenamefont {Brice\~no}\ \emph {et~al.}(2015)\citenamefont
  {Brice\~no}, \citenamefont {Hansen},\ and\ \citenamefont
  {Walker-Loud}}]{Briceno:2014uqa}%
  \BibitemOpen
  \bibfield  {author} {\bibinfo {author} {\bibfnamefont {R.~A.}\ \bibnamefont
  {Brice\~no}}, \bibinfo {author} {\bibfnamefont {M.~T.}\ \bibnamefont
  {Hansen}},\ and\ \bibinfo {author} {\bibfnamefont {A.}~\bibnamefont
  {Walker-Loud}},\ }\href {https://doi.org/10.1103/PhysRevD.91.034501}
  {\bibfield  {journal} {\bibinfo  {journal} {Phys.Rev.}\ }\textbf {\bibinfo
  {volume} {D91}},\ \bibinfo {pages} {034501} (\bibinfo {year} {2015})},\
  \Eprint {https://arxiv.org/abs/1406.5965} {arXiv:1406.5965 [hep-lat]}
  \BibitemShut {NoStop}%
\bibitem [{\citenamefont {Brice\~no}\ and\ \citenamefont
  {Hansen}(2015)}]{Briceno:2015csa}%
  \BibitemOpen
  \bibfield  {author} {\bibinfo {author} {\bibfnamefont {R.~A.}\ \bibnamefont
  {Brice\~no}}\ and\ \bibinfo {author} {\bibfnamefont {M.~T.}\ \bibnamefont
  {Hansen}},\ }\href {https://doi.org/10.1103/PhysRevD.92.074509} {\bibfield
  {journal} {\bibinfo  {journal} {Phys.Rev.}\ }\textbf {\bibinfo {volume}
  {D92}},\ \bibinfo {pages} {074509} (\bibinfo {year} {2015})},\ \Eprint
  {https://arxiv.org/abs/1502.04314} {arXiv:1502.04314 [hep-lat]} \BibitemShut
  {NoStop}%
\bibitem [{\citenamefont {Polejaeva}\ and\ \citenamefont
  {Rusetsky}(2012)}]{Polejaeva:2012ut}%
  \BibitemOpen
  \bibfield  {author} {\bibinfo {author} {\bibfnamefont {K.}~\bibnamefont
  {Polejaeva}}\ and\ \bibinfo {author} {\bibfnamefont {A.}~\bibnamefont
  {Rusetsky}},\ }\href {https://doi.org/10.1140/epja/i2012-12067-8} {\bibfield
  {journal} {\bibinfo  {journal} {Eur.Phys.J.}\ }\textbf {\bibinfo {volume}
  {A48}},\ \bibinfo {pages} {67} (\bibinfo {year} {2012})},\ \Eprint
  {https://arxiv.org/abs/1203.1241} {arXiv:1203.1241 [hep-lat]} \BibitemShut
  {NoStop}%
%%CITATION = ARXIV:1203.1241;%%
\bibitem [{\citenamefont {Hansen}\ and\ \citenamefont
  {Sharpe}(2015)}]{Hansen:2015zga}%
  \BibitemOpen
  \bibfield  {author} {\bibinfo {author} {\bibfnamefont {M.~T.}\ \bibnamefont
  {Hansen}}\ and\ \bibinfo {author} {\bibfnamefont {S.~R.}\ \bibnamefont
  {Sharpe}},\ }\href {https://doi.org/10.1103/PhysRevD.92.114509} {\bibfield
  {journal} {\bibinfo  {journal} {Phys.Rev.}\ }\textbf {\bibinfo {volume}
  {D92}},\ \bibinfo {pages} {114509} (\bibinfo {year} {2015})},\ \Eprint
  {https://arxiv.org/abs/1504.04248} {arXiv:1504.04248 [hep-lat]} \BibitemShut
  {NoStop}%
%%CITATION = ARXIV:1504.04248;%%
\bibitem [{\citenamefont {Mai}\ and\ \citenamefont
  {D\"oring}(2017)}]{Mai:2017bge}%
  \BibitemOpen
  \bibfield  {author} {\bibinfo {author} {\bibfnamefont {M.}~\bibnamefont
  {Mai}}\ and\ \bibinfo {author} {\bibfnamefont {M.}~\bibnamefont {D\"oring}},\
  }\href {https://doi.org/10.1140/epja/i2017-12440-1} {\bibfield  {journal}
  {\bibinfo  {journal} {Eur.Phys.J.}\ }\textbf {\bibinfo {volume} {A53}},\
  \bibinfo {pages} {240} (\bibinfo {year} {2017})},\ \Eprint
  {https://arxiv.org/abs/1709.08222} {arXiv:1709.08222 [hep-lat]} \BibitemShut
  {NoStop}%
%%CITATION = ARXIV:1709.08222;%%
\bibitem [{\citenamefont {Hammer}\ \emph
  {et~al.}(2017{\natexlab{a}})\citenamefont {Hammer}, \citenamefont {Pang},\
  and\ \citenamefont {Rusetsky}}]{Hammer:2017uqm}%
  \BibitemOpen
  \bibfield  {author} {\bibinfo {author} {\bibfnamefont {H.-W.}\ \bibnamefont
  {Hammer}}, \bibinfo {author} {\bibfnamefont {J.-Y.}\ \bibnamefont {Pang}},\
  and\ \bibinfo {author} {\bibfnamefont {A.}~\bibnamefont {Rusetsky}},\ }\href
  {https://doi.org/10.1007/JHEP09(2017)109} {\bibfield  {journal} {\bibinfo
  {journal} {JHEP}\ }\textbf {\bibinfo {volume} {09}},\ \bibinfo {pages}
  {109}},\ \Eprint {https://arxiv.org/abs/1706.07700} {arXiv:1706.07700
  [hep-lat]} \BibitemShut {NoStop}%
%%CITATION = ARXIV:1706.07700;%%
\bibitem [{\citenamefont {Hammer}\ \emph
  {et~al.}(2017{\natexlab{b}})\citenamefont {Hammer}, \citenamefont {Pang},\
  and\ \citenamefont {Rusetsky}}]{Hammer:2017kms}%
  \BibitemOpen
  \bibfield  {author} {\bibinfo {author} {\bibfnamefont {H.~W.}\ \bibnamefont
  {Hammer}}, \bibinfo {author} {\bibfnamefont {J.~Y.}\ \bibnamefont {Pang}},\
  and\ \bibinfo {author} {\bibfnamefont {A.}~\bibnamefont {Rusetsky}},\ }\href
  {https://doi.org/10.1007/JHEP10(2017)115} {\bibfield  {journal} {\bibinfo
  {journal} {JHEP}\ }\textbf {\bibinfo {volume} {10}},\ \bibinfo {pages}
  {115}},\ \Eprint {https://arxiv.org/abs/1707.02176} {arXiv:1707.02176
  [hep-lat]} \BibitemShut {NoStop}%
%%CITATION = ARXIV:1707.02176;%%
\bibitem [{\citenamefont {Jackura}\ \emph
  {et~al.}(2019{\natexlab{b}})\citenamefont {Jackura}, \citenamefont {Dawid},
  \citenamefont {Fernández-Ramírez}, \citenamefont {Mathieu}, \citenamefont
  {Mikhasenko}, \citenamefont {Pilloni}, \citenamefont {Sharpe},\ and\
  \citenamefont {Szczepaniak}}]{Jackura:2019bmu}%
  \BibitemOpen
  \bibfield  {author} {\bibinfo {author} {\bibfnamefont {A.~W.}\ \bibnamefont
  {Jackura}}, \bibinfo {author} {\bibfnamefont {S.~M.}\ \bibnamefont {Dawid}},
  \bibinfo {author} {\bibfnamefont {C.}~\bibnamefont {Fernández-Ramírez}},
  \bibinfo {author} {\bibfnamefont {V.}~\bibnamefont {Mathieu}}, \bibinfo
  {author} {\bibfnamefont {M.}~\bibnamefont {Mikhasenko}}, \bibinfo {author}
  {\bibfnamefont {A.}~\bibnamefont {Pilloni}}, \bibinfo {author} {\bibfnamefont
  {S.~R.}\ \bibnamefont {Sharpe}},\ and\ \bibinfo {author} {\bibfnamefont
  {A.~P.}\ \bibnamefont {Szczepaniak}},\ }\href
  {https://doi.org/10.1103/PhysRevD.100.034508} {\bibfield  {journal} {\bibinfo
   {journal} {Phys.Rev.}\ }\textbf {\bibinfo {volume} {D100}},\ \bibinfo
  {pages} {034508} (\bibinfo {year} {2019}{\natexlab{b}})},\ \Eprint
  {https://arxiv.org/abs/1905.12007} {arXiv:1905.12007 [hep-ph]} \BibitemShut
  {NoStop}%
%%CITATION = ARXIV:1905.12007;%%
\bibitem [{\citenamefont {Blanton}\ and\ \citenamefont
  {Sharpe}(2020)}]{Blanton:2020gha}%
  \BibitemOpen
  \bibfield  {author} {\bibinfo {author} {\bibfnamefont {T.~D.}\ \bibnamefont
  {Blanton}}\ and\ \bibinfo {author} {\bibfnamefont {S.~R.}\ \bibnamefont
  {Sharpe}},\ }\href {https://doi.org/10.1103/PhysRevD.102.054520} {\bibfield
  {journal} {\bibinfo  {journal} {Phys.Rev.}\ }\textbf {\bibinfo {volume}
  {D102}},\ \bibinfo {pages} {054520} (\bibinfo {year} {2020})},\ \Eprint
  {https://arxiv.org/abs/2007.16188} {arXiv:2007.16188 [hep-lat]} \BibitemShut
  {NoStop}%
\bibitem [{\citenamefont {Mai}\ \emph {et~al.}(2020)\citenamefont {Mai},
  \citenamefont {D\"oring}, \citenamefont {Culver},\ and\ \citenamefont
  {Alexandru}}]{Mai:2019fba}%
  \BibitemOpen
  \bibfield  {author} {\bibinfo {author} {\bibfnamefont {M.}~\bibnamefont
  {Mai}}, \bibinfo {author} {\bibfnamefont {M.}~\bibnamefont {D\"oring}},
  \bibinfo {author} {\bibfnamefont {C.}~\bibnamefont {Culver}},\ and\ \bibinfo
  {author} {\bibfnamefont {A.}~\bibnamefont {Alexandru}},\ }\href
  {https://doi.org/10.1103/PhysRevD.101.054510} {\bibfield  {journal} {\bibinfo
   {journal} {Phys.Rev.}\ }\textbf {\bibinfo {volume} {D101}},\ \bibinfo
  {pages} {054510} (\bibinfo {year} {2020})},\ \Eprint
  {https://arxiv.org/abs/1909.05749} {arXiv:1909.05749 [hep-lat]} \BibitemShut
  {NoStop}%
\bibitem [{\citenamefont {Alexandru}\ \emph {et~al.}(2020)\citenamefont
  {Alexandru}, \citenamefont {Brett}, \citenamefont {Culver}, \citenamefont
  {D\"oring}, \citenamefont {Guo}, \citenamefont {Lee},\ and\ \citenamefont
  {Mai}}]{Alexandru:2020xqf}%
  \BibitemOpen
  \bibfield  {author} {\bibinfo {author} {\bibfnamefont {A.}~\bibnamefont
  {Alexandru}}, \bibinfo {author} {\bibfnamefont {R.}~\bibnamefont {Brett}},
  \bibinfo {author} {\bibfnamefont {C.}~\bibnamefont {Culver}}, \bibinfo
  {author} {\bibfnamefont {M.}~\bibnamefont {D\"oring}}, \bibinfo {author}
  {\bibfnamefont {D.}~\bibnamefont {Guo}}, \bibinfo {author} {\bibfnamefont
  {F.~X.}\ \bibnamefont {Lee}},\ and\ \bibinfo {author} {\bibfnamefont
  {M.}~\bibnamefont {Mai}},\ }\href
  {https://doi.org/10.1103/PhysRevD.102.114523} {\bibfield  {journal} {\bibinfo
   {journal} {Phys.Rev.}\ }\textbf {\bibinfo {volume} {D102}},\ \bibinfo
  {pages} {114523} (\bibinfo {year} {2020})},\ \Eprint
  {https://arxiv.org/abs/2009.12358} {arXiv:2009.12358 [hep-lat]} \BibitemShut
  {NoStop}%
\bibitem [{\citenamefont {Brett}\ \emph {et~al.}(2021)\citenamefont {Brett},
  \citenamefont {Culver}, \citenamefont {Mai}, \citenamefont {Alexandru},
  \citenamefont {D\"oring},\ and\ \citenamefont {Lee}}]{Brett:2021wyd}%
  \BibitemOpen
  \bibfield  {author} {\bibinfo {author} {\bibfnamefont {R.}~\bibnamefont
  {Brett}}, \bibinfo {author} {\bibfnamefont {C.}~\bibnamefont {Culver}},
  \bibinfo {author} {\bibfnamefont {M.}~\bibnamefont {Mai}}, \bibinfo {author}
  {\bibfnamefont {A.}~\bibnamefont {Alexandru}}, \bibinfo {author}
  {\bibfnamefont {M.}~\bibnamefont {D\"oring}},\ and\ \bibinfo {author}
  {\bibfnamefont {F.~X.}\ \bibnamefont {Lee}},\ }\href
  {https://doi.org/10.1103/PhysRevD.104.014501} {\bibfield  {journal} {\bibinfo
   {journal} {Phys.Rev.}\ }\textbf {\bibinfo {volume} {D104}},\ \bibinfo
  {pages} {014501} (\bibinfo {year} {2021})},\ \Eprint
  {https://arxiv.org/abs/2101.06144} {arXiv:2101.06144 [hep-lat]} \BibitemShut
  {NoStop}%
\bibitem [{\citenamefont {Donnachie}\ \emph {et~al.}(2005)\citenamefont
  {Donnachie}, \citenamefont {Dosch}, \citenamefont {Nachtmann},\ and\
  \citenamefont {Landshoff}}]{Donnachie:2002en}%
  \BibitemOpen
  \bibfield  {author} {\bibinfo {author} {\bibfnamefont {S.}~\bibnamefont
  {Donnachie}}, \bibinfo {author} {\bibfnamefont {H.~G.}\ \bibnamefont
  {Dosch}}, \bibinfo {author} {\bibfnamefont {O.}~\bibnamefont {Nachtmann}},\
  and\ \bibinfo {author} {\bibfnamefont {P.}~\bibnamefont {Landshoff}},\ }\href
  {https://books.google.it/books?id=WpGHPwAACAAJ} {\emph {\bibinfo {title}
  {{Pomeron physics and QCD}}}},\ Cambridge Monographs on Particle Physics,
  Nuclear Physics and Cosmology\ (\bibinfo  {publisher} {Cambridge University
  Press},\ \bibinfo {year} {2005})\BibitemShut {NoStop}%
\bibitem [{\citenamefont {Abazov}\ \emph {et~al.}(2021)\citenamefont {Abazov}
  \emph {et~al.}}]{TOTEM:2020zzr}%
  \BibitemOpen
  \bibfield  {author} {\bibinfo {author} {\bibfnamefont {V.~M.}\ \bibnamefont
  {Abazov}} \emph {et~al.} (\bibinfo {collaboration} {TOTEM, D0}),\ }\href
  {https://doi.org/10.1103/PhysRevLett.127.062003} {\bibfield  {journal}
  {\bibinfo  {journal} {Phys.Rev.Lett.}\ }\textbf {\bibinfo {volume} {127}},\
  \bibinfo {pages} {062003} (\bibinfo {year} {2021})},\ \Eprint
  {https://arxiv.org/abs/2012.03981} {arXiv:2012.03981 [hep-ex]} \BibitemShut
  {NoStop}%
\bibitem [{\citenamefont {Mathieu}\ \emph
  {et~al.}(2015{\natexlab{a}})\citenamefont {Mathieu}, \citenamefont {Fox},\
  and\ \citenamefont {Szczepaniak}}]{Mathieu:2015eia}%
  \BibitemOpen
  \bibfield  {author} {\bibinfo {author} {\bibfnamefont {V.}~\bibnamefont
  {Mathieu}}, \bibinfo {author} {\bibfnamefont {G.}~\bibnamefont {Fox}},\ and\
  \bibinfo {author} {\bibfnamefont {A.~P.}\ \bibnamefont {Szczepaniak}},\
  }\href {https://doi.org/10.1103/PhysRevD.92.074013} {\bibfield  {journal}
  {\bibinfo  {journal} {Phys.Rev.}\ }\textbf {\bibinfo {volume} {D92}},\
  \bibinfo {pages} {074013} (\bibinfo {year} {2015}{\natexlab{a}})},\ \Eprint
  {https://arxiv.org/abs/1505.02321} {arXiv:1505.02321 [hep-ph]} \BibitemShut
  {NoStop}%
%%CITATION = ARXIV:1505.02321;%%
\bibitem [{\citenamefont {Nys}\ \emph {et~al.}(2018{\natexlab{a}})\citenamefont
  {Nys}, \citenamefont {Mathieu}, \citenamefont {Fern\'andez-Ram\'\i{}rez},
  \citenamefont {Jackura}, \citenamefont {Mikhasenko}, \citenamefont {Pilloni},
  \citenamefont {Sherrill}, \citenamefont {Ryckebusch}, \citenamefont
  {Szczepaniak},\ and\ \citenamefont
  {Fox}}]{JointPhysicsAnalysisCenter:2017del}%
  \BibitemOpen
  \bibfield  {author} {\bibinfo {author} {\bibfnamefont {J.}~\bibnamefont
  {Nys}}, \bibinfo {author} {\bibfnamefont {V.}~\bibnamefont {Mathieu}},
  \bibinfo {author} {\bibfnamefont {C.}~\bibnamefont
  {Fern\'andez-Ram\'\i{}rez}}, \bibinfo {author} {\bibfnamefont
  {A.}~\bibnamefont {Jackura}}, \bibinfo {author} {\bibfnamefont
  {M.}~\bibnamefont {Mikhasenko}}, \bibinfo {author} {\bibfnamefont
  {A.}~\bibnamefont {Pilloni}}, \bibinfo {author} {\bibfnamefont
  {N.}~\bibnamefont {Sherrill}}, \bibinfo {author} {\bibfnamefont
  {J.}~\bibnamefont {Ryckebusch}}, \bibinfo {author} {\bibfnamefont {A.~P.}\
  \bibnamefont {Szczepaniak}},\ and\ \bibinfo {author} {\bibfnamefont
  {G.}~\bibnamefont {Fox}} (\bibinfo {collaboration} {Joint Physics Analysis
  Center}),\ }\href {https://doi.org/10.1016/j.physletb.2018.01.075} {\bibfield
   {journal} {\bibinfo  {journal} {Phys.Lett.}\ }\textbf {\bibinfo {volume}
  {B779}},\ \bibinfo {pages} {77} (\bibinfo {year} {2018}{\natexlab{a}})},\
  \Eprint {https://arxiv.org/abs/1710.09394} {arXiv:1710.09394 [hep-ph]}
  \BibitemShut {NoStop}%
%%CITATION = ARXIV:1710.09394;%%
\bibitem [{\citenamefont {Mathieu}\ \emph {et~al.}(2017)\citenamefont
  {Mathieu}, \citenamefont {Nys}, \citenamefont {Fernández-Ramírez},
  \citenamefont {Jackura}, \citenamefont {Mikhasenko}, \citenamefont {Pilloni},
  \citenamefont {Szczepaniak},\ and\ \citenamefont {Fox}}]{Mathieu:2017jjs}%
  \BibitemOpen
  \bibfield  {author} {\bibinfo {author} {\bibfnamefont {V.}~\bibnamefont
  {Mathieu}}, \bibinfo {author} {\bibfnamefont {J.}~\bibnamefont {Nys}},
  \bibinfo {author} {\bibfnamefont {C.}~\bibnamefont {Fernández-Ramírez}},
  \bibinfo {author} {\bibfnamefont {A.}~\bibnamefont {Jackura}}, \bibinfo
  {author} {\bibfnamefont {M.}~\bibnamefont {Mikhasenko}}, \bibinfo {author}
  {\bibfnamefont {A.}~\bibnamefont {Pilloni}}, \bibinfo {author} {\bibfnamefont
  {A.~P.}\ \bibnamefont {Szczepaniak}},\ and\ \bibinfo {author} {\bibfnamefont
  {G.}~\bibnamefont {Fox}},\ }\href
  {https://doi.org/10.1016/j.physletb.2017.09.081} {\bibfield  {journal}
  {\bibinfo  {journal} {Phys.Lett.}\ }\textbf {\bibinfo {volume} {B774}},\
  \bibinfo {pages} {362} (\bibinfo {year} {2017})},\ \Eprint
  {https://arxiv.org/abs/1704.07684} {arXiv:1704.07684 [hep-ph]} \BibitemShut
  {NoStop}%
%%CITATION = ARXIV:1704.07684;%%
\bibitem [{\citenamefont {Mathieu}\ \emph
  {et~al.}(2018{\natexlab{a}})\citenamefont {Mathieu}, \citenamefont {Nys},
  \citenamefont {Fern\'andez-Ram\'irez}, \citenamefont {Jackura}, \citenamefont
  {Pilloni}, \citenamefont {Sherrill}, \citenamefont {Szczepaniak},\ and\
  \citenamefont {Fox}}]{Mathieu:2018xyc}%
  \BibitemOpen
  \bibfield  {author} {\bibinfo {author} {\bibfnamefont {V.}~\bibnamefont
  {Mathieu}}, \bibinfo {author} {\bibfnamefont {J.}~\bibnamefont {Nys}},
  \bibinfo {author} {\bibfnamefont {C.}~\bibnamefont {Fern\'andez-Ram\'irez}},
  \bibinfo {author} {\bibfnamefont {A.}~\bibnamefont {Jackura}}, \bibinfo
  {author} {\bibfnamefont {A.}~\bibnamefont {Pilloni}}, \bibinfo {author}
  {\bibfnamefont {N.}~\bibnamefont {Sherrill}}, \bibinfo {author}
  {\bibfnamefont {A.~P.}\ \bibnamefont {Szczepaniak}},\ and\ \bibinfo {author}
  {\bibfnamefont {G.}~\bibnamefont {Fox}} (\bibinfo {collaboration} {JPAC}),\
  }\href {https://doi.org/10.1103/PhysRevD.97.094003} {\bibfield  {journal}
  {\bibinfo  {journal} {Phys.Rev.}\ }\textbf {\bibinfo {volume} {D97}},\
  \bibinfo {pages} {094003} (\bibinfo {year} {2018}{\natexlab{a}})},\ \Eprint
  {https://arxiv.org/abs/1802.09403} {arXiv:1802.09403 [hep-ph]} \BibitemShut
  {NoStop}%
%%CITATION = ARXIV:1802.09403;%%
\bibitem [{\citenamefont {Mathieu}\ \emph {et~al.}(2020)\citenamefont
  {Mathieu}, \citenamefont {Pilloni}, \citenamefont {Albaladejo}, \citenamefont
  {Bibrzycki}, \citenamefont {Celentano}, \citenamefont
  {Fern\'andez-Ram\'irez},\ and\ \citenamefont
  {Szczepaniak}}]{Mathieu:2020zpm}%
  \BibitemOpen
  \bibfield  {author} {\bibinfo {author} {\bibfnamefont {V.}~\bibnamefont
  {Mathieu}}, \bibinfo {author} {\bibfnamefont {A.}~\bibnamefont {Pilloni}},
  \bibinfo {author} {\bibfnamefont {M.}~\bibnamefont {Albaladejo}}, \bibinfo
  {author} {\bibfnamefont {L.}~\bibnamefont {Bibrzycki}}, \bibinfo {author}
  {\bibfnamefont {A.}~\bibnamefont {Celentano}}, \bibinfo {author}
  {\bibfnamefont {C.}~\bibnamefont {Fern\'andez-Ram\'irez}},\ and\ \bibinfo
  {author} {\bibfnamefont {A.}~\bibnamefont {Szczepaniak}} (\bibinfo
  {collaboration} {JPAC}),\ }\href
  {https://doi.org/10.1103/PhysRevD.102.014003} {\bibfield  {journal} {\bibinfo
   {journal} {Phys.Rev.}\ }\textbf {\bibinfo {volume} {D102}},\ \bibinfo
  {pages} {014003} (\bibinfo {year} {2020})},\ \Eprint
  {https://arxiv.org/abs/2005.01617} {arXiv:2005.01617 [hep-ph]} \BibitemShut
  {NoStop}%
\bibitem [{\citenamefont {Nys}\ \emph {et~al.}(2018{\natexlab{b}})\citenamefont
  {Nys}, \citenamefont {Hiller~Blin}, \citenamefont {Mathieu}, \citenamefont
  {Fern\'andez-Ram\'irez}, \citenamefont {Jackura}, \citenamefont {Pilloni},
  \citenamefont {Ryckebusch}, \citenamefont {Szczepaniak},\ and\ \citenamefont
  {Fox}}]{Nys:2018vck}%
  \BibitemOpen
  \bibfield  {author} {\bibinfo {author} {\bibfnamefont {J.}~\bibnamefont
  {Nys}}, \bibinfo {author} {\bibfnamefont {A.}~\bibnamefont {Hiller~Blin}},
  \bibinfo {author} {\bibfnamefont {V.}~\bibnamefont {Mathieu}}, \bibinfo
  {author} {\bibfnamefont {C.}~\bibnamefont {Fern\'andez-Ram\'irez}}, \bibinfo
  {author} {\bibfnamefont {A.}~\bibnamefont {Jackura}}, \bibinfo {author}
  {\bibfnamefont {A.}~\bibnamefont {Pilloni}}, \bibinfo {author} {\bibfnamefont
  {J.}~\bibnamefont {Ryckebusch}}, \bibinfo {author} {\bibfnamefont
  {A.}~\bibnamefont {Szczepaniak}},\ and\ \bibinfo {author} {\bibfnamefont
  {G.}~\bibnamefont {Fox}} (\bibinfo {collaboration} {JPAC}),\ }\href
  {https://doi.org/10.1103/PhysRevD.98.034020} {\bibfield  {journal} {\bibinfo
  {journal} {Phys.Rev.}\ }\textbf {\bibinfo {volume} {D98}},\ \bibinfo {pages}
  {034020} (\bibinfo {year} {2018}{\natexlab{b}})},\ \Eprint
  {https://arxiv.org/abs/1806.01891} {arXiv:1806.01891 [hep-ph]} \BibitemShut
  {NoStop}%
\bibitem [{\citenamefont {Mathieu}\ \emph
  {et~al.}(2015{\natexlab{b}})\citenamefont {Mathieu}, \citenamefont
  {Danilkin}, \citenamefont {Fern\'andez-Ram\'irez}, \citenamefont
  {Pennington}, \citenamefont {Schott}, \citenamefont {Szczepaniak},\ and\
  \citenamefont {Fox}}]{Mathieu:2015gxa}%
  \BibitemOpen
  \bibfield  {author} {\bibinfo {author} {\bibfnamefont {V.}~\bibnamefont
  {Mathieu}}, \bibinfo {author} {\bibfnamefont {I.~V.}\ \bibnamefont
  {Danilkin}}, \bibinfo {author} {\bibfnamefont {C.}~\bibnamefont
  {Fern\'andez-Ram\'irez}}, \bibinfo {author} {\bibfnamefont {M.~R.}\
  \bibnamefont {Pennington}}, \bibinfo {author} {\bibfnamefont
  {D.}~\bibnamefont {Schott}}, \bibinfo {author} {\bibfnamefont {A.~P.}\
  \bibnamefont {Szczepaniak}},\ and\ \bibinfo {author} {\bibfnamefont
  {G.}~\bibnamefont {Fox}},\ }\href
  {https://doi.org/10.1103/PhysRevD.92.074004} {\bibfield  {journal} {\bibinfo
  {journal} {Phys.Rev.}\ }\textbf {\bibinfo {volume} {D92}},\ \bibinfo {pages}
  {074004} (\bibinfo {year} {2015}{\natexlab{b}})},\ \Eprint
  {https://arxiv.org/abs/1506.01764} {arXiv:1506.01764 [hep-ph]} \BibitemShut
  {NoStop}%
%%CITATION = ARXIV:1506.01764;%%
\bibitem [{\citenamefont {Nys}\ \emph {et~al.}(2017)\citenamefont {Nys},
  \citenamefont {Mathieu}, \citenamefont {Fern\'andez-Ram\'irez}, \citenamefont
  {Hiller~Blin}, \citenamefont {Jackura}, \citenamefont {Mikhasenko},
  \citenamefont {Pilloni}, \citenamefont {Szczepaniak}, \citenamefont {Fox},\
  and\ \citenamefont {Ryckebusch}}]{Nys:2016vjz}%
  \BibitemOpen
  \bibfield  {author} {\bibinfo {author} {\bibfnamefont {J.}~\bibnamefont
  {Nys}}, \bibinfo {author} {\bibfnamefont {V.}~\bibnamefont {Mathieu}},
  \bibinfo {author} {\bibfnamefont {C.}~\bibnamefont {Fern\'andez-Ram\'irez}},
  \bibinfo {author} {\bibfnamefont {A.~N.}\ \bibnamefont {Hiller~Blin}},
  \bibinfo {author} {\bibfnamefont {A.}~\bibnamefont {Jackura}}, \bibinfo
  {author} {\bibfnamefont {M.}~\bibnamefont {Mikhasenko}}, \bibinfo {author}
  {\bibfnamefont {A.}~\bibnamefont {Pilloni}}, \bibinfo {author} {\bibfnamefont
  {A.~P.}\ \bibnamefont {Szczepaniak}}, \bibinfo {author} {\bibfnamefont
  {G.}~\bibnamefont {Fox}},\ and\ \bibinfo {author} {\bibfnamefont
  {J.}~\bibnamefont {Ryckebusch}} (\bibinfo {collaboration} {JPAC}),\ }\href
  {https://doi.org/10.1103/PhysRevD.95.034014} {\bibfield  {journal} {\bibinfo
  {journal} {Phys.Rev.}\ }\textbf {\bibinfo {volume} {D95}},\ \bibinfo {pages}
  {034014} (\bibinfo {year} {2017})},\ \Eprint
  {https://arxiv.org/abs/1611.04658} {arXiv:1611.04658 [hep-ph]} \BibitemShut
  {NoStop}%
%%CITATION = ARXIV:1611.04658;%%
\bibitem [{\citenamefont {Mathieu}\ \emph
  {et~al.}(2018{\natexlab{b}})\citenamefont {Mathieu}, \citenamefont {Nys},
  \citenamefont {Pilloni}, \citenamefont {Fern\'andez-Ram\'irez}, \citenamefont
  {Jackura}, \citenamefont {Mikhasenko}, \citenamefont {Pauk}, \citenamefont
  {Szczepaniak},\ and\ \citenamefont {Fox}}]{Mathieu:2017but}%
  \BibitemOpen
  \bibfield  {author} {\bibinfo {author} {\bibfnamefont {V.}~\bibnamefont
  {Mathieu}}, \bibinfo {author} {\bibfnamefont {J.}~\bibnamefont {Nys}},
  \bibinfo {author} {\bibfnamefont {A.}~\bibnamefont {Pilloni}}, \bibinfo
  {author} {\bibfnamefont {C.}~\bibnamefont {Fern\'andez-Ram\'irez}}, \bibinfo
  {author} {\bibfnamefont {A.}~\bibnamefont {Jackura}}, \bibinfo {author}
  {\bibfnamefont {M.}~\bibnamefont {Mikhasenko}}, \bibinfo {author}
  {\bibfnamefont {V.}~\bibnamefont {Pauk}}, \bibinfo {author} {\bibfnamefont
  {A.}~\bibnamefont {Szczepaniak}},\ and\ \bibinfo {author} {\bibfnamefont
  {G.}~\bibnamefont {Fox}},\ }\href
  {https://doi.org/10.1209/0295-5075/122/41001} {\bibfield  {journal} {\bibinfo
   {journal} {Europhys.Lett.}\ }\textbf {\bibinfo {volume} {122}},\ \bibinfo
  {pages} {41001} (\bibinfo {year} {2018}{\natexlab{b}})},\ \Eprint
  {https://arxiv.org/abs/1708.07779} {arXiv:1708.07779 [hep-ph]} \BibitemShut
  {NoStop}%
%%CITATION = ARXIV:1708.07779;%%
\bibitem [{\citenamefont {Mathieu}\ \emph
  {et~al.}(2018{\natexlab{c}})\citenamefont {Mathieu}, \citenamefont {Nys},
  \citenamefont {Fern\'andez-Ram\'irez}, \citenamefont {Hiller~Blin},
  \citenamefont {Jackura}, \citenamefont {Pilloni}, \citenamefont
  {Szczepaniak},\ and\ \citenamefont {Fox}}]{Mathieu:2018mjw}%
  \BibitemOpen
  \bibfield  {author} {\bibinfo {author} {\bibfnamefont {V.}~\bibnamefont
  {Mathieu}}, \bibinfo {author} {\bibfnamefont {J.}~\bibnamefont {Nys}},
  \bibinfo {author} {\bibfnamefont {C.}~\bibnamefont {Fern\'andez-Ram\'irez}},
  \bibinfo {author} {\bibfnamefont {A.~N.}\ \bibnamefont {Hiller~Blin}},
  \bibinfo {author} {\bibfnamefont {A.}~\bibnamefont {Jackura}}, \bibinfo
  {author} {\bibfnamefont {A.}~\bibnamefont {Pilloni}}, \bibinfo {author}
  {\bibfnamefont {A.~P.}\ \bibnamefont {Szczepaniak}},\ and\ \bibinfo {author}
  {\bibfnamefont {G.}~\bibnamefont {Fox}} (\bibinfo {collaboration} {JPAC}),\
  }\href@noop {} {\bibfield  {journal} {\bibinfo  {journal} {Phys.Rev.}\
  }\textbf {\bibinfo {volume} {D98}},\ \bibinfo {pages} {014041} (\bibinfo
  {year} {2018}{\natexlab{c}})},\ \Eprint {https://arxiv.org/abs/1806.08414}
  {arXiv:1806.08414 [hep-ph]} \BibitemShut {NoStop}%
%%CITATION = ARXIV:1806.08414;%%
\bibitem [{\citenamefont {Shi}\ \emph {et~al.}(2015)\citenamefont {Shi},
  \citenamefont {Danilkin}, \citenamefont {Fern\'andez-Ram\'irez},
  \citenamefont {Mathieu}, \citenamefont {Pennington}, \citenamefont {Schott},\
  and\ \citenamefont {Szczepaniak}}]{Shi:2014nea}%
  \BibitemOpen
  \bibfield  {author} {\bibinfo {author} {\bibfnamefont {M.}~\bibnamefont
  {Shi}}, \bibinfo {author} {\bibfnamefont {I.~V.}\ \bibnamefont {Danilkin}},
  \bibinfo {author} {\bibfnamefont {C.}~\bibnamefont {Fern\'andez-Ram\'irez}},
  \bibinfo {author} {\bibfnamefont {V.}~\bibnamefont {Mathieu}}, \bibinfo
  {author} {\bibfnamefont {M.~R.}\ \bibnamefont {Pennington}}, \bibinfo
  {author} {\bibfnamefont {D.}~\bibnamefont {Schott}},\ and\ \bibinfo {author}
  {\bibfnamefont {A.~P.}\ \bibnamefont {Szczepaniak}},\ }\href
  {https://doi.org/10.1103/PhysRevD.91.034007} {\bibfield  {journal} {\bibinfo
  {journal} {Phys.Rev.}\ }\textbf {\bibinfo {volume} {D91}},\ \bibinfo {pages}
  {034007} (\bibinfo {year} {2015})},\ \Eprint
  {https://arxiv.org/abs/1411.6237} {arXiv:1411.6237 [hep-ph]} \BibitemShut
  {NoStop}%
%%CITATION = ARXIV:1411.6237;%%
\bibitem [{\citenamefont {Bibrzycki}\ \emph {et~al.}(2021)\citenamefont
  {Bibrzycki}, \citenamefont {Fern\'andez-Ram\'\i{}rez}, \citenamefont
  {Mathieu}, \citenamefont {Mikhasenko}, \citenamefont {Albaladejo},
  \citenamefont {Hiller~Blin}, \citenamefont {Pilloni},\ and\ \citenamefont
  {Szczepaniak}}]{Bibrzycki:2021rwh}%
  \BibitemOpen
  \bibfield  {author} {\bibinfo {author} {\bibfnamefont {{\L}.}~\bibnamefont
  {Bibrzycki}}, \bibinfo {author} {\bibfnamefont {C.}~\bibnamefont
  {Fern\'andez-Ram\'\i{}rez}}, \bibinfo {author} {\bibfnamefont
  {V.}~\bibnamefont {Mathieu}}, \bibinfo {author} {\bibfnamefont
  {M.}~\bibnamefont {Mikhasenko}}, \bibinfo {author} {\bibfnamefont
  {M.}~\bibnamefont {Albaladejo}}, \bibinfo {author} {\bibfnamefont {A.~N.}\
  \bibnamefont {Hiller~Blin}}, \bibinfo {author} {\bibfnamefont
  {A.}~\bibnamefont {Pilloni}},\ and\ \bibinfo {author} {\bibfnamefont {A.~P.}\
  \bibnamefont {Szczepaniak}},\ }\href
  {https://doi.org/10.1140/epjc/s10052-021-09420-1} {\bibfield  {journal}
  {\bibinfo  {journal} {Eur.Phys.J.}\ }\textbf {\bibinfo {volume} {C81}},\
  \bibinfo {pages} {647} (\bibinfo {year} {2021})}\BibitemShut {NoStop}%
\bibitem [{\citenamefont {Albaladejo}\ \emph
  {et~al.}(2020{\natexlab{c}})\citenamefont {Albaladejo}, \citenamefont
  {Hiller~Blin}, \citenamefont {Pilloni}, \citenamefont {Winney}, \citenamefont
  {Fern\'andez-Ram\'irez}, \citenamefont {Mathieu},\ and\ \citenamefont
  {Szczepaniak}}]{Albaladejo:2020tzt}%
  \BibitemOpen
  \bibfield  {author} {\bibinfo {author} {\bibfnamefont {M.}~\bibnamefont
  {Albaladejo}}, \bibinfo {author} {\bibfnamefont {A.~N.}\ \bibnamefont
  {Hiller~Blin}}, \bibinfo {author} {\bibfnamefont {A.}~\bibnamefont
  {Pilloni}}, \bibinfo {author} {\bibfnamefont {D.}~\bibnamefont {Winney}},
  \bibinfo {author} {\bibfnamefont {C.}~\bibnamefont {Fern\'andez-Ram\'irez}},
  \bibinfo {author} {\bibfnamefont {V.}~\bibnamefont {Mathieu}},\ and\ \bibinfo
  {author} {\bibfnamefont {A.}~\bibnamefont {Szczepaniak}} (\bibinfo
  {collaboration} {JPAC}),\ }\href
  {https://doi.org/10.1103/PhysRevD.102.114010} {\bibfield  {journal} {\bibinfo
   {journal} {Phys.Rev.}\ }\textbf {\bibinfo {volume} {D102}},\ \bibinfo
  {pages} {114010} (\bibinfo {year} {2020}{\natexlab{c}})},\ \Eprint
  {https://arxiv.org/abs/2008.01001} {arXiv:2008.01001 [hep-ph]} \BibitemShut
  {NoStop}%
\bibitem [{\citenamefont {Hiller~Blin}\ \emph {et~al.}(2016)\citenamefont
  {Hiller~Blin}, \citenamefont {Fern\'andez-Ram\'\i{}rez}, \citenamefont
  {Jackura}, \citenamefont {Mathieu}, \citenamefont {Mokeev}, \citenamefont
  {Pilloni},\ and\ \citenamefont {Szczepaniak}}]{HillerBlin:2016odx}%
  \BibitemOpen
  \bibfield  {author} {\bibinfo {author} {\bibfnamefont {A.~N.}\ \bibnamefont
  {Hiller~Blin}}, \bibinfo {author} {\bibfnamefont {C.}~\bibnamefont
  {Fern\'andez-Ram\'\i{}rez}}, \bibinfo {author} {\bibfnamefont
  {A.}~\bibnamefont {Jackura}}, \bibinfo {author} {\bibfnamefont
  {V.}~\bibnamefont {Mathieu}}, \bibinfo {author} {\bibfnamefont {V.~I.}\
  \bibnamefont {Mokeev}}, \bibinfo {author} {\bibfnamefont {A.}~\bibnamefont
  {Pilloni}},\ and\ \bibinfo {author} {\bibfnamefont {A.~P.}\ \bibnamefont
  {Szczepaniak}},\ }\href {https://doi.org/10.1103/PhysRevD.94.034002}
  {\bibfield  {journal} {\bibinfo  {journal} {Phys.Rev.}\ }\textbf {\bibinfo
  {volume} {D94}},\ \bibinfo {pages} {034002} (\bibinfo {year} {2016})},\
  \Eprint {https://arxiv.org/abs/1606.08912} {arXiv:1606.08912 [hep-ph]}
  \BibitemShut {NoStop}%
%%CITATION = ARXIV:1606.08912;%%
\bibitem [{\citenamefont {Winney}\ \emph {et~al.}(2019)\citenamefont {Winney},
  \citenamefont {Fanelli}, \citenamefont {Pilloni}, \citenamefont
  {Hiller~Blin}, \citenamefont {Fern\'andez-Ram\'irez}, \citenamefont
  {Albaladejo}, \citenamefont {Mathieu}, \citenamefont {Mokeev},\ and\
  \citenamefont {Szczepaniak}}]{Winney:2019edt}%
  \BibitemOpen
  \bibfield  {author} {\bibinfo {author} {\bibfnamefont {D.}~\bibnamefont
  {Winney}}, \bibinfo {author} {\bibfnamefont {C.}~\bibnamefont {Fanelli}},
  \bibinfo {author} {\bibfnamefont {A.}~\bibnamefont {Pilloni}}, \bibinfo
  {author} {\bibfnamefont {A.~N.}\ \bibnamefont {Hiller~Blin}}, \bibinfo
  {author} {\bibfnamefont {C.}~\bibnamefont {Fern\'andez-Ram\'irez}}, \bibinfo
  {author} {\bibfnamefont {M.}~\bibnamefont {Albaladejo}}, \bibinfo {author}
  {\bibfnamefont {V.}~\bibnamefont {Mathieu}}, \bibinfo {author} {\bibfnamefont
  {V.~I.}\ \bibnamefont {Mokeev}},\ and\ \bibinfo {author} {\bibfnamefont
  {A.~P.}\ \bibnamefont {Szczepaniak}} (\bibinfo {collaboration} {JPAC}),\
  }\href {https://doi.org/10.1103/PhysRevD.100.034019} {\bibfield  {journal}
  {\bibinfo  {journal} {Phys.Rev.}\ }\textbf {\bibinfo {volume} {D100}},\
  \bibinfo {pages} {034019} (\bibinfo {year} {2019})},\ \Eprint
  {https://arxiv.org/abs/1907.09393} {arXiv:1907.09393 [hep-ph]} \BibitemShut
  {NoStop}%
%%CITATION = ARXIV:1907.09393;%%
\bibitem [{\citenamefont {Abdul~Khalek}\ \emph {et~al.}(2021)\citenamefont
  {Abdul~Khalek} \emph {et~al.}}]{AbdulKhalek:2021gbh}%
  \BibitemOpen
  \bibfield  {author} {\bibinfo {author} {\bibfnamefont {R.}~\bibnamefont
  {Abdul~Khalek}} \emph {et~al.},\ }\Eprint {https://arxiv.org/abs/2103.05419}
  {arXiv:2103.05419 [physics.ins-det]}  (\bibinfo {year} {2021})\BibitemShut
  {NoStop}%
\bibitem [{\citenamefont {Bodwin}\ \emph {et~al.}(1995)\citenamefont {Bodwin},
  \citenamefont {Braaten},\ and\ \citenamefont {Lepage}}]{Bodwin:1994jh}%
  \BibitemOpen
  \bibfield  {author} {\bibinfo {author} {\bibfnamefont {G.~T.}\ \bibnamefont
  {Bodwin}}, \bibinfo {author} {\bibfnamefont {E.}~\bibnamefont {Braaten}},\
  and\ \bibinfo {author} {\bibfnamefont {G.~P.}\ \bibnamefont {Lepage}},\
  }\href {https://doi.org/10.1103/PhysRevD.55.5853, 10.1103/PhysRevD.51.1125}
  {\bibfield  {journal} {\bibinfo  {journal} {Phys.Rev.}\ }\textbf {\bibinfo
  {volume} {D51}},\ \bibinfo {pages} {1125} (\bibinfo {year} {1995})},\
  \bibinfo {note} {[Erratum: Phys.Rev.D55,5853(1997)]},\ \Eprint
  {https://arxiv.org/abs/hep-ph/9407339} {arXiv:hep-ph/9407339 [hep-ph]}
  \BibitemShut {NoStop}%
%%CITATION = HEP-PH/9407339;%%
\bibitem [{\citenamefont {Bignamini}\ \emph {et~al.}(2009)\citenamefont
  {Bignamini}, \citenamefont {Grinstein}, \citenamefont {Piccinini},
  \citenamefont {Polosa},\ and\ \citenamefont {Sabelli}}]{Bignamini:2009sk}%
  \BibitemOpen
  \bibfield  {author} {\bibinfo {author} {\bibfnamefont {C.}~\bibnamefont
  {Bignamini}}, \bibinfo {author} {\bibfnamefont {B.}~\bibnamefont
  {Grinstein}}, \bibinfo {author} {\bibfnamefont {F.}~\bibnamefont
  {Piccinini}}, \bibinfo {author} {\bibfnamefont {A.}~\bibnamefont {Polosa}},\
  and\ \bibinfo {author} {\bibfnamefont {C.}~\bibnamefont {Sabelli}},\ }\href
  {https://doi.org/10.1103/PhysRevLett.103.162001} {\bibfield  {journal}
  {\bibinfo  {journal} {Phys.Rev.Lett.}\ }\textbf {\bibinfo {volume} {103}},\
  \bibinfo {pages} {162001} (\bibinfo {year} {2009})},\ \Eprint
  {https://arxiv.org/abs/0906.0882} {arXiv:0906.0882 [hep-ph]} \BibitemShut
  {NoStop}%
%%CITATION = ARXIV:0906.0882;%%
\bibitem [{\citenamefont {Artoisenet}\ and\ \citenamefont
  {Braaten}(2010)}]{Artoisenet:2009wk}%
  \BibitemOpen
  \bibfield  {author} {\bibinfo {author} {\bibfnamefont {P.}~\bibnamefont
  {Artoisenet}}\ and\ \bibinfo {author} {\bibfnamefont {E.}~\bibnamefont
  {Braaten}},\ }\href {https://doi.org/10.1103/PhysRevD.81.114018} {\bibfield
  {journal} {\bibinfo  {journal} {Phys.Rev.}\ }\textbf {\bibinfo {volume}
  {D81}},\ \bibinfo {pages} {114018} (\bibinfo {year} {2010})},\ \Eprint
  {https://arxiv.org/abs/0911.2016} {arXiv:0911.2016 [hep-ph]} \BibitemShut
  {NoStop}%
%%CITATION = ARXIV:0911.2016;%%
\bibitem [{\citenamefont {Cho}\ \emph {et~al.}(2017)\citenamefont {Cho} \emph
  {et~al.}}]{ExHIC:2017smd}%
  \BibitemOpen
  \bibfield  {author} {\bibinfo {author} {\bibfnamefont {S.}~\bibnamefont
  {Cho}} \emph {et~al.} (\bibinfo {collaboration} {ExHIC}),\ }\href
  {https://doi.org/10.1016/j.ppnp.2017.02.002} {\bibfield  {journal} {\bibinfo
  {journal} {Prog.Part.Nucl.Phys.}\ }\textbf {\bibinfo {volume} {95}},\
  \bibinfo {pages} {279} (\bibinfo {year} {2017})},\ \Eprint
  {https://arxiv.org/abs/1702.00486} {arXiv:1702.00486 [nucl-th]} \BibitemShut
  {NoStop}%
\bibitem [{\citenamefont {Zhang}\ \emph {et~al.}(2020)\citenamefont {Zhang},
  \citenamefont {Liao}, \citenamefont {Wang}, \citenamefont {Wang},\ and\
  \citenamefont {Xing}}]{Zhang:2020dwn}%
  \BibitemOpen
  \bibfield  {author} {\bibinfo {author} {\bibfnamefont {H.}~\bibnamefont
  {Zhang}}, \bibinfo {author} {\bibfnamefont {J.}~\bibnamefont {Liao}},
  \bibinfo {author} {\bibfnamefont {E.}~\bibnamefont {Wang}}, \bibinfo {author}
  {\bibfnamefont {Q.}~\bibnamefont {Wang}},\ and\ \bibinfo {author}
  {\bibfnamefont {H.}~\bibnamefont {Xing}},\ }\href
  {https://doi.org/10.1103/PhysRevLett.126.012301} {\bibfield  {journal}
  {\bibinfo  {journal} {Phys.Rev.Lett.}\ }\textbf {\bibinfo {volume} {126}},\
  \bibinfo {pages} {012301} (\bibinfo {year} {2020})},\ \Eprint
  {https://arxiv.org/abs/2004.00024} {arXiv:2004.00024 [hep-ph]} \BibitemShut
  {NoStop}%
\bibitem [{\citenamefont {Wu}\ \emph {et~al.}(2021)\citenamefont {Wu},
  \citenamefont {Du}, \citenamefont {Sibila},\ and\ \citenamefont
  {Rapp}}]{Wu:2020zbx}%
  \BibitemOpen
  \bibfield  {author} {\bibinfo {author} {\bibfnamefont {B.}~\bibnamefont
  {Wu}}, \bibinfo {author} {\bibfnamefont {X.}~\bibnamefont {Du}}, \bibinfo
  {author} {\bibfnamefont {M.}~\bibnamefont {Sibila}},\ and\ \bibinfo {author}
  {\bibfnamefont {R.}~\bibnamefont {Rapp}},\ }\href
  {https://doi.org/10.1140/epja/s10050-021-00435-6} {\bibfield  {journal}
  {\bibinfo  {journal} {Eur.Phys.J.}\ }\textbf {\bibinfo {volume} {A57}},\
  \bibinfo {pages} {122} (\bibinfo {year} {2021})},\ \Eprint
  {https://arxiv.org/abs/2006.09945} {arXiv:2006.09945 [nucl-th]} \BibitemShut
  {NoStop}%
\bibitem [{\citenamefont {Esposito}\ \emph {et~al.}(2021)\citenamefont
  {Esposito}, \citenamefont {Ferreiro}, \citenamefont {Pilloni}, \citenamefont
  {Polosa},\ and\ \citenamefont {Salgado}}]{Esposito:2020ywk}%
  \BibitemOpen
  \bibfield  {author} {\bibinfo {author} {\bibfnamefont {A.}~\bibnamefont
  {Esposito}}, \bibinfo {author} {\bibfnamefont {E.~G.}\ \bibnamefont
  {Ferreiro}}, \bibinfo {author} {\bibfnamefont {A.}~\bibnamefont {Pilloni}},
  \bibinfo {author} {\bibfnamefont {A.~D.}\ \bibnamefont {Polosa}},\ and\
  \bibinfo {author} {\bibfnamefont {C.~A.}\ \bibnamefont {Salgado}},\ }\href
  {https://doi.org/10.1140/epjc/s10052-021-09425-w} {\bibfield  {journal}
  {\bibinfo  {journal} {Eur.Phys.J.}\ }\textbf {\bibinfo {volume} {C81}},\
  \bibinfo {pages} {669} (\bibinfo {year} {2021})},\ \Eprint
  {https://arxiv.org/abs/2006.15044} {arXiv:2006.15044 [hep-ph]} \BibitemShut
  {NoStop}%
\bibitem [{\citenamefont {Braaten}\ \emph {et~al.}(2021)\citenamefont
  {Braaten}, \citenamefont {He}, \citenamefont {Ingles},\ and\ \citenamefont
  {Jiang}}]{Braaten:2020iqw}%
  \BibitemOpen
  \bibfield  {author} {\bibinfo {author} {\bibfnamefont {E.}~\bibnamefont
  {Braaten}}, \bibinfo {author} {\bibfnamefont {L.-P.}\ \bibnamefont {He}},
  \bibinfo {author} {\bibfnamefont {K.}~\bibnamefont {Ingles}},\ and\ \bibinfo
  {author} {\bibfnamefont {J.}~\bibnamefont {Jiang}},\ }\href
  {https://doi.org/10.1103/PhysRevD.103.L071901} {\bibfield  {journal}
  {\bibinfo  {journal} {Phys.Rev.}\ }\textbf {\bibinfo {volume} {D103}},\
  \bibinfo {pages} {L071901} (\bibinfo {year} {2021})},\ \Eprint
  {https://arxiv.org/abs/2012.13499} {arXiv:2012.13499 [hep-ph]} \BibitemShut
  {NoStop}%
\bibitem [{\citenamefont {Workman}\ \emph {et~al.}(2012)\citenamefont
  {Workman}, \citenamefont {Paris}, \citenamefont {Briscoe},\ and\
  \citenamefont {Strakovsky}}]{Workman:2012jf}%
  \BibitemOpen
  \bibfield  {author} {\bibinfo {author} {\bibfnamefont {R.~L.}\ \bibnamefont
  {Workman}}, \bibinfo {author} {\bibfnamefont {M.~W.}\ \bibnamefont {Paris}},
  \bibinfo {author} {\bibfnamefont {W.~J.}\ \bibnamefont {Briscoe}},\ and\
  \bibinfo {author} {\bibfnamefont {I.~I.}\ \bibnamefont {Strakovsky}},\ }\href
  {https://doi.org/10.1103/PhysRevC.86.015202} {\bibfield  {journal} {\bibinfo
  {journal} {Phys.Rev.}\ }\textbf {\bibinfo {volume} {C86}},\ \bibinfo {pages}
  {015202} (\bibinfo {year} {2012})},\ \Eprint
  {https://arxiv.org/abs/1202.0845} {arXiv:1202.0845 [hep-ph]} \BibitemShut
  {NoStop}%
%%CITATION = ARXIV:1202.0845;%%
\bibitem [{\citenamefont {Efron}\ and\ \citenamefont
  {Tibshirani}(1994)}]{EfroTibs93}%
  \BibitemOpen
  \bibfield  {author} {\bibinfo {author} {\bibfnamefont {B.}~\bibnamefont
  {Efron}}\ and\ \bibinfo {author} {\bibfnamefont {R.}~\bibnamefont
  {Tibshirani}},\ }\href
  {https://www.crcpress.com/An-Introduction-to-the-Bootstrap/Efron-Tibshirani/p/book/9780412042317}
  {\emph {\bibinfo {title} {An Introduction to the Bootstrap}}},\ Chapman \&
  Hall/CRC Monographs on Statistics \& Applied Probability\ (\bibinfo
  {publisher} {Taylor \& Francis},\ \bibinfo {year} {1994})\BibitemShut
  {NoStop}%
\bibitem [{\citenamefont {James}\ \emph {et~al.}(2017)\citenamefont {James},
  \citenamefont {Witten}, \citenamefont {Hastie},\ and\ \citenamefont
  {Tibshirani}}]{StatisticalLearning}%
  \BibitemOpen
  \bibfield  {author} {\bibinfo {author} {\bibfnamefont {G.}~\bibnamefont
  {James}}, \bibinfo {author} {\bibfnamefont {D.}~\bibnamefont {Witten}},
  \bibinfo {author} {\bibfnamefont {T.}~\bibnamefont {Hastie}},\ and\ \bibinfo
  {author} {\bibfnamefont {R.}~\bibnamefont {Tibshirani}},\ }\href
  {https://faculty.marshall.usc.edu/gareth-james/ISL/} {\emph {\bibinfo {title}
  {An Introduction to Statistical Learning}}}\ (\bibinfo  {publisher}
  {Springer},\ \bibinfo {year} {2017})\BibitemShut {NoStop}%
\bibitem [{\citenamefont {Guegan}\ \emph {et~al.}(2015)\citenamefont {Guegan},
  \citenamefont {Hardin}, \citenamefont {Stevens},\ and\ \citenamefont
  {Williams}}]{Guegan:2015mea}%
  \BibitemOpen
  \bibfield  {author} {\bibinfo {author} {\bibfnamefont {B.}~\bibnamefont
  {Guegan}}, \bibinfo {author} {\bibfnamefont {J.}~\bibnamefont {Hardin}},
  \bibinfo {author} {\bibfnamefont {J.}~\bibnamefont {Stevens}},\ and\ \bibinfo
  {author} {\bibfnamefont {M.}~\bibnamefont {Williams}},\ }\href
  {https://doi.org/10.1088/1748-0221/10/09/P09002} {\bibfield  {journal}
  {\bibinfo  {journal} {JINST}\ }\textbf {\bibinfo {volume} {10}}\bibfield
  {number} {\bibinfo  {number} { (09)},\ \bibinfo {pages} {P09002}},\ }\Eprint
  {https://arxiv.org/abs/1505.05133} {arXiv:1505.05133 [physics.data-an]}
  \BibitemShut {NoStop}%
\bibitem [{\citenamefont {Landay}\ \emph {et~al.}(2017)\citenamefont {Landay},
  \citenamefont {D\"oring}, \citenamefont {Fern\'andez-Ram\'irez},
  \citenamefont {Hu},\ and\ \citenamefont {Molina}}]{Landay:2016cjw}%
  \BibitemOpen
  \bibfield  {author} {\bibinfo {author} {\bibfnamefont {J.}~\bibnamefont
  {Landay}}, \bibinfo {author} {\bibfnamefont {M.}~\bibnamefont {D\"oring}},
  \bibinfo {author} {\bibfnamefont {C.}~\bibnamefont {Fern\'andez-Ram\'irez}},
  \bibinfo {author} {\bibfnamefont {B.}~\bibnamefont {Hu}},\ and\ \bibinfo
  {author} {\bibfnamefont {R.}~\bibnamefont {Molina}},\ }\href
  {https://doi.org/10.1103/PhysRevC.95.015203} {\bibfield  {journal} {\bibinfo
  {journal} {Phys.Rev.}\ }\textbf {\bibinfo {volume} {C95}},\ \bibinfo {pages}
  {015203} (\bibinfo {year} {2017})},\ \Eprint
  {https://arxiv.org/abs/1610.07547} {arXiv:1610.07547 [nucl-th]} \BibitemShut
  {NoStop}%
%%CITATION = ARXIV:1610.07547;%%
\bibitem [{\citenamefont {Sombillo}\ \emph {et~al.}(2020)\citenamefont
  {Sombillo}, \citenamefont {Ikeda}, \citenamefont {Sato},\ and\ \citenamefont
  {Hosaka}}]{Sombillo:2020ccg}%
  \BibitemOpen
  \bibfield  {author} {\bibinfo {author} {\bibfnamefont {D.~L.~B.}\
  \bibnamefont {Sombillo}}, \bibinfo {author} {\bibfnamefont {Y.}~\bibnamefont
  {Ikeda}}, \bibinfo {author} {\bibfnamefont {T.}~\bibnamefont {Sato}},\ and\
  \bibinfo {author} {\bibfnamefont {A.}~\bibnamefont {Hosaka}},\ }\href
  {https://doi.org/10.1103/PhysRevD.102.016024} {\bibfield  {journal} {\bibinfo
   {journal} {Phys.Rev.}\ }\textbf {\bibinfo {volume} {D102}},\ \bibinfo
  {pages} {016024} (\bibinfo {year} {2020})},\ \Eprint
  {https://arxiv.org/abs/2003.10770} {arXiv:2003.10770 [hep-ph]} \BibitemShut
  {NoStop}%
\bibitem [{\citenamefont {Sombillo}\ \emph
  {et~al.}(2021{\natexlab{a}})\citenamefont {Sombillo}, \citenamefont {Ikeda},
  \citenamefont {Sato},\ and\ \citenamefont {Hosaka}}]{Sombillo:2021rxv}%
  \BibitemOpen
  \bibfield  {author} {\bibinfo {author} {\bibfnamefont {D.~L.~B.}\
  \bibnamefont {Sombillo}}, \bibinfo {author} {\bibfnamefont {Y.}~\bibnamefont
  {Ikeda}}, \bibinfo {author} {\bibfnamefont {T.}~\bibnamefont {Sato}},\ and\
  \bibinfo {author} {\bibfnamefont {A.}~\bibnamefont {Hosaka}},\ }\href
  {https://doi.org/10.1103/PhysRevD.104.036001} {\bibfield  {journal} {\bibinfo
   {journal} {Phys.Rev.}\ }\textbf {\bibinfo {volume} {D104}},\ \bibinfo
  {pages} {036001} (\bibinfo {year} {2021}{\natexlab{a}})},\ \Eprint
  {https://arxiv.org/abs/2105.04898} {arXiv:2105.04898 [hep-ph]} \BibitemShut
  {NoStop}%
\bibitem [{\citenamefont {Sombillo}\ \emph
  {et~al.}(2021{\natexlab{b}})\citenamefont {Sombillo}, \citenamefont {Ikeda},
  \citenamefont {Sato},\ and\ \citenamefont {Hosaka}}]{Sombillo:2021yxe}%
  \BibitemOpen
  \bibfield  {author} {\bibinfo {author} {\bibfnamefont {D.~L.~B.}\
  \bibnamefont {Sombillo}}, \bibinfo {author} {\bibfnamefont {Y.}~\bibnamefont
  {Ikeda}}, \bibinfo {author} {\bibfnamefont {T.}~\bibnamefont {Sato}},\ and\
  \bibinfo {author} {\bibfnamefont {A.}~\bibnamefont {Hosaka}},\ }\Eprint
  {https://arxiv.org/abs/2104.14182} {arXiv:2104.14182 [hep-ph]}  (\bibinfo
  {year} {2021}{\natexlab{b}})\BibitemShut {NoStop}%
\bibitem [{\citenamefont {Yoon}\ \emph {et~al.}(2018)\citenamefont {Yoon},
  \citenamefont {Sim},\ and\ \citenamefont {Han}}]{Yoon:2018}%
  \BibitemOpen
  \bibfield  {author} {\bibinfo {author} {\bibfnamefont {H.}~\bibnamefont
  {Yoon}}, \bibinfo {author} {\bibfnamefont {J.-H.}\ \bibnamefont {Sim}},\ and\
  \bibinfo {author} {\bibfnamefont {M.~J.}\ \bibnamefont {Han}},\ }\href
  {https://doi.org/10.1103/PhysRevB.98.245101} {\bibfield  {journal} {\bibinfo
  {journal} {Phys.Rev.}\ }\textbf {\bibinfo {volume} {B98}},\ \bibinfo {pages}
  {245101} (\bibinfo {year} {2018})},\ \Eprint
  {https://arxiv.org/abs/1806.03841} {arXiv:1806.03841 [cond-mat.str-el]}
  \BibitemShut {NoStop}%
\bibitem [{\citenamefont {Fournier}\ \emph {et~al.}(2018)\citenamefont
  {Fournier}, \citenamefont {Wang}, \citenamefont {Yazyev},\ and\ \citenamefont
  {Wu}}]{Fournier:2018}%
  \BibitemOpen
  \bibfield  {author} {\bibinfo {author} {\bibfnamefont {R.}~\bibnamefont
  {Fournier}}, \bibinfo {author} {\bibfnamefont {L.}~\bibnamefont {Wang}},
  \bibinfo {author} {\bibfnamefont {O.~V.}\ \bibnamefont {Yazyev}},\ and\
  \bibinfo {author} {\bibfnamefont {Q.}~\bibnamefont {Wu}},\ }\href
  {https://doi.org/10.1103/PhysRevLett.124.056401} {\bibfield  {journal}
  {\bibinfo  {journal} {Phys.Rev.Lett.}\ }\textbf {\bibinfo {volume} {124}},\
  \bibinfo {pages} {056401} (\bibinfo {year} {2018})},\ \Eprint
  {https://arxiv.org/abs/1810.00913} {arXiv:1810.00913 [physics.comp-ph]}
  \BibitemShut {NoStop}%
\bibitem [{\citenamefont {Rojo}\ and\ \citenamefont
  {Latorre}(2004)}]{Rojo:2004iq}%
  \BibitemOpen
  \bibfield  {author} {\bibinfo {author} {\bibfnamefont {J.}~\bibnamefont
  {Rojo}}\ and\ \bibinfo {author} {\bibfnamefont {J.~I.}\ \bibnamefont
  {Latorre}},\ }\href {https://doi.org/10.1088/1126-6708/2004/01/055}
  {\bibfield  {journal} {\bibinfo  {journal} {JHEP}\ }\textbf {\bibinfo
  {volume} {01}},\ \bibinfo {pages} {055}},\ \Eprint
  {https://arxiv.org/abs/hep-ph/0401047} {arXiv:hep-ph/0401047} \BibitemShut
  {NoStop}%
\bibitem [{\citenamefont {Ball}\ \emph {et~al.}(2021)\citenamefont {Ball} \emph
  {et~al.}}]{NNPDF:2021uiq}%
  \BibitemOpen
  \bibfield  {author} {\bibinfo {author} {\bibfnamefont {R.~D.}\ \bibnamefont
  {Ball}} \emph {et~al.} (\bibinfo {collaboration} {NNPDF}),\ }\href
  {https://doi.org/10.1140/epjc/s10052-021-09747-9} {\bibfield  {journal}
  {\bibinfo  {journal} {Eur.Phys.J.}\ }\textbf {\bibinfo {volume} {C81}},\
  \bibinfo {pages} {958} (\bibinfo {year} {2021})},\ \Eprint
  {https://arxiv.org/abs/2109.02671} {arXiv:2109.02671 [hep-ph]} \BibitemShut
  {NoStop}%
\end{thebibliography}%

\end{document}